\documentclass[11pt]{article}
\usepackage{jcapmod}

\usepackage{mathtools}
\usepackage{booktabs}
\usepackage[english]{babel}
\usepackage{amsmath,amssymb,amsbsy,amstext, amsthm, amsfonts}
\usepackage{hyperref}
\usepackage{graphicx}
\usepackage[small]{caption}
\usepackage{siunitx}
\usepackage{framed}
\usepackage{wrapfig}
\usepackage{multirow}
\usepackage[svgnames,dvipsnames,x11names]{xcolor}
\usepackage{selinput}
\usepackage{bm}
\usepackage{float}
\usepackage{geometry}
\usepackage{yfonts}
\usepackage{caption}
\usepackage{subcaption}
\usepackage{subfloat}
\usepackage{sidecap}
\usepackage{longtable}
\usepackage{physics}
\usepackage[compat=1.0.0]{tikz-feynman}
\usepackage{tikz}
\usetikzlibrary{shapes.geometric}
\renewcommand{\d}{\partial}

\newcommand{\m}{\mu}
\newcommand{\n}{\nu}

\usepackage{colortbl}
\definecolor{summersky}{cmyk}{0.71,0.33,0,0.5}
\definecolor{flamingo}{cmyk}{0,0.51,0.71,0.5}
\definecolor{rp}{cmyk}{0.2, 1, 0.6, 0}
\definecolor{pacificblue}{cmyk}{0.95,0.3,0, 0.5}
\definecolor{gray60}{cmyk}{0.4,0.4,0,0.8}

\usepackage[framemethod=default]{mdframed}
\newmdenv[skipabove=7pt,
skipbelow=7pt,
rightline=false,
leftline=false,
topline=false,
bottomline=false,
backgroundcolor=pacificblue!10,
linecolor=gray,
innerleftmargin=5pt,
innerrightmargin=5pt,
innertopmargin=2pt,
innerbottommargin=10pt,
leftmargin=0cm,
rightmargin=0cm,
linewidth=4pt]{eBox}
\newmdenv[skipabove=7pt,
skipbelow=7pt,
rightline=false,
leftline=false,
topline=false,
bottomline=false,
backgroundcolor=gray!10,
linecolor=gray,
innerleftmargin=5pt,
innerrightmargin=5pt,
innertopmargin=-5pt,
innerbottommargin=5pt,
leftmargin=0cm,
rightmargin=0cm,
linewidth=4pt]{eBox2}

\definecolor{blue3}{RGB}{31, 119, 180}
\definecolor{red3}{RGB}{	214, 39, 40}
\definecolor{orange3}{RGB}{255, 127, 14}
\definecolor{green3}{RGB}{44, 160, 44}
\definecolor{repBlue}{RGB}{31, 119, 180}
\definecolor{repRed}{RGB}{	214, 39, 40}
\definecolor{repGreen}{RGB}{44, 160, 44}
\definecolor{MyRed}{RGB}{208,57,75}
\definecolor{MyBlue}{RGB}{68,123,178}
\definecolor{MyYellow}{RGB}{238,140,59}

\renewcommand{\(}{\left(}
\renewcommand{\)}{\right)}
\renewcommand{\[}{\left[}
\renewcommand{\]}{\right]}

\def\be{\begin{equation}}
\def\ee{\end{equation}}
\newcommand{\bea}{\begin{eqnarray}}
\newcommand{\eea}{\end{eqnarray}}

\newcommand{\rom}[1]{\uppercase\expandafter{\romannumeral #1\relax}}%Roman number

\makeatletter

\newcommand{\Rmnum}[1]{\expandafter\@slowromancap\romannumeral #1@}
\makeatother

\usepackage{colortbl}
\definecolor{lightgreen}{cmyk}{0.2, 0, 0.2, 0.2}
\definecolor{lightgray}{cmyk}{0.1,0.2,0,0.1}
\definecolor{lightgray2}{cmyk}{0.1,0.1,0,0.1}

\makeatletter
\newlength{\apb@width}
\newcommand{\autoparbox}[2][c]{\settowidth{\apb@width}{#2}\parbox[#1]{\apb@width}{#2}}

\makeatother


\def\beq{\begin{equation}}
\def\eeq{\end{equation}}

\def\k{\vec k}

\allowdisplaybreaks[1]
\title{
Revisiting  the
matching of black hole tidal responses: 
a systematic study of relativistic and 
logarithmic corrections
}

\author[1a]{Mikhail M. Ivanov,\note{ivanov@ias.edu}}
\author[2b]{Zihan Zhou\note{zihanz@princeton.edu}}

\affiliation[a]{School of Natural Sciences, Institute for Advanced Study, 1 Einstein Drive, Princeton, NJ 08540, USA}
\affiliation[b]{Department of Physics, Princeton University, Princeton, NJ 08540, USA}

\date{}

\abstract{
The worldline effective field theory (EFT) 
gives a gauge-invariant definition 
of black hole conservative tidal responses (Love 
numbers), dissipation numbers, and their spin-0 and spin-1 analogs. 
In the first part of this paper 
we show how the EFT allows us to circumvent 
the source/response ambiguity 
without 
having to use the analytic continuation prescription. 
The source/response ambiguity 
appears if relativistic
corrections to external sources overlap 
with the response. 
However, these corrections can be 
clearly 
identified and isolated  
using the EFT.
We illustrate that by
explicitly 
computing static one-point functions of various external fields 
perturbing the four-dimensional Schwarzschild geometry.
Upon resumming all relevant Feynman diagrams,
we find that the relativistic terms that may mimic the response
actually vanish for static black holes. 
Thus, the extraction of Love numbers
from matching the EFT and general relativity (GR) calculations
is completely unambiguous,
and it confirms 
previous results 
that the Love numbers vanish identically 
for all types of perturbations. 
We also study in detail another type of fine-tuning in the EFT,
the absence of Love numbers' running. 
We show that logarithmic 
corrections to Love numbers do stem from individual loop diagrams
in generic gauges, but cancel after all diagrams are summed over.
In the particular cases of spin-0 and spin-2 fields the logarithms 
are completely absent if one uses the Kaluza-Klein metric decomposition.
In the second part of the paper we compute frequency-dependent 
dissipative response contributions to the one-point functions
using the Schwinger-Keldysh formalism. 
We extract black hole
dissipation numbers by comparing the one-point functions
in the EFT and GR. 
Our results
are in perfect agreement with those 
obtained from a manifestly gauge-invariant
matching of absorption cross sections.  
}

\begin{document}

\maketitle

\section{Introduction and Main Results}
\label{sec:int}

\subsection*{Background}

The detection of gravitational wave signals 
with the LIGO/VIRGO interferometer have started the era of 
precision strong field gravity~\cite{LIGOScientific:2016aoc}. This remarkable experimental 
success has also motivated many new theoretical studies. 
One of the key parameters 
affecting the shape of the gravitational wave
signal are tidal deformability coefficients, called
Love numbers. In the context of neutron stars, the 
measurement of Love numbers offers a way to probe the 
neutron star equation of state~\cite{Flanagan:2007ix, Vines:2011ud, Bini:2012gu,LIGOScientific:2017vwq}.
As far as black holes (BHs) are concerned, their Love 
numbers have been found to vanish 
identically in four dimensions, which has 
interesting phenomenological and 
theoretical implications.
In particular, Love numbers appear as Wilson 
coefficients in the point-particle worldline effective field
theory. 
Hence, their vanishing implies a fine-tuning 
problem that is reminiscent of the notorious cosmological 
constant problem~\cite{Porto:2016zng}. 

The effective field theory (EFT) of gravitational wave sources
is a theoretical tool for systematic calculations 
of gravitational waveforms~\cite{Goldberger:2004jt,Goldberger:2007hy,Porto:2005ac,Porto:2016pyg,Levi:2018nxp,Goldberger:2022ebt}. 
Within the EFT each compact  
object of an inspiraling binary is represented as 
an effective point particle. The finite-size
structure is then captured by means of higher-derivative 
effective worldline couplings. This approach is similar to the 
multipole expansion in classical electrodynamics. 
The leading finite-size effects of compact objects are
captured by worldline operators quadratic in curvature.
The corresponding Wilson coefficients
can be shown to reduce to ``classical'' Love numbers in 
the Newtonian limit. As mentioned above, the black hole Love numbers
are zero, which means that the worldline EFT 
exhibits a strong fine-tuning when applied to black holes. 

% The
% Love numbers appear as Wilson coefficients
% in the EFT finite-size action. 
% Their vanishing thus implies 
% that the EFT exhibits a strong fine-tuning  
% when applied to black holes.
% the EFT exhibits a 
% strong fine tuning

In four dimensions, the Love numbers were shown to vanish 
for both the Schwarzschild  
\mbox{(static)~\cite{Fang:2005qq,Damour:2009vw,Binnington:2009bb,Kol:2011vg}}
and rotating (Kerr) 
BHs~\cite{Landry:2015zfa,LeTiec:2020bos,Chia:2020yla,Charalambous:2021mea}. 
The situation is more intricate 
for higher dimensional static BHs, 
where Love numbers can vanish, be order-one constant, 
or exhibit classical renormalization group running
depending on the multipole index $\ell$ and the number of spacetime 
dimensions~ \cite{Kol:2011vg,Hui:2020xxx}. This behavior is also 
quite unnatural from the 
perspective of Wilsonian naturalness. 
Importantly, spin-0 and spin-1 analogs of Love numbers
also follow the same patterns as the spin-2 gravitational perturbations.
This hints that the vanishing of Love numbers
should have a general geometric origin.
Recently, the naturalness paradox associated with the strange behavior
of Love numbers has been addressed by a new symmetry of 
general relativity called 
the Love symmetry~\cite{Charalambous:2021kcz}
(see~\cite{Hui:2021vcv,Hui:2022vbh} for alternative proposals).

The literature on BH Love numbers is vast, 
but there are certain conceptual and technical 
difficulties that are yet to be addressed.
Broadly, these are the problems with the definition
of Love numbers in general relativity, and the
extraction of Love numbers from black hole 
perturbation theory (BHPT) calculations.
The main goal of this paper is to show how these problems 
can be resolved in the worldline EFT approach.
Let us describe these problems in more detail. 

\textbf{Definition of Love numbers in GR. Source/response ambiguity.} Love numbers were originally 
defined in the context of Newtonian gravity. 
Imagine a nonrotating fluid star of mass $M$ perturbed by an external tidal  
field of a small body. In the absence of the external perturbation
the star would be spherical. The tidal forces, however, deform 
the star, and it acquires internal multipole moments $I_L$.
The total gravitational potential around the star will look like~\cite{poisson2014gravity} 
\be 
      \phi(\boldsymbol{x}) = \frac{M}{r} 
      - \sum_{\ell = 2} 
       \left[   \frac{(\ell-1)!}{\ell!}\mathcal{E}_L n^L r^\ell
       -\frac{(2\ell-1)!!}{\ell!} \frac{I_L n^L}{r^{\ell+1}}
       \right]
      \,,
\ee
where $\mathcal{E}_L$ are the multipole moments of the tidal potential, 
$L=i_1\cdots i_\ell$ is the multi-index,
and $n^L = n^{i_1}...n^{i_\ell}$ is the tensor product of unit 
direction vectors $n^i = x^i/r$. In linear response theory 
the induced mass multipoles must be proportional to external 
perturbations, $I_L = -k_\ell R^{2\ell+1}\frac{(\ell-2)!}{(2\ell-1)!!} \mathcal{E}_L$, where $R$ is the size of the star that we insert 
in accordance with dimensional analysis. 
The total potential 
then takes the form
\be \label{eq:Lnew}
      \phi^{\rm pert.}(\boldsymbol{x}) = 
      % \frac{M}{r} 
      - \sum_{\ell = 2} 
      \frac{(\ell-1)!}{\ell!}   \mathcal{E}_L n^L r^\ell \left[   
      \textcolor{red}{\underbrace{
   1
      }_{\text{source}}}\quad 
      +\quad  \textcolor{blue}{\underbrace{k_\ell \frac{R^{2\ell+1} }{r^{2\ell+1}}}_{\text{response}}}
       \right]
      \,,
\ee
where we have subtracted the monopole component.
We stress that there is a clear 
separation between the source and response contributions
in the Newtonian theory. The Love number is a coefficient 
in front of the $r^{-\ell-1}$ term in the Newtonian potential 
profile. Since we will be discussing BHs, we replace $R\to r_s$ (Schwarzschild radius)
in what follows.

There are several ways to define the Love number
in general relativity. Ideally, one wants a definition that would be 
gauge and coordinate invariant, and that would also reproduce 
Eq.~\eqref{eq:Lnew} in the Newtonian limit. One common way is to 
extend the expression~\eqref{eq:Lnew}
to full general relativity. 
For instance, one may look at the temporal metric component
$h_{00}=(g_{00}-1)/2$ in the body's local asymptotic rest frame~\cite{Thorne:1984mz},
which takes the following form~\cite{Kol:2011vg}:
\be 
\label{eq:PNprof}
    \begin{aligned}
        h_{00}^{\rm pert}(\boldsymbol{x}) =  
        \sum_{\ell = 2 }
        \frac{(\ell-1)!}{\ell!}   \mathcal{E}_L n^L r^\ell
        \Bigg[&
        \textcolor{red}{
        \underbrace{\left(1+ c_1 \left(\frac{r_s}{r}\right) + \cdots \right)}_{\text{source}}
        }
       \\
       & + 
        \textcolor{blue}{
         \underbrace{
k_{\ell }\left(\frac{R}{r}\right)^{2\ell +1 }
        \left(1+ b_1 \left(\frac{r_s}{r}\right) + \cdots \right)
        }_{\text{response}}
        }
        % +k_{\ell }
        % \(\frac{R}{r}\)^{2\ell +1 }}\( 1 + b_1 \(\frac{r_s}{r}\) + \cdots \)
     %    \red{ \underbrace{\}_{source}} \\
     %    & \quad +
     % \underbrace{ k_{\ell }\(\frac{R}{r}\)^{2\ell +1 }}\( 1 + b_1 \(\frac{r_s}{r}\) + \cdots \)}_{response}
        \Bigg]\,,     
    \end{aligned}
\ee
where $c_1,b_1$ are some calculable  $\mathcal{O}(1)$ coefficients.
The terms in the first line above represent post-Newtonian (PN)
corrections\footnote{Physically, we are interested in a situation when the external source is a companion object in the binary. In this case $\frac{r_s}{r}\sim v^2$ is a PN parameter.
By this reason, we will call an expansion in $r_s/r$ ``Post-Newtonian''
in this paper,  although we never explicitly assume
that the external source of tides and the BH are bound objects.  } to the source generated by 
gravitational nonlinearity. We 
call them the ``source series.'' 
The second line above contains the response contribution
plus PN corrections to it. We call them the ``response series.''
An obvious problem with the above 
definition is a possible ambiguity due to an overlap 
between the source and response series~\cite{Kol:2011vg,LeTiec:2020spy,Charalambous:2021mea}. 
If this is the case, 
the coefficient in front of the 
$r^{-\ell - 1}$ term in the generalized Newtonian potential is actually given by 
\be 
\label{eq:c2l1}
\textcolor{red}{c_{2\ell + 1}}+\textcolor{blue}{ k_{\ell m} }\,.
\ee
Now it is not uniquely defined by the Love number.
A popular way to get around this ambiguity is to do an 
analytic continuation for $\ell$ from the physical region $\ell \in \mathbb{N}$ to the unphysical region $\ell \in \mathbb{R}$ \cite{Kol:2011vg,LeTiec:2020spy,Chia:2020yla,Charalambous:2021mea}. This is motivated by
the observation that 
for general noninteger $\ell$ the source and response series 
in Eq.~\eqref{eq:PNprof}
do not overlap.
This is generically true for 
BH perturbations in a number of 
spacetime dimension
greater than four~\cite{Kol:2011vg}.
Some physical interpretation of this procedure in four 
dimensions is given 
in Refs.~\cite{Chia:2020yla,Charalambous:2021mea}
in the context of the renormalized angular momentum~\cite{Mano:1996mf,Mano:1996vt,Mano:1996gn}. 
The analytic continuation of the angular multipole number is, however, 
still
an ad hoc prescription whose validity is not under rigorous 
theoretical control. 
In addition, the analytic continuation
can work only when it is possible to 
obtain a closed perturbative solution for a generic $\ell$. 
This may not be the case
for some modified gravity theories,  e.g.,~\cite{Cardoso:2017cfl,Cai:2019npx}.
This makes it desirable to develop
a systematic and controlled approach that does not 
rely on the analytic continuation prescription. 

\textbf{Gauge invariance.}
The second problem with the definition~\eqref{eq:PNprof} 
is that it is
given in a particular coordinate system. 
Hence, it is not obvious that this definition is gauge invariant~\cite{Gralla:2017djj}.

\textbf{Logarithmic corrections.}
In principle,
the functional form of the relevant field profiles
can be more complicated than~\eqref{eq:PNprof}. In particular, there could be 
logs multiplying the $r^{-\ell-1}$ term~\cite{Kol:2011vg,Charalambous:2021mea}.
This additionally obscures the non-EFT definitions of the Love numbers.

\textbf{Dissipation numbers.}
The final ambiguity associated with~\eqref{eq:PNprof} 
is the interpretation of the coefficient 
in front of the $r^{-\ell-1}$ term. In general, the tidal response
has conservative (time-reversal even) and dissipative (time-reversal odd)
corrections. For Schwarzschild BHs this means that 
the total time-dependent response can be written as 
\be\label{eq:kfull}
k_{\ell } (\omega)\equiv k_{\ell } +  i\nu_{\ell }r_s\omega + \mathcal{O}(r_s^2\omega^2)\,,
\ee
where $\nu_{\ell }$ is the dissipation number (dissipative response coefficient) and $\omega$
is the frequency in the BH's rest frame. 
In the Newtonian theory of fluid stars $\nu_{\ell }$ is proportional 
to the fluid's viscosity~\cite{poisson2014gravity}.
Note that this contribution vanishes in the static regime. 
The situation is more
complicated in the case of Kerr black holes, where 
due to frame dragging, one needs to replace 
$\omega \to \omega- m\Omega$ ($\Omega$ and $m$ being the angular velocity 
of the black hole horizon and the magnetic number, respectively), so that the dissipation is present 
even for static external sources~\cite{Chia:2020yla,Goldberger:2020fot,Charalambous:2021mea}.
This effect has caused some confusion 
in the previous literature; 
cf.~\cite{Poisson:2014gka,Landry:2015zfa,LeTiec:2020spy,LeTiec:2020bos,Chia:2020yla,Poisson:2020vap}. 

However, many of these ambiguities can be addressed within the context of worldline EFT~\cite{Goldberger:2004jt,Goldberger:2007hy,Porto:2005ac,Porto:2016pyg,Levi:2018nxp,Goldberger:2022ebt}.
In this theory, Love numbers as defined as Wilson coefficients of the static finite-size action
\be
\label{eq:LN1}
\begin{split}
& S_{\text{finite size}}=\sum_{\ell=2}\frac{\lambda_\ell}{2\ell!}\int d\tau ~E_{a_1\cdots a_\ell} E^{a_1\cdots a_\ell}\,, \\ 
& E^{a_1\cdots a_\ell} =
% \frac{1}{(\ell-2)!} 
e^{a_1}_{\mu_1}\cdots e^{a_\ell}_{\mu_\ell}  
\nabla^{\langle \mu_2} \cdots \nabla^{\mu_\ell}C^{\mu_1 |\alpha| \mu_2  |\beta|\rangle}v_\alpha v_\beta\,,
\end{split}
\ee 
where $C_{\mu\alpha \nu\beta}$
is the Weyl tensor, $v^\alpha$ is the point particle four-velocity,
and $e^a_\mu$ are vectors defining a frame orthogonal to $v^\alpha$.
$\langle \cdots \rangle$ denotes the procedure of symmetrization 
and subtracting 
traces.
In the definition~\eqref{eq:LN1} $E^{a_1\cdots a_\ell}$
are multipole moments measured in the BH
frame, $a=1,2,3$ are the $SO(3)$ indices. One can perform a linear response calculation 
with the action~\eqref{eq:LN1} and find that 
upon identification $\lambda_\ell = (-1)^\ell r_s^{2\ell+1} k_\ell (\pi^{1/2}2^\ell/\Gamma(1/2-\ell))$,
it precisely 
reproduces Eq.~\eqref{eq:Lnew} in the Newtonian limit~\cite{Charalambous:2021mea}.

Within the EFT all gravitational corrections 
are computed as relativistic perturbations around the flat background~\cite{Goldberger:2004jt,Kol:2007rx,Kol:2009mj}. 
In this regard the EFT is  sometimes referred to as nonrelativistic 
general relativity. 
In this approach, the PN corrections in the 
source series of~\eqref{eq:PNprof} are just classical 
nonlinear
graviton corrections 
to the external source profile. 
These corrections, i.e.,
the coefficient $c_{2\ell+1}$, 
can be computed explicitly.
After that the whole external field profile 
can be matched to a corresponding BHPT calculation. 
% In other words, 
Therefore, the EFT 
allows us to get around the source/response ambiguity
without having to use the analytic continuation of the multipole index.

To extract the Love numbers, one needs to match EFT 
and full general relativity (GR) (or BHPT) calculations. 
The cleanest way to do so is to compare two 
gauge-invariant observables, such as a cross section of 
the elastic scattering of gravitational waves off a BH geometry. 
For that one needs to know this cross section at least 
at the 5 post-Minkowskian (PM) order~\cite{Goldberger:2004jt,Porto:2016pyg}.\footnote{See~\cite{Ivanov:2022qqt} for a recent discussion and simplifications 
in the context of the near/far zone factorization 
of the scattering amplitude.}
However, there is a simper way to obtain the Love numbers: 
one can match static graviton one-point functions such as ~\eqref{eq:PNprof}.  
This procedure is delicate as it is done 
in a particular coordinate system.

To overcome possible issue with coordinate dependence, one needs to make sure that 
the one-point function calculations
on both EFT and UV sides 
are carried out in consistent gauges. 
In this paper, we define an EFT gauge to be consistent with the 
background geometry if the background EFT  
one-point function of gravitons
(i.e., without external fields)
coincides with a perturbatively expanded 
full geometry. 
Once we have specified 
a consistent gauge, we can match the full one-point
function including external perturbations, 
which will then give us the Love numbers. 
Since the EFT Wilson coefficients are universal, 
results would be gauge-independent even if some specific
one-point functions are used for the matching are not.

As far as logarithmic corrections are concerned, 
they can easily be incorporated within the EFT 
and interpreted as a classical
renormalization group
(RG) running. In particular, the authors of \cite{Kol:2011vg} have carried out such a matching calculation for a scalar (dilaton)
perturbation with a quadrupolar source. In this work, 
we build on the ideas of~\cite{Kol:2011vg}
and investigate the logarithmic running 
for a general multipolar index $\ell$.
As we discuss later, we also find some diagrams that were omitted in~\cite{Kol:2011vg}, and we argue why this did not affect their 
results. The authors of \cite{Kol:2011vg}
also proposed a 
symmetry explanation of the absence of logarithmic 
running of Schwarzschild Love numbers in D=4,
which we thoroughly scrutinize in our work.
% Despite such pioneer work, some ingredients still need to be added. Part of the goal of this paper is to generalize the ideas in \cite{Kol:2011vg} to do rigorous matching with GR, including RG runnings. 
% First, in \cite{Kol:2011vg}, several diagrams are missing in the EFT loop calculations. The authors only consider a particular type of loop formed by dilatons and tensorial fields, which is insufficient to match the full GR one-point functions. In this paper, we fix this problem by considering all possible diagrams. 
% Secondly, \cite{Kol:2011vg} only considers the scalar field perturbations with quadrupole $
% \ell = 2$ source. 
% This is insufficient to conclude vanishing Love numbers in 4D for all $\ell$. Here, we use the Feynman diagram recurrence relation techniques to generalize to all $\ell \in \mathbb{N}$, including spin-0 scalar, spin-1 electric, and spin-2 dilaton perturbations. Thirdly, the authors of \cite{Kol:2011vg} conjectured a wrong low-energy (IR) $Z_2$ symmetry for dilaton field $\phi \rightarrow - \phi$, which seems to protect the Love number running. However, in this paper, we shall point out that this is no longer true for spin-1 perturbations. The vanishing of the Love number running can only be seen after summing all possible loops.

\subsection*{Summary of Main Results}

The result of this paper is summarized as follows.

\begin{itemize}
    \item \textbf{Systematic study of PN corrections.}: We generalize the formalism of \cite{Kol:2011vg} by computing static one-point functions (profiles)
of generic external fields perturbing the Schwarzschild geometry
in the EFT. We carry out explicit calculations in terms of EFT 
Feynman diagrams for the 
electric-type (parity even) 
spin-0 and scalar graviton\footnote{With some abuse of notation, we will refer to the scalar graviton (or dilaton, or the generalized Newtonian potential) field as a ``spin-2 field.'' This is because the  Love number for the scalar graviton 
is the same as the Love number for the actual 
spin-2 metric field.} fluctuations 
in a general multipole sector $\ell$.
For the spin-1 case, we limit ourselves to the dipole
sector $\ell=1$. We show that in all these cases
$c_{2\ell+1}=0$,
which implies that there is no mixing between the source and 
response contributions in the Schwarzschild
background. 
By comparing our EFT expressions with the BHPT results,
we confirm that the Schwarzschild BH Love numbers vanish identically.
This is yet another confirmation of the gauge independence 
of the vanishing of Love numbers.

We have discovered a diagrammatic recurrence relation
that haw allowed us to resum all PN external source diagrams
for a generic multipolar index $\ell$. Our EFT diagrammatic 
recurrence relation matches 
the Frobenius series expansion of the 
relevant BHPT solutions. 
This allows us to completely reconstruct 
the BHPT results for 
one-point functions using the EFT. 
This extends and generalizes 
the result previously obtained in~\cite{Kol:2011vg}, which 
gave an EFT interpretation of the 
vanishing of 
quadrupole-type spin-0 Love numbers
and their RG running.

\item  \textbf{Detailed study of nonrenormalization of Love numbers}: With the Feynman diagram techniques mentioned above, we conduct a detailed analysis of logarithmic corrections to the Love numbers (i.e., their RG running).
In the context of Wilsonian naturalness, 
one may expect that the classical RG running of 
Wilson coefficients should be a generic phenomenon. 
Indeed, we will confirm this expectation by showing that generic individual EFT loop corrections
to the Love numbers do
produce some logarithmic running for arbitrary gauge choices.
However, the logarithms cancel once we sum over 
all loop diagrams. We interpret this miraculous cancellation 
as a consequence of the recently discovered Love symmetries~\cite{Charalambous:2022rre}.
For spin-0 and spin-2 perturbations the logarithmic 
corrections to Love numbers 
are 
completely absent if we use the Kaluza-Klein (KK) metric 
split~\cite{Kol:2011vg, Kol:2007rx,Kol:2009mj} 
where there 
are no interaction vertices that 
could produce the logs. 
To the best of our knowledge, 
the absence of the RG running of Love numbers has not yet been 
explicitly demonstrated 
in the literature in full generality,
although this fact is  
known in the EFT community~\cite{unp1,unp2,unp3} (see also~\cite{Kol:2011vg} for 
the spin-0 quadruple case results).
Our results imply 
that the structure of 
perturbations in the
isotropic KK
gauge, 
or the apparent $Z_2$ symmetry for dilaton field $\phi \rightarrow - \phi$~\cite{Kol:2011vg}, 
in fact
~\textit{does not provide}
a general IR symmetry explanation to
the nonrenormalization of Love numbers. 
Rather, the KK split and the 
isotropic gauge simply appear as convenient tools to obtain this result
in the particular cases of spin-0 and spin-2 fields.
% In other words, the $Z_2$ dilaton symmetry 
% is a restatement of the fact that there are no logs, 
% but not the reason for it. 
We demonstrate this explicitly in the case of spin-1 perturbations,
which do not have the $Z_2$ dilaton symmetry, but whose 
worldline Wilson coefficient 
still possesses the 
nonrenormalization property.  
Working in the same isotropic KK gauge,
we find that individual loop
diagrams do produce log corrections 
to Love numbers, 
but these corrections cancel 
in an intricate manner when 
all contributions are 
summed together. 
This is a clear example showing that 
the $Z_2$ dilaton symmetry, 
in general, cannot be 
interpreted as an IR symmetry 
enforcing the nonrenormalization of Love numbers. 

\item \textbf{Off-shell matching of dissipation numbers}: We also study in detail the dissipative response
of Schwarzschild black holes, especially, the off-shell 1-pt function matching, without using the analytic continuation techniques.
%Previous works~\cite{Goldberger:2005cd,Goldberger:2020wbx,Goldberger:2020fot,Charalambous:2021mea}
%pointed out that the conservative and dissipative 
%effects can be clearly distinguished 
%in the point particle EFT. 
%There, the Love numbers are generated by local counterterms in the 
%finite-size aciton
In the EFT, the dissipation numbers are generated by internal degrees of freedom, which are encapsulated in composite mass multipole moments on the worldline~\cite{Goldberger:2005cd}.
In this paper we establish the explicit connection between 
this approach 
and the recent off-shell GR calculations of dissipation numbers 
in Refs.~\cite{Chia:2020yla,Charalambous:2021mea}.
To that end we compute the imaginary time-dependent 
part of the graviton one-point function using 
the Schwinger-Keldysh in-in approach~\cite{schwinger1961brownian, bakshi1963expectation, Bakshi:1963bn, keldysh1965diagram, jordan1986effective} 
(also see \cite{landau1981course,chou1985equilibrium,Haehl:2016pec, Galley:2009px} for reviews). 
In the EFT, this contribution is produced by the imaginary part of the 
retarded two-point correlator of composite mass multipole
operators. 
With this calculation we confirm that the imaginary part of the coefficient 
in front of the $r^{-\ell-1}$ term \eqref{eq:kfull}
is indeed produced by the dissipation of the BH horizon. 
This helps resolve some confusion about the conservative and dissipative response 
terms, which was especially acute in the case of Kerr BHs~\cite{LeTiec:2020bos, LeTiec:2020spy}.

We explicitly match the dissipation numbers in the EFT and GR 
for spin-0, spin-1, and spin-2 
external fields in a genetic multipole sector.
Unlike Refs.~\cite{Chia:2020yla,Charalambous:2021mea}, our 
results here do not rely on the analytic 
continuation prescription.

Our calculation 
explicitly demonstrates the
equivalence of off-shell and on-shell extraction of dissipative responses:
the dissipation numbers we extract 
from a graviton one-point function agree with the results
of the matching at the level of the absorption cross sections~\cite{starobinskii1973amplification, page1976particle}.
This serves as a consistency check of the EFT approach and solidly confirms interpretations 
of recent GR response function calculations~\cite{LeTiec:2020bos, LeTiec:2020spy,Chia:2020yla,Charalambous:2021mea}.

\end{itemize}

\subsection*{Outline}

Our paper is structured as follows. We start with a recap of 
the nonrelativistic general relativity and the point-particle EFT 
in Section \ref{Worldline EFT for Schwarzschild BH}.
Then we discuss the EFT diagrammatic structure and power
counting rules in Section~\ref{sec:PNgen}.
There we show that logarithmic corrections to Love numbers are
expected from the EFT, in general. 
In Section.~\ref{sec:spec} we explicitly compute the scalar, photon,
and scalar graviton static one-point functions in the EFT in the isotropic Kaluza-Klein gauge.
For the spin-1 field we show that the logarithmic corrections actually cancel,
while for the spin-0 and spin-2 cases the logs are actually not present at all 
in the isotropic Kaluza-Klein gauge. 
We also explicitly resumm the spin-0, spin-2 dilaton and spin-1 dipole
one-point functions in the EFT to all PN orders. 
In Section~\ref{sec:bhpt} we compute the same 
spin-0, spin-1, and spin-2 one-point functions
and compare them with the EFT expressions. This way 
we establish that Love numbers vanish identically 
without any source/response ambiguity. 
In Section~\ref{sec:dissN} we match the dissipation numbers
by comparing the time-dependent one-point functions computed in the EFT
and in BHPT. We draw conclusions in Section~\ref{sec:concl}.

Some additional material is presented in several appendices. In Appendix~\ref{Feynman Rule}, we provide Feynman rules used in our diagrammatic EFT computation. 
In Appendix~\ref{Useful Mathematical Relations} we collect some useful mathematical relations. 
In Appendix~\ref{Reproducing Schwarzschild Metric} we show that our EFT setup correctly reproduces the Schwarzschild metric perturbatively. 
In Appendix~\ref{Spin-1 1-pt Function} we provide the details of the spin-1 electric dipole one-point function calculation. 
In Appendix \ref{Teukolsky Equation in Schwarzschild BH} we derive 
the static spin-s Teukolsky equations in Schwarzschild coordinates 
and isotropic coordinates. We also show that the dissipation number is the same
in both coordinates, which confirms its gauge invariance.
Finally, in Appendix~\ref{app:abs} we derive the 
dissipation-fluctuation relation for Schwarzschild BHs
and compute the graviton absorption cross section in the EFT.

\section{Worldline EFT for Schwarzschild BHs}\label{Worldline EFT for Schwarzschild BH}

In this section we introduce the EFT 
for Schwarzschild black holes in a long-wavelength tidal environment. 
We systematically describe the tidal response of a black hole 
to spin-0, spin-1, and spin-2 electric-type external perturbations.
We start with a general EFT for GR
in the Newtonian limit, and then we discuss an effective 
description of black holes. 
Importantly, we will show that the EFT clearly separates between
the conservative and dissipative contributions.

% We will first try to be as general as possible to study the scalar type perturbations as whole, i.e. spin-0, spin-1 electric and spin-2 electric perturbations, without specifying the background gravitational field gauge.

\subsection{Perturbative General Relativity}\label{sec:GRact}

 % \cite{feynman1963quantum, duff1973quantum, goldberger2006effective,goldberger2006houches}

% As originally studied in, the metric of Schwarzschild BH can be reproduced by coupling the bulk gravity with the point particle worldline action. 

% Here, in order to study the scalar type perturbations, we introduce another generic scalar field $\varphi$. The total EFT is of the form

Let us consider gravity coupled to a source, which can be approximated 
as a point particle at leading order. They are described by the following 
action:
\begin{equation}
    S = S_{\rm EH}+ S_{\rm pp} \,,
\end{equation}
where $S_{\rm EH}$ is the standard Einstein-Hilbert (EH) action, 
whilst $S_{\rm pp}$ is the point-particle action that depends on both the black hole worldline $x^\m(\tau)$ and the metric,\footnote{We 
work in a unit system where $\hbar=G=c=1$.}
\be 
\label{eq:EH}
    S_{\rm EH}  = -\frac{1}{16 \pi} \int d^4 x \sqrt{-g}R ~,\quad 
    S_{\rm pp} = -m \int d\tau \sqrt{\frac{d x^\mu}{d\tau} \frac{d x^\nu }{d\tau} g_{\mu\nu}} ~,
\ee
where $\tau$ is a worldline parameter.
We also consider the bulk scalar and electromagnetic fields,
which are described by the standard actions
\be
\label{eq:EMS}
 S_{\varphi}  =\frac{1}{2}\int d^4 x \sqrt{-g} g^{\mu\nu}\d_\mu\varphi\d_\nu\varphi\,,\quad 
 S_{\rm EM}=-\frac{1}{4}\int d^4 x \sqrt{-g} g^{\mu\nu}g^{\lambda\rho}{F}_{\mu\lambda}
 {F}_{\nu\rho}\,,
\ee
where ${F}_{\m\n}=\d_\m {A}_\n - \d_\n {A}_\m$.
To reproduce a Schwarzschild black hole, we do not couple 
these test fields to our point mass. 
The black hole will still be affected by these fields through 
polarization effects.

We will use the ``static gauge'' choice 
for the worldline parameter $\tau = t$.
We decompose the spin-0, spin-1, spin-2 fields, and the 
center of mass coordinate into the background part and 
long-wavelength fluctuating parts,
\be
\label{eq:backfluc}
\begin{split}
     & g_{\mu\nu}  = \eta_{\mu\nu} + h_{\mu\nu} \,, \quad  
     \frac{dx^\mu}{dt}= (1,v^i)\,, \\
     &  \varphi = \bar \varphi + \delta \varphi\,, \quad 
     {A}_\mu  =\bar{{A}}_\mu + \delta   {A}_\mu\,.
\end{split}
\ee
where $h_{\mu\nu},~\delta   {A}_\mu,~\delta \varphi$ are fluctuations 
of the fields, and $v^i$ is the point-particle spatial velocity component. 
% where $h_{\mu\nu}$, $v^i$ and $\delta\varphi$ are the fluctuating part. 
% $\bar\varphi$ plays the role of the external source. 
In the rest frame of the BH (an equivalent of static gauge), 
the computation can be further simplified by setting $v^i = 0$. 

Within the EFT both the background BH geometry and
the fluctuations around it are computed perturbatively starting with a 
Minkowski background~\cite{Goldberger:2004jt,Goldberger:2007hy,Kol:2007rx,Kol:2009mj,Porto:2016pyg,Donoghue:2017pgk}. 
To that end we expand the EH and the point particle 
actions over perturbations in 
$h_{\m\n}$, and solve them as an expansion in $m/r$ in 
an appropriate gauge~\cite{feynman1963quantum, duff1973quantum, Goldberger:2004jt,Goldberger:2007hy}.
The one-point function calculation of the metric field $h_{\mu\nu}$
perturbatively recovers the Schwarzschild metric.
We perform this computation explicitly in Appendix~\ref{Reproducing Schwarzschild Metric}.
Once we have recovered the Schwarzschild metric, 
we can consider fluctuations of the test spin-$s$ fields ($s=0,1,2$ here),
and compute nonlinear, post-Newtonian corrections to their profiles. 
The actions~\eqref{eq:EH},~\eqref{eq:EMS}, however, do not
capture finite-size effects of the black hole. 
We discuss them in detail now.

\subsection{Finite Size Effects}\label{sec:fs}

% In the previous section, we discuss the gravitational nonlinear corrections to the source, now we are going to discuss the BH responce which corresponds to the finite size effects 

To incorporate BH finite-size effects,
such as responses to spin-$s$
test fields, we use the approach of the point-particle 
EFT~\cite{Goldberger:2007hy,Goldberger:2005cd,Porto:2016pyg, Goldberger:2020fot, Charalambous:2021mea}.
In the limit when the size of the body ($R$) is parametrically 
smaller than the wavelength of external perturbations, i.e., $|\k| R\ll 1$, 
conservative 
finite-size effects can be captured by a most general worldline action built out of 
long-distance degrees of freedom , i.e., long-wavelength test fields 
and the center of mass position $x^\mu$, and satisfying symmetries of the 
problem. In the case of spherically symmetric spacetimes, such 
as Schwarzschild, these symmetries are the diffeomorphism invariance, 
gauge invariance for the Maxwell field,
worldline reparametrization invariance, and local rotation symmetry. 
To explicitly realize these symmetries, it is 
convenient to use the four-velocity $v^\mu = \frac{dx^\mu}{d\tau}$,
the covariant derivate along it, $D \equiv v^\mu\nabla_\mu$, and
a set of tetrads $e^\mu_a$ carrying $SO(3)$ indices
$a=1,2,3$,
and defining a frame orthogonal to $v^\mu$.
They satisfy $g_{\mu\nu}e^\mu_a e^\nu_b =\delta_{ab}$ 
and define the projector 
% the projector onto the space
% transverse to the worldline, 
\be
P_{\mu\nu} =\delta_{ab} e_\mu^a e_\nu^b  = g_{\mu\nu} +v^\mu v_\nu\,.
\ee
In what follows we will consider only the electric-type (parity-even)
perturbations. Generalization to the magnetic (parity-odd) sector is straightforward.
Note that thanks to the electric-magnetic duality of the Schwarzschild
spacetime in four dimensions~\cite{Porto:2007qi}, 
the Love numbers for the magnetic perturbations 
must coincide with the electric ones~\cite{Hui:2020xxx}.
Thus, for the purposes of our work it will be sufficient 
to consider the electric sector only.

We can use the above geometric objects to define field multipole moments,
\be 
E^{(s)}_{L} \equiv E^{(s)}_{\mu_1 \mu_2 \cdots \mu_\ell}e^{\mu_1}_{\langle a_1} e^{\mu_2}_{a_2} \cdots e^{\mu_\ell}_{a_\ell\rangle}\,,
\ee 
where $L$ denotes the multi-index $a_1,a_2,\ldots a_n$,
and 
$\langle ... \rangle$ denotes 
the symmetric trace-free (STF) part. 
Explicitly, we have
\be
\label{eq:tidalgen}
\begin{split}
&E^{(s=0)}_{a_1...a_\ell} = 
\nabla_{\langle a_1}...\nabla_{a_\ell \rangle } \varphi\\
&E^{(s=1)}_{a_1...a_\ell} = \nabla_{\langle a_1}...\nabla_{a_{\ell-1} } E_{a_\ell \rangle} \,,\quad E_a = e_a^\mu v^\nu F_{\mu \nu}  \\
& E^{(s=2)}_{a_1...a_\ell} = 
\nabla_{\langle a_1}...\nabla_{a_{\ell-2} } E_{a_{\ell-1} a_\ell \rangle} 
\,,\quad E_{ab} = v^\alpha v^\mu e_a^\beta e_b^\nu C_{\alpha \beta \mu \nu}\,,  \\
\end{split} 
\ee
where $C_{\alpha \beta \mu \nu}$ is the Weyl tensor.
The effective action is naturally built from 
the fields' multipole moments $E^{(s)}_{L}$.
In particular, the leading order (quadratic in perturbations)
effective point-particle action is given by:
\be 
\label{eq:eft1}
\begin{split}
S^{(s)~\text{local}}_{\rm finite\; size}& =  
\sum_{\ell} \frac{\lambda^{(s)}_{\ell}}{2\ell!} \int d\tau~
E^{(s)}{}^{L}(x(\tau))
E^{(s)}{}_{L}(x(\tau)) \\
& + \sum_{\ell} \frac{\lambda^{(s)}_{\ell(\omega^2)}}{2\ell!} \int d\tau~
DE^{(s)}{}^{L}(x(\tau))
DE^{(s)}{}_{L}(x(\tau)) +  ...
\,.
\end{split}
\ee
As usual in the EFT, perturbation theory is organized 
in terms of the field strength and the derivative expansion.
Note that since the theory is nonrelativistic, 
spatial and temporal derivatives in the body's rest frame
enter effective operators on different footing;
e.g., the first term in~\eqref{eq:eft1} does not 
have time derivatives at all. 
The frequency of the 
perturbation should also be smaller than 
the object's inverse size 
in order for the EFT to be valid, $\omega R \ll 1$.
We will suppress the index $(s)$ in what follows.

Note that due to isomorphism between the STF tensors and spherical 
harmonics~\cite{Thorne:1980ru}, the number of indices in $L$ corresponds to a multipole index $\ell$;
i.e.,  $L=(a)$ describes the dipole (not present for spin-2 perturbations), 
$L=(a_1 a_2)$ describes the quadrupole, 
$L=(a_1 a_2 a_3)$ - the octupole moment, etc. 
The monopole moment ($\ell=0$)
may be present only for spin-0 fluctuations.

The local action~\eqref{eq:eft1} cannot reproduce 
absorption. To incorporate this effect we need 
to take into account unknown gapless degrees of freedom $X$ 
on the worldline.
To that end one introduces composite operators 
$Q_{L}(X)$ that correspond to body's multipole moments,
including internal degrees of freedom. 
% Here, as before, we project everything
% onto the tetrades $Q^{(s)}_{L} \equiv Q^{(s)}_{\mu_1 \mu_2 \cdots \mu_n}e^{\mu_1}_{\langle i} e^{\mu_2}_j \cdots e^{\mu_n}_{n\rangle}$.
Then we add
a new coupling between the composite internal 
moments $Q_{L}$ and the long-wavelength tidal moments 
of perturbing fields $E_L$,
which yield the following additional action \cite{Goldberger:2005cd}:
\be 
\label{eq:eft2}
S^{(s)}_{\rm finite\; size} 
= -   \sum_{\ell} \int d\tau ~ Q^{(s)}_{L}(X,\tau) E^{(s)}{}^{L}(x(\tau)) \,,
\ee
Although we do not know the explicit form of the operator $Q_L$, we can still analyze the structure of its 
correlation functions by making use of symmetry 
and parameterizing it with some unknown 
coefficients that are determined through matching to the UV theory.
In general, correlation functions of $Q_{L}$ contain both conservative and dissipative effects; i.e. they are nonlocal in time in general. 

In this paper we focus on the matching of the one-point functions. 
Such a matching can be performed for different physical 
observables, i.e., correlation functions. 
In general, these correlation
functions have to be of the Schwinger-Keldysh (in-in) type;
i.e., the corresponding path integral 
has to satisfy the in-in boundary conditions. 
However, certain observables can be 
extracted from the usual in-out
path integral; i.e., 
scattering amplitudes \cite{Cheung:2018wkq,Bern:2019nnu,Cheung:2020gyp,Cheung:2020sdj,Damour:2017zjx,Bern:2020uwk}, conservative forces \cite{Kalin:2020fhe,Bini:2019nra} and the total radiated power \cite{Goldberger:2009qd,Ross:2012fc}.
We will discuss this approach in detail shortly. 

% Boundary conditions should be in-in. 
% The in-out transition probablity 
% In principle, there are two types of correlation functions that we can compute, i.e. the Feynman (in-out) correlation functions and the Schwinger-Keldysh (in-in) correlators. 
% These two options differ by boundary conditions. 
% % Both types of correlators can be used to match Love numbers 
% % and dissipation coefficients. 
%  In the in-out formalism, we fix the scattering boundary condition, which is sufficient for the computation of 

%   However, this formalism is not convenient for the analysis of the time-dependent 
%  1-point function because it does not follow the causal evolution.
% \begin{equation}
%     \langle Q_{L}(\tau) \rangle_{\rm in-out} = \int d\tau' {G_{\rm Fey}}^{L'}_L(\tau-\tau') \mathcal{E}_{L'}(x(\tau')) ~,
% \end{equation}
% where the Feynman Green function is defined as 
% \begin{equation}
%     {G_{\rm Fey}}^{L'}_L = \theta(\tau-\tau')\langle Q_{L}(\tau)Q^{L'}(\tau')\rangle + \theta(\tau'-\tau) \langle Q^{L'}(\tau') Q_{L}(\tau) \rangle ~.   
% \end{equation}

To compute the perturbative 
one-point function, we fix the retarded boundary condition when computing the Green function. 
This is equivalent to using the following 
linear response theory expression for $Q_L$,
% The best way is to compute the real expectation value, i.e. the in-in expectation value
\be 
    \langle Q_{L}(\tau) \rangle_{\rm in-in} = \int d\tau' ~{G_{\rm ret}}^{L'}_{L}(\tau-\tau')E_{L'}(x(\tau')) ~,
\ee
where we have introduced the retarded Green function,
\be 
    {G_{\rm ret}}^{L'}_{L}(\tau-\tau') =  i \langle[Q_{L}(\tau),Q^{L'}(\tau')]\rangle \theta(\tau-\tau') \,,
\ee
where $\theta(\tau)$ is the Heaviside step function. 
Since the retarded Green function in the complex frequency domain is analytic around $\omega=0$, we can parametrize its Taylor expansion around the origin as~\cite{Goldberger:2020fot,Charalambous:2021mea},
\be 
\label{eq:Gret}
{G_{\rm ret}}^{L'}_{L}(\omega) = (\lambda_{0}^{\rm loc.} + 
i\lambda_{1}^{\rm non-loc.}(r_s\omega) 
+\lambda_{2}^{\rm loc.}(r_s\omega)^2 
+  ...)\delta^{\langle L \rangle}_{\langle L' \rangle}\,,
\ee
where $...$ denote terms higher order in frequency and $\lambda_{0}^{\rm loc.}$, $\lambda_{1}^{\rm non-loc.}$,
and $\lambda_{2}^{\rm loc.}$ are free parameters (Wilson 
coefficients).
They all have the dimensionality $\mathcal{O}(r_s^{2\ell+1})$.
There are three comments in order. 

% so for low frequency tidal perturbations, we can do taylor expansion around $\omega=0$ \footnote{Here, we need to emphasize that the Feynman Green function is not analytic around $\omega=0$.}.  
% Now, we can parametrize the retarded Green function as the following 
% \begin{equation}\label{Kerr Green Function}
%     {G_{\rm ret}}^{L'}_{L}(\omega) = \sum_{n=0} \omega^{2n} \Big( {\lambda_{2n}^{\rm loc.}}^{L'}_{L} + {\epsilon_{2n}^{\rm non-loc.}}^{L'}_L +  i {\lambda^{\rm non-loc.}_{2n+1}}^{L'}_{L} \omega + i {\epsilon^{\rm loc.}_{2n+1}}^{L'}_L \omega \Big) ~,
% \end{equation}
% where we have introduces the phenomenological tensors ${\lambda_{2n}^{\rm loc.}}^{L'}_{L}$, ${\lambda^{\rm non-loc.}_{2n+1}}^{L'}_{L}$, ${\epsilon_{2n}^{\rm non-loc.}}^{L'}_L$ and ${\epsilon^{\rm loc.}_{2n+1}}^{L'}_L$. Now, let me explain the reason for this parametrization:

\begin{itemize}
    \item \textbf{Tensorial Structure.} 
    The rotational symmetry of the Schwarzschild background 
    dictates that the retarded Green function can only 
    be an STF version of the Kronecker symbol.
    % The tensorial structure between $L$ and $L'$ forms the rank 2 tensor. We can naturally decompose it into the symmetric part denoted by $\lambda$, and the anti-symmetric part denoted by $\epsilon$. Also, based on the STF nature of tidal fields $\mathcal{E}_{L}$, all the index structures inside $L$ and $L'$ must be STF. For Kerr BH, the tensorial building block we have are the Dirac-Delta symbol $\delta_{ij}$, the Levi-Civita antisymmetric symbol $\epsilon_{ijk}$ which can be contracted with the spin operator $s^i$. But for Schwarzschild, we only have $\delta_{ij}$.
    
    \item \textbf{Time Reversal Symmetry.} 
    The terms with even 
    and odd powers of frequency in
    the retarded Green function
    transform differently under time reversal. 
    The part which is even under the exchange $\omega\to -\omega$
    describes conservative effects, while the time-reversal
    odd part captures dissipation. 

    % The time reversal even part has even power of $\omega$, while the odd part has odd power of $\omega$. We know that both the dissipative operator and the spin operator can break time reversal symmetry. So, dissipative effects can appear both in the even and odd power of $\omega$. But in the $\omega^{2n}$ part, the dissipative operator mush combine with the spin operator in order to restore the time reversal symmetry. This tells us that dissipative effect in $\omega^{2n}$ part corresponds to the antisymmetric tensor $\epsilon_{2n}$, while in $\omega^{2n+1}$ part corresponds to the symmetric tensor $\lambda_{2n+1}$.  
    
    \item \textbf{Locality.} 
  The time-reversal invariant terms can be absorbed 
  into local counterterms in the point-particle action~\eqref{eq:eft1}. 
  In this sense they 
  just renormalize 
  the Wilson coefficients that we already had 
  in Eq.~\eqref{eq:eft1}.
  In contrast, the time-reversal odd terms, 
  cannot be recast into a local worldline action,
  and therefore we call them ``nonlocal.''
    % "Local" means this composite object can be written as the local worldline operators in the EFT, while "non-local" means it can only be written as non-local (non-local in time) worldline operators in the EFT. It will become clear when we specifically write down the effective action in the in-out formalism.
\end{itemize}

It is instructive to
compare Eq.~\eqref{eq:Gret} with the 
Feynman time-ordered Green function,
\be 
    {G_{\rm Fey}}^{L'}_{L}(\tau-\tau') =  
    \langle T Q_{L}(\tau)Q^{L'}(\tau')\rangle   ~.
\ee
Its Fourier transform 
is symmetric under $\omega\rightarrow -\omega$ but not analytic around $\omega=0$. The second relevant observation 
is that 
the Feynman and retarded Green functions are equivalent 
for conservative effects (i.e., off-shell modes)~\cite{Porto:2016pyg}.
The third important observation is 
a variant of the fluctuation-dissipation theorem for 
static BHs~\cite{Goldberger:2005cd}, 
\be 
\label{eq:dissfluct}
\int_{-\infty}^{+\infty} d\tau e^{i\omega \tau}~
\langle Q_{L}(\tau)Q^{L'}(0)\rangle
=2\text{Im}\left(i\int^{+\infty}_{-\infty} d\tau e^{i\omega \tau}~
\langle T Q_{L}(\tau)Q^{L'}(0)\rangle\right)\,,
\ee
valid for $\omega>0$.
All together, the above facts completely fix the 
form of the Feynman propagator in terms of the Wilson coefficients
that we had in the EFT expansion 
of the retarded Green function.
If we restrict the latter 
to the form~\eqref{eq:Gret}, 
the Feynman propagator 
would take the form
\be 
\label{eq:Fey}
 {G_{\rm Fey} }^{L'}_{L}(\omega) 
 =   \Big( -i {\lambda}_{0}^{\rm loc.} 
 + {\lambda}^{\rm non-loc.}_{1} r_s|\omega| 
-i {\lambda}_{2}^{\rm loc.}(r_s\omega)^2 + ...
 \Big)  \delta^{\langle L' \rangle}_{\langle L \rangle} ~.
\ee 
The modulus of frequency next to the ${\lambda}^{\rm non-loc.}_{1}$
term above explicitly reflects the nonanalyticity 
of the Feynman propagator at $\omega=0$.
% which is non-analytic around $\omega=0$ but symmetric under $\omega\rightarrow -\omega$. 
% More precisely, the term with $\omega$ even power is analytic while the term with $\omega$ odd power is non-analytic.
 % $\tilde{\lambda}$ coefficients are some different unknown coefficients than $\lambda$, which will be fixed by matching to the UV theory. As we will be able to see, in the in-out formalism, the analytic part can be formulated into the worldline local operator which corresponds to conservative effects while the non-analytic part cannot be formulated into the local operator which denotes the dissipative part.

% For the special situation of Schwarzschild BH, the tensorial building block can only be $\delta_{ij}$, so it is impossible to produce the antisymmetric rank 2 tensor between $L$ and $L'$. More concretely, we can write the retarded Green function as  
% \begin{equation}\label{Schwarz Green Function}
%     {G_{\rm ret}^{\rm Sch}}^{L'}_{L}(\omega) = \sum_{n=0} \omega^{2n} \Big( {\lambda_{2n}^{\rm loc.}}+ i {\lambda^{\rm non-loc.}_{2n+1}} \omega \Big)  \delta^{\langle L' \rangle}_{\langle L \rangle}~.
% \end{equation}

\subsubsection*{Love Numbers}\label{sec:EFTlove}

Let us focus on conservative effects. 
It is instructive to start with the
finite-size action Eq.~\eqref{eq:eft2}, and use the standard EFT 
definition for the
point-particle
in-out effective action, 
\begin{equation}\label{eq:eff}
    \exp\Big( i S_{\rm eff}^{\rm in-out}(x^\mu,F)\Big)
     \equiv \int \mathcal{D}X ~e^{iS[X,x^\mu,F]} ~,
\end{equation}
% \textcolor{blue}{Check the measure!}
where $X$ is the unknown degrees of freedom on the worldline, 
and $F=(h_{\mu\n},A_\m,\varphi)$ is the collective 
notation of long-wavelength probe
fields. The effective action~\eqref{eq:eff} results from 
integrating out all relevant short scale degrees of freedom plus
the internal degrees of freedom $X$.
We will suppress the explicit dependence on $F$ in what follows.
The leading order interaction term for a spin-$s$ field for 
an individual orbital sector $\ell$ is given by  
\begin{equation}\label{Finite Size Effective Action}
    S_{\rm int}^{\rm in-out}(x) = \frac{i}{2}\int d\tau d\tau' \langle T Q_{L}(\tau)Q^{L'}(\tau')\rangle E^L(\tau) E_{L'}(\tau') ~.
\end{equation}
In the static case it is sufficient to consider the $O(\omega^0)$ conservative part. After performing the Fourier transform, we 
get a local in time operator
\begin{equation}
    \langle T Q_{L}(\tau)Q^{L'}(\tau')\rangle   = - i  {\lambda}_{0}^{\rm loc.} \delta(\tau-\tau') \delta^{\langle L' \rangle}_{\langle L \rangle} ~.
\end{equation}
As anticipated, after plugging this into Eq.~\eqref{Finite Size Effective Action}, we find that conservative effects can be captured by the local action 
\begin{equation}
\label{eq:love1}
    S_{\text{finite size}}^{\rm local} =  \frac{\lambda_\ell}{2\ell !} \int d\tau  E_{L} E^L ~,
\end{equation}
where $\lambda_\ell \equiv \ell !\lambda_0^{\rm loc.}$. The Wilson coefficients 
$\lambda_\ell$ define static Love numbers in the point-particle
EFT. 

Let us focus now on the scalar-type
Love numbers, for which $E_L = \partial_{\langle L \rangle} \varphi$. 
To extract the Love numbers, we decompose the 
profile $\varphi$ into a background and fluctuation parts 
as in Eq.~\eqref{eq:backfluc}, and assume the background
that solves the static bulk Klein-Gordon equation at $r\to \infty$,
\be
\label{eq:scalsource}
\bar \varphi =  \mathcal{E}_{i_1\ldots i_\ell} x^{i_1} \cdots x^{i_\ell}\propto r^\ell Y_{\ell m}(\theta,\phi)\,,
\ee
where $\mathcal{E}_{i_1\ldots i_\ell}$
is a constant tidal moments' tensor.
% \begin{equation}
%     S_{\rm int}^{\rm love} \supset \frac{1}{\ell!} \int dt \lambda_\ell \partial_{\langle L \rangle} \bar\varphi \partial^{\langle L \rangle}\delta\varphi
% \end{equation}
% To extract the love number, we only need the leading order interaction
Plugging this ansatz into the action Eq.~\eqref{eq:love1},
and taking into account the bulk scalar field action,
we can perform a simple linear response calculation,
whose diagrammatic representation is given below:
\begin{equation}
    \vcenter{\hbox{\begin{tikzpicture}[scale=0.7]
        \begin{feynman}
            \vertex (i) at (0,0);
            \vertex (e) at (0,3);
            \node[circle, draw=repGreen, fill = repGreen, scale=1, label=left:$\lambda_\ell$] (w1) at (0, 1.5);
            \vertex[dot, MyBlue, label= $\boldsymbol{x}$] (f1) at (1.5,2.8) {};
            \vertex[crossed dot, MyBlue] (fs) at (1.5,0.2) {}; 

            \diagram*{
                (i) -- [double, double, thick] (w1),
                (w1) -- [double, double, thick] (e),
                (f1) -- [MyBlue, ultra thick] (w1),
                (fs) -- [MyBlue, ultra thick, edge label' = $r^\ell$] (w1)
            };
        \end{feynman}
    \end{tikzpicture}}}
    =   \frac{(2\ell - 1) !!}{4\pi} \lambda_{\ell} \mathcal{E}_{i_1\ldots i_\ell} x^{i_1} \cdots x^{i_\ell} \frac{1}{r^{2\ell + 1}} ~.
\end{equation}
Thus, we see that the Wilson coefficient 
$\lambda_\ell$ exactly coincides with the classic
Love number in the Newtonian limit. The generalization to 
spin-1 and spin-2 perturbations is straightforward,
and the corresponding Feynman rules are given in Appendix~\ref{Feynman Rule},
see also~\cite{Hui:2020xxx,Charalambous:2021mea}.

\subsection*{Dissipation Numbers}

The dissipation can be analyzed both within
the in-out approach by matching to the total BH absorption cross section, 
or in the in-in formalism by matching to the frequency dependent one-point function of the external test field. 

From the in-out formalism point of view, the $O(\omega)$ dissipative effect cannot be described by a local Lagrangian. Indeed, the term $ \dot{E}_{L} E^L$ is a total derivative and thus can be removed from the Lagrangian \footnote{More generally, all the terms involving odd number of derivatives are actually total derivatives and cannot be inserted into the local Lagrangian.}. 
To compute the cumulative 
power loss due to BH absorption, one can 
introduce a nonlocal in time action~\cite{Goldberger:2005cd}. 
Indeed, plugging the Feynman Green function~\eqref{eq:Fey}
into~\eqref{eq:eft2}, we get:
% But we can do it non-locally. Recall that the formula Eq.\eqref{Schwarz Feynman Green Function}, pluging into the Eq.\eqref{Finite Size Effective Action}, we get 
\be 
    S^{\rm diss.} = \frac{i}{2}\int d\tau d\tau' \(\int \frac{d\omega}{(2\pi)} \(\lambda_1^{\rm non-loc.} r_s|\omega|\) e^{-i\omega(\tau-\tau')}\) E_{L}(\tau)E^L(\tau') \,.
\ee
The parameter $\lambda_1^{\rm non-loc.}$
can be determined from matching to the absorption
cross section. 
For instance, 
the EFT absorption cross section in the 
sector $s=\ell$ is given by~\cite{Goldberger:2005cd}
(see Appendix~\ref{app:abs} for a derivation in our unit system)
% \begin{equation}
%     \sigma_{\rm abs}(\omega) = \frac{\omega}{8} \int d\tau 
%     e^{i\omega \tau} \omega^2 
%     \epsilon^{*ab}\epsilon_{cd} \langle Q_{ab}(\tau) Q^{cd}(0) \rangle ~.
% \end{equation}
% where $\epsilon_{ij}$ is the graviton polarization vector.
% Using Eq.~\eqref{eq:dissfluct} we arrive at
\be 
\sigma^{(\ell=s)}_{\text{abs, EFT}}(\omega) = 
2^s \ell!
{\omega^{2s} r_s\lambda^{\text{non-loc.}}_{1}}|_{s=\ell}\,.
\ee
Comparing this with the BH perturbation theory result 
in general relativity
\cite{starobinskii1973amplification, page1976particle},
\begin{equation}
    \sigma^{(\ell=s)}_{\text{abs, GR}}(\omega) = 
    \frac{8\pi (\ell+s)!2^s}{(2\ell+1)!!((2\ell-1)!!)^22^{2\ell+1}}  r_s^2 (r_s\omega)^{2\ell} ~,
\end{equation}
we obtain the following dissipation number
\begin{eBox}
\be 
\label{eq:absmatch}
  \lambda^{\text{non-loc.}}_1|_{s=\ell} =   \frac{8\pi (\ell+s)!}{(2\ell+1)!!((2\ell-1)!!)^2 2^{2\ell+1}\ell!}  r_s^{2\ell+1} \,.
\ee
\end{eBox}

 % by matching to the BH absorption cross section \cite{goldberger2006dissipative}, see also Appendix~\ref{}. 
% This action can be used to extract the total absorption
% cross-section.
% To extract the dissipation effect, we only need the leading order interaction in the number of fields. After reparametrizing $\tilde\lambda_1^{\rm non-loc.}$ to be $\lambda_{\ell(|\omega|)}$, we get 
% \begin{equation}
%     S_{\rm int}^{\rm diss.} \supset \frac{i}{\ell!} \int dt dt'  \(\int \frac{d\omega}{(2\pi)} \lambda_{\ell (|\omega|)} |\omega| e^{-i\omega(t-t')}\) \partial^{\langle L \rangle }\delta\varphi(t)\partial_{\langle L \rangle}\bar\varphi(t') ~.
% \end{equation}
% The BH dissipation number is proportional to . 

Alternatively, the dissipation number 
can be read off from the one-point function of the external field
computed within the Schwinger-Keldysh formalism.
To that end, we compute the in-in effective action $\Gamma_{\rm eff}^{\rm in-in}$ defined through
\begin{equation}
    \exp(i\Gamma_{\rm eff}^{\rm in-in}(\boldsymbol{x_1},F_1;\boldsymbol{x_2},F_1)) = \int \mathcal{D}X_1 \mathcal{D}X_2 e^{iS[X_1,x_1,F_1]-iS[X_2,x_2,F_2]} ~,
\end{equation}
where $\boldsymbol{x_1}$ and $\boldsymbol{x}_2$ are the worldline coordinates 
and the subscripts $1,2$ denote forward and backward indices in the closed time path (CTP) (we follow the notation of \cite{Galley:2009px}). 
The path integral here is done over two copies of the $X$ field.
All field satisfy the boundary condition  $X_1 = X_2$,
$F_1 = F_2$, and 
$\boldsymbol{x_1}=\boldsymbol{x_2}=\boldsymbol{x_0}$
at the final time slice $t=+\infty$ and the vacuum boundary condition at the initial time slice $t=-\infty$.
We also use our freedom to choose $\boldsymbol{x_0}$
and place the BH at the origin.
It is convenient to work in the Keldysh representation~\cite{keldysh1965diagram}, 
\begin{equation}
    \begin{aligned}
    X_- &\equiv X_1 - X_2, \quad X_+ \equiv \frac{1}{2}(X_1+X_2) ~,\\
    F_- &\equiv F_1 -F_2, \quad F_+ \equiv \frac{1}{2} (F_1 +F_2) ~,\\
    \boldsymbol{x_-} &\equiv \boldsymbol{x_1} - \boldsymbol{x_2}, \quad \boldsymbol{x}_+ \equiv \frac{1}{2}(\boldsymbol{x_1}+\boldsymbol{x_2}) \,.        
    \end{aligned}
\end{equation}
The $2\times 2$ Green function of each field in $F=(h_{\mu\nu},A_\mu,\varphi)$
is given by 
\begin{equation}
    G^{AB} = 
    \begin{pmatrix}
        0 &  -i G_{\rm adv} \\
        -i G_{\rm ret} & \frac{1}{2} G_{\rm H}
    \end{pmatrix} \,,
\end{equation}
where $A,B = \pm$, and $G_{\rm adv}$, $G_{\rm H}$ are the advanced and Hadamard two-point functions respectively. The Feynman rules in the in-in formalism are very similar to 
those of the in-out formalism, but with contractions made over all closed time path indices $A,B$ with
the ``effective metric''
\begin{equation}
    c^{AB} = c_{AB} = 
    \begin{pmatrix}
        0 & 1 \\
        1 & 0
    \end{pmatrix} \,,
\end{equation}
Now we can write down the leading order interaction term in the in-in action, 
\begin{equation}
        \Gamma_{\rm int}^{\rm in-in}(\boldsymbol{x}_\pm,F_\pm)  = \frac{i}{2} \int d\tau d\tau' \langle Q_{L}^A(\tau) Q^{L'B}(\tau') \rangle E^L_A(\tau) E_{L'} {}_B(\tau') \,.
\end{equation}

Now let us focus on the scalar field case, $F=\varphi$.
To account for the external source component $\bar\varphi(\boldsymbol{x},t)$, we choose $\bar\varphi_1 = \bar\varphi_2 = \bar\varphi$ so that $\bar\varphi_- = 0$, $\bar\varphi_+ = \bar\varphi$. It is now straightforward to compute the one-point function, 
% \be 
%     \langle \delta\varphi(\boldsymbol{x},t) \rangle_{\rm in-in} =  \frac{1}{\ell!}\int dt_1 dt_1' \langle Q(t_1) Q(t_1') \rangle_{\rm ret} \partial^{\langle L \rangle} \langle \delta\varphi(\boldsymbol{x},t) \delta\varphi (\boldsymbol{x_0},t_1) \rangle_{\rm ret}  \partial_{\langle L \rangle}\bar\varphi(t_1') ~,
% \ee
\begin{equation}
\label{eq:1pfinin}
    \begin{aligned}
            \langle \delta\varphi(\boldsymbol{x},t) \rangle_{\rm in-in} &=  \int \mathcal{D}\delta\varphi_+ \mathcal{D}\delta\varphi_- \(\delta\varphi_+(\boldsymbol{x},t) e^{i\Gamma_{\rm int}^{\rm in-in}} e^{iS[\delta\varphi_1]-i S[\delta\varphi_2]}\) \\
            & \approx  - \int dt_1 dt_1' \langle Q^A(t_1) Q^B(t_1') \rangle  \partial^{\langle L \rangle} \langle \delta\varphi_+(\boldsymbol{x},t) \delta\varphi_A (\boldsymbol{x_0},t_1) \rangle \partial_{\langle L \rangle}\bar\varphi_B(t_1') \\
            & =  \int dt_1 dt_1' \langle Q(t_1) Q(t_1') \rangle_{\rm ret} \partial^{\langle L \rangle} \langle \delta\varphi(\boldsymbol{x},t) \delta\varphi (\boldsymbol{x_0},t_1) \rangle_{\rm ret}  \partial_{\langle L \rangle}\bar\varphi(t_1') \,.
    \end{aligned}
\end{equation}
Let us compute now the dissipative contribution to the one-point function generated by 
the time-dependent generalization of the profile~\eqref{eq:scalsource},
\be 
\label{eq:scalsource2}
\bar \varphi =  e^{-i\omega t} \mathcal{E}_{i_1\ldots i_\ell} x^{i_1} \cdots x^{i_\ell}\,.
\ee 
We insert this into Eq.~\eqref{eq:1pfinin},
and use the static (instantaneous) propagator for $\delta \varphi$,
which is sufficient to obtain the 
response at $\mathcal{O}(\omega)$.
This calculation 
is almost identical to the above Love number calculation,
and it can be represented by the following Feynman
diagram:
\begin{equation}\label{eq:disdiag}
    \vcenter{\hbox{\begin{tikzpicture}[scale=0.7]
        \begin{feynman}
            \vertex (i) at (0,0);
            \vertex (e) at (0,3);
            \node[circle, draw=Orange, fill = Orange, scale=0.5, label=left:$Q^a$] (w1) at (0, 1.0);
            \node[circle, draw=Orange, fill = Orange, scale=0.5, label=left:$Q^b$] (w2) at (0, 2.0);
            \vertex[dot, MyBlue, label= $x$] (f1) at (1.5,2.8) {};
            \vertex[crossed dot, MyBlue] (fs) at (1.5,0.2) {}; 

            \diagram*{
                (i) -- [double, double, thick] (w1),
                (w1) -- [dashed, double, double, thick] (w2),
                (w2) -- [double, double, thick] (e),
                (f1) -- [MyBlue, ultra thick] (w2),
                (fs) -- [MyBlue, ultra thick, edge label' = $r^\ell$] (w1)
            };
        \end{feynman}
    \end{tikzpicture}}}
    = i (r_s \omega) \frac{(2\ell-1)!!}{4\pi} \lambda_{\ell (\omega)}
     \mathcal{E}_{i_1\cdots i_\ell} x^{i_1}\cdots x^{i_\ell} \frac{1}{r^{2\ell+1}} e^{-i\omega t} \,,
\end{equation}
where we showed results in time-domain 
Fourier space. For future convenience, 
we have also redefined the coefficient $\lambda_{1}^{\rm non-loc.}$ as
\be \label{eq:redef}
\lambda_{\ell(\omega)} = \ell!~\lambda_{1}^{\rm non-loc.} \,.
\ee  
% Note that definition of the dissipation number in the EFT 
% ~\eqref{eq:Gret}
% is gauge-invariant. 
% This will be explicitly confirmed by the fact that
% matching to 
% the 1-point function and to the absorption 
% cross-section give identical results.

From~\eqref{eq:disdiag}, we explicitly see that the dissipation effect corresponds to the imaginary part of the one-point function of the external fields. 
The calculations for spin-1 and spin-2 perturbations
are identical to the scalar field case, and the
Feynman rules are given in Appendix~\ref{Feynman Rule}.
We will perform an explicit matching to the UV theory 
in Section~\ref{sec:dissN}.

% To make the story complete, we shall match this dissipation number with the full black hole perturbation theory, which has the near horizon ${\rm SL}(2,\mathbb{R})$ symmetry in the near horizon region, with the analytic continuation method in Section \ref{Schwarzschild Dissipation Number}.

\section{PN Corrections to External Fields: Generalities}\label{sec:PNgen}

In the previous section
we have introduced the BH response
to external probes. If the external field 
profile has an asymptotic behavior $\propto r^\ell$
at spatial infinity, the response function
would generate a correction to the 
field profile scaling as $\sim r^{-\ell-1}$.
These corrections may be degenerate with 
the $2\ell+1$PN corrections to the source due to gravitational
nonlinearities.
Calculation of these corrections may be 
quite laborious given that GR is an effective 
theory with an infinite number of interaction vertices. 
In this section we show that this calculation is still
possible thanks to 
a particular structure of the worldline EFT.

First, we introduce power counting rules 
and single out the relevant type of diagrams
producing PN corrections. 
Second, we show that every such PN diagram 
can be presented as a ``ladder'' diagram built out of 
basic building blocks, which we call ``pyramids.''
This decomposition structure naturally leads to a 
recurrence relation between ladder diagrams of different PN 
order. In passing, we introduce an off-shell amplitude 
approach that allows us to estimate the momentum 
dependence of each PN diagram, which will be important 
for their future evaluation.

The discussion of this section is quite general,
so we do not specify any gauge in GR calculations at the moment. 
For simplicity we focus on the case of the scalar test field
profile here, although all results can be carried over 
to a general spin-$s$ case. We do that explicitly in the next 
section. 

% In this section 
% we discuss these corrections 
% in detail and 
% outline their general structure. 
% The key observation of this section 
% is that  

\subsection{Power Counting}\label{Power Counting}

The relevant small parameter in our discussion
is the PN expansion parameter $m/r \ll 1$.
The point-particle mass $m$ is present only in the worldline
point-particle action, so if we insert 
$N_m$ point masses in our wordline we expect 
that it should modulate the static field profile by 
a factor $(m/r)^{N_m}$.
We assume that the vacuum scalar field profile, i.e. 
in the absence of point masses and graviton, it simply 
equal to $\bar\varphi(\boldsymbol{x}) \propto r^{\ell}$.
Then we are interested in diagrams which produce the following 
corrections,
\begin{equation}\label{Power Counting Law}
    \delta\varphi(\boldsymbol{x})  =
    \vcenter{\hbox{\begin{tikzpicture}
        \begin{feynman}
          \vertex (i) at (0, 0);
          \vertex (e) at (0, 3);
          \vertex[dot, MyYellow] (w1) at (0, 2.5) {};
          \vertex[dot, MyYellow] (w2) at (0, 2.0) {};
          \vertex[dot, MyYellow] (w3) at (0, 0.5) {};
          \node at (0.5,1.1) {$\cdot$};
          \node at (0.5,1.3) {$\cdot$};
          \node at (0.5,1.5) {$\cdot$};
          \vertex[blob, scale =2] (c1) at ( 2, 1.5) {};
          \node[circle, fill=white, scale = 0.8] at (2,1.5) {static};
          \vertex[dot, MyBlue, label= $\boldsymbol{x}$] (f1) at (3.5,2.8) {};
          \vertex[crossed dot, MyBlue] (fs) at (3.5,0.2) {}; 
    
          \diagram*{
            (i) -- [double,double, thick, edge label = $N_m$] (e), 
            (w1) -- [boson, MyYellow, ultra thick] (c1) -- [boson, MyYellow, ultra thick] (w2),  
            (w3) -- [boson, MyYellow, ultra thick] (c1),
            (c1) -- [ultra thick, MyBlue] (f1), 
            (fs) -- [ultra thick, MyBlue, edge label' = $r^\ell$] (c1),  
          };
        \end{feynman}
    \end{tikzpicture}}}
    \sim O\( r^{\ell} \times \(\frac{m}{r}\)^{N_m}\) \,,
\end{equation}
where the blob diagram inside is made of 
various GR vertices connected by propagators in the static 
limit. 

The first relevant observation is that the scaling above 
is valid for \textit{any} type of the blob diagram inside 
\eqref{Power Counting Law}, provided that there are no
quantum graviton loops. These loops are irrelevant in 
classical GR calculations that we do here and hence can be 
ignored.\footnote{Note that in principle we can use the worldline EFT for quantum calculations as well.} In general, the number of bulk quantum loops $L_{\rm Bulk}$ is given by
\begin{equation}
    L_{\rm Bulk} = P_h + P_{\delta\varphi} -V +1 \,,
\end{equation}
where $P_h$ is the number of corresponding bulk graviton propagators, $P_{\delta\varphi}$ is the number of $\delta\varphi$ propagators and $V$ is the number of bulk vertices. 
Let us see now that $L_{\rm Bulk}=0$ ensures that the 
scaling~\eqref{Power Counting Law} is always correct. 
Each propagator scales as $1/r$. 
Each bulk vertex scales as $r$
since in gravity we only have the derivative couplings 
producing the 
$\int d^3 x \partial \partial$ contribution. 
Now we can compute the total correction to the 
one-point function $\delta\varphi(\boldsymbol{x})$,
from a diagram with $N_m$ worldline point mass insertions, $V$ vertices, $P_h$ graviton 
propagators, and $P_{\delta\varphi}$ scalar propagators,
% If we want to evaluate the 1-point function $\delta\varphi(\boldsymbol{x})$, it is easy to count its 
\begin{equation}
    P_h + P_{\delta\varphi} - V + (N_m + 1) = L_{\rm Bulk} + N_m = N_m ~,
\end{equation}
where in the last equality above we used $L_{\rm Bulk}=0$. 

Two comments are in order here. First, the power counting above
does not capture logarithmic corrections,
which can give rise to the renormalization group (RG)
flow of Love numbers. In this sense it is appropriate 
to call it a ``naive power counting.''
Second, this power counting is gauge invariant, 
as it is based on the requirement that there are 
no quantum loops, which is clearly a gauge-independent 
statement.

\subsection{EFT Diagramatic Structure}\label{EFT Diagramatic Structure}

Let us discuss now the general 
form of the EFT worldline diagrams. 
We first show that each diagram has a typical 
``ladder'' structure made of ``pyramids.''
These pyramid graphs are similar to the one-particle 
irreducible (1PI) diagrams 
% in what they 
% are the simplest building blocks of our Feynman graphs.
% They are similar to 1PI diagrams in QFT 
because they cannot be
reduced by cutting an external test field line. The 1PI diagrams are the simplest building 
blocks in our expansion.
Any complicated ``reducible'' worldline graph can be presented 
as a product of these ``1PI'' diagrams.

It is also convenient to discuss diagrams 
assuming that the worldline is ``amputated.''
In this case each PN EFT diagrams can be thought of 
as an off-shell scattering amplitude. 

% How to evaluate such EFT diagrams? Actually, there are two methods: one is analyzing the EFT diagramatic structure, and the other one is directly matching to the UV theory, i.e. the BH perturbation theory through Frobenius method, which we tend to study in Section \ref{Frobenius Method}. It is crucial to notice that these two methods are not completely equivalent for 4D Schwarzschild BH $\ell \in \mathbb{N}$. The EFT diagramatic analysis can be extended to beyond $2\ell + 1$ PM order, while the Frobenius method cannot due to the series mixing mentioned in Section \ref{GR Nonlinear Correction BH Response}.  In this section, we try the first method to discuss the EFT diagramatic structure for the gravitational nonlinear correction. The key point is to study the single off-shell scattering process.

\subsubsection*{Pyramids \& Ladders}\label{Ladder and Pyramid}

% The single off-shell scattering process involves only two off-shell scalars and the gravitational dressed particle with mass $m$. Actually, this is very similar to the one particle irreducible (1PI) diagram for $\varphi$ in quantum field theory

The basic building block in our diagrammatic
expansion is a diagram that corresponds to an off-shell
scattering process between a scalar and 
a gravitationally dressed worldline,

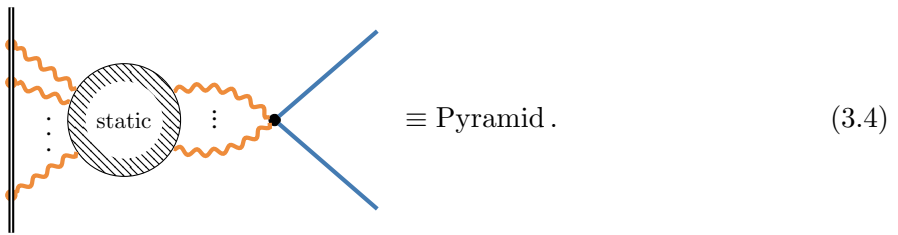
\begin{equation}
    \vcenter{\hbox{\begin{tikzpicture}
        \begin{feynman}
          \vertex (i) at (0, 0);
          \vertex (e) at (0, 3);
          \vertex[dot, MyYellow] (w1) at (0, 2.5) {};
          \vertex[dot, MyYellow] (w2) at (0, 2.0) {};
          \vertex[dot, MyYellow] (w3) at (0, 0.5) {};
          \node at (0.5,1.1) {$\cdot$};
          \node at (0.5,1.3) {$\cdot$};
          \node at (0.5,1.5) {$\cdot$};
          \node at (2.7,1.5) {$\cdot$};
          \node at (2.7,1.6) {$\cdot$};
          \node at (2.7,1.4) {$\cdot$};
          \vertex[blob, scale =2] (c1) at (1.5, 1.5) {};
          \node[circle, fill=white, scale = 0.8] (c2) at (1.5,1.5) {static};
          \vertex[dot, black] (c3) at (3.5, 1.5) {};
          \vertex (f1) at (5,2.8) {};
          \vertex (f2) at (5,0.2) {}; 
    
          \diagram*{
            (c1) -- [boson, ultra thick, MyYellow, bend right=30] (c3),
            (c1) -- [boson, ultra thick, MyYellow, bend left=30] (c3),
            (i) -- [double,double, thick] (e), 
            (w1) -- [boson, MyYellow, ultra thick] (c1) -- [boson, MyYellow, ultra thick] (w2),  
            (w3) -- [boson, MyYellow, ultra thick] (c1),
            (c3) -- [ultra thick, MyBlue] (f1), 
            (f2) -- [ultra thick, MyBlue] (c3),  
          };
        \end{feynman}
    \end{tikzpicture}}}
    \equiv \text{Pyramid} \,.
\end{equation}
The above dots denote any number of graviton propagators
that can be inside the diagram. 

We call 
this diagram a ``pyramid'' because characteristic 
diagrams of this type have a pyramid shape. 
In what follows we use the 
term ``connecting point'' for the interaction vertex between $\varphi\varphi$ and bulk gravitons.
We denote it with a black dot.
Since we do not have quantum loops 
and the external source field
cannot become virtual, the only diagrammatic 
structure compatible with bulk Feynman rules 
is the one where different pyramid graphs are connected
to each other by scalar leg links between the connecting points.
Such graphs are called ``uncrossed ladder diagrams'' \footnote{The crossed ladder diagrams are not possible in the static limit}. Thus, 
a general PN diagram can be represented as a ladder graph
made of pyramids,
% How can these 1PI diagrams connect with each other? The requirement for vanishing bulk quantum loop diagrams and the free field nature of the external perturbations tell us that the only reasonable way is to use $\varphi$ legs as links bettwen connecting point.

\begin{equation}
    \vcenter{\hbox{\begin{tikzpicture}[scale=0.6]
        \begin{feynman}
          \vertex (i) at (0, 0);
          \vertex (e) at (0, 6);
          \vertex[dot, MyYellow] (w1) at (0, 5) {};
          \vertex[dot, MyYellow] (w2) at (0, 4) {};
          \vertex[dot, MyYellow] (w3) at (0, 1) {};
          \node at (0.25,2.2) {$\cdot$};
          \node at (0.25,2.6) {$\cdot$};
          \node at (0.25,3) {$\cdot$};
          \vertex[blob, scale =2.4] (c1) at ( 2, 3) {};
          \vertex[dot, MyBlue, label= $\boldsymbol{x}$] (f1) at (3.5,5.6) {};
          \vertex[crossed dot, MyBlue] (fs) at (3.5,0.4) {}; 
    
          \diagram*{
            (i) -- [double,double, thick] (e), 
            (w1) -- [boson, MyYellow, ultra thick] (c1) -- [boson, MyYellow, ultra thick] (w2),  
            (w3) -- [boson, MyYellow, ultra thick] (c1),
            (c1) -- [ultra thick, MyBlue] (f1), 
            (fs) -- [ultra thick, MyBlue] (c1),  
          };
        \end{feynman}
    \end{tikzpicture}}}
    \quad =  \quad 
    \vcenter{\hbox{\begin{tikzpicture}[scale=0.6]
        \begin{feynman}
            \vertex (i) at (0, 0);
            \vertex (e) at (0, 6);
            \vertex[dot, MyYellow] (w1) at (0, 5.5) {};
            \vertex[dot, MyYellow] (w2) at (0, 5.2) {};
            \vertex[dot, MyYellow] (w3) at (0, 4.5) {};
            \vertex[blob, scale =0.7] (c1) at ( 1, 5) {};
            \vertex[dot, MyYellow] (w4) at (0, 4) {};
            \vertex[dot, MyYellow] (w5) at (0, 3.7) {};
            \vertex[dot, MyYellow] (w6) at (0, 3) {};
            \vertex[blob, scale =0.7] (c2) at ( 1, 3.5) {};
            \vertex[dot, MyYellow] (w7) at (0, 1.5) {};
            \vertex[dot, MyYellow] (w8) at (0, 1.2) {};
            \vertex[dot, MyYellow] (w9) at (0, 0.5) {};
            \vertex[blob, scale =0.7] (c3) at ( 1, 1) {};
            \vertex[dot, MyBlue, label= $\boldsymbol{x}$] (f1) at (3.5,5.6) {};
            \vertex[crossed dot, MyBlue] (fs) at (3.5,0.4) {};
            \vertex[dot, black] (d1) at (2.5, 5) {};
            \vertex[dot, black] (d2) at (2.5, 3.5) {};
            \vertex[dot, black] (d3) at (2.5, 1) {};
            \node at (0.25,0.95) {$\cdot$};
            \node at (0.25,0.85) {$\cdot$};
            \node at (0.25,0.75) {$\cdot$};
            \node at (0.25,3.45) {$\cdot$};
            \node at (0.25,3.35) {$\cdot$};
            \node at (0.25,3.25) {$\cdot$};
            \node at (0.25,4.95) {$\cdot$};
            \node at (0.25,4.85) {$\cdot$};
            \node at (0.25,4.75) {$\cdot$};
            \node at (1.9,1.05) {$\cdot$};
            \node at (1.9,0.95) {$\cdot$};
            \node at (1.9,0.85) {$\cdot$};
            \node at (1.9,3.55) {$\cdot$};
            \node at (1.9,3.45) {$\cdot$};
            \node at (1.9,3.35) {$\cdot$};
            \node at (1.9,5.05) {$\cdot$};
            \node at (1.9,4.95) {$\cdot$};
            \node at (1.9,4.85) {$\cdot$};
            \node at (1.5,2.2) {$\cdot$};
            \node at (1.5,2.7) {$\cdot$};
            \node at (1.5,1.7) {$\cdot$};
      
            \diagram*{
              (i) -- [double,double, thick] (e), 
              (w1) -- [boson, MyYellow, ultra thick] (c1) -- [boson, MyYellow, ultra thick] (w2),  
              (w3) -- [boson, MyYellow, ultra thick] (c1),
              (w4) -- [boson, MyYellow, ultra thick] (c2) -- [boson, MyYellow, ultra thick] (w5),  
              (w6) -- [boson, MyYellow, ultra thick] (c2),
              (w7) -- [boson, MyYellow, ultra thick] (c3) -- [boson, MyYellow, ultra thick] (w8),  
              (w9) -- [boson, MyYellow, ultra thick] (c3),
              (c1) -- [boson, ultra thick, MyYellow, bend left=30] (d1),
              (c1) -- [boson, ultra thick, MyYellow, bend right=30] (d1),
              (c2) -- [boson, ultra thick, MyYellow, bend left=30] (d2),
              (c2) -- [boson, ultra thick, MyYellow, bend right=30] (d2),
              (c3) -- [boson, ultra thick, MyYellow, bend left=30] (d3),
              (c3) -- [boson, ultra thick, MyYellow, bend right=30] (d3),
              (d1) -- [ultra thick, MyBlue] (f1), 
              (fs) -- [ultra thick, MyBlue] (d3),  
              (d1) -- [ultra thick, MyBlue] (d2) -- [ultra thick, MyBlue] (d3),
            };
          \end{feynman}        
        
    \end{tikzpicture}}}
    ~.
\end{equation}
Physically, this ladder structure has the interpretation 
that the gravitational nonlinear corrections to the source are  generated by multiple  
scatterings between the off-shell scalars and gravitationally dressed point masses. 
As we will show shortly, the repetition of the same pyramid diagrams in a more complicated graph 
indicates that there should be a diagrammatic 
recurrence relation between corrections of different PN order.

Now, let us consider the pyramid diagram in detail. 
Starting from a single connecting point, 
diagrams connecting a scalar with the worldline 
have the following typical pyramid structure:

\begin{equation}\label{Pyramid Structure}
    \begin{aligned}
        \vcenter{\hbox{\begin{tikzpicture}[scale=0.7]
            \begin{feynman}
              \vertex (i) at (0, 0);
              \vertex (e) at (0, 3);
              \vertex[dot, MyYellow] (w1) at (0, 2.5) {};
              \vertex[dot, MyYellow] (w2) at (0, 2.0) {};
              \vertex[dot, MyYellow] (w3) at (0, 0.5) {};
              \node at (0.5,1.1) {$\cdot$};
              \node at (0.5,1.3) {$\cdot$};
              \node at (0.5,1.5) {$\cdot$};
              \node at (2.7,1.5) {$\cdot$};
              \node at (2.7,1.6) {$\cdot$};
              \node at (2.7,1.4) {$\cdot$};
              \vertex[blob, scale =1.4] (c1) at (1.5, 1.5) {};
              \node[circle, fill=white, scale = 0.56] (c2) at (1.5,1.5) {static};
              \vertex[dot, black] (c3) at (3.5, 1.5) {};
              \vertex (f1) at (5,2.8) {};
              \vertex (f2) at (5,0.2) {}; 
        
              \diagram*{
                (c1) -- [boson, ultra thick, MyYellow, bend right=30] (c3),
                (c1) -- [boson, ultra thick, MyYellow, bend left=30] (c3),
                (i) -- [double,double, thick] (e), 
                (w1) -- [boson, MyYellow, ultra thick] (c1) -- [boson, MyYellow, ultra thick] (w2),  
                (w3) -- [boson, MyYellow, ultra thick] (c1),
                (c3) -- [ultra thick, MyBlue] (f1), 
                (f2) -- [ultra thick, MyBlue] (c3),  
              };
            \end{feynman}
        \end{tikzpicture}}}
        = 
        \vcenter{\hbox{
            \begin{tikzpicture}[scale=0.7]
                \begin{feynman}
                  \vertex (i) at (0, 0);
                  \vertex (e) at (0, 3);
                  \vertex[dot, MyYellow] (w1) at (0, 1.5) {};
                  \vertex (c1) at (1.5,1.5);
                  \vertex (f1) at (3.0,2.8) {};
                  \vertex (f2) at (3.0,0.2) {}; 
            
                  \diagram*{
                    (i) -- [double,double, thick] (e), 
                    (w1) -- [boson, MyYellow, ultra thick] (c1),
                    (c1) -- [ultra thick, MyBlue] (f1), 
                    (f2) -- [ultra thick, MyBlue] (c1),  
                  };
                \end{feynman}
            \end{tikzpicture}}} 
            + 
        \vcenter{\hbox{
            \begin{tikzpicture}[scale=0.7]
                \begin{feynman}
                  \vertex (i) at (0, 0);
                  \vertex (e) at (0, 3);
                  \vertex[dot, MyYellow] (w1) at (0, 2.75) {};
                  \vertex[dot, MyYellow] (w4) at (0, 0.5) {};
                  \vertex (c3) at (1.0, 1.5);
                  \vertex (c4) at (2.5, 1.5);
                  \vertex (f1) at (3.5,2.8) {};
                  \vertex (f2) at (3.5,0.2) {}; 
            
                  \diagram*{
                    (i) -- [double,double, thick] (e), 
                    (w1) -- [boson, MyYellow, ultra thick] (c3),
                    (w4) -- [boson, MyYellow, ultra thick] (c3),
                    (c4) -- [ultra thick, MyBlue] (f1), 
                    (f2) -- [ultra thick, MyBlue] (c4), 
                    (c3) -- [boson, ultra thick, MyYellow] (c4), 
                  };
                \end{feynman}
            \end{tikzpicture}}}  
                +
                \vcenter{\hbox{
                    \begin{tikzpicture}[scale=0.7]
                        \begin{feynman}
                          \vertex (i) at (0, 0);
                          \vertex (e) at (0, 3);
                          \vertex[dot, MyYellow] (w1) at (0, 2.75) {};
                          \vertex[dot, MyYellow] (w4) at (0, 0.5) {};
                          \vertex (c3) at (1.5, 1.5);
                          \vertex (f1) at (3,2.8) {};
                          \vertex (f2) at (3,0.2) {}; 
            
                        \diagram*{
                          (i) -- [double,double, thick] (e), 
                          (w1) -- [boson, MyYellow, ultra thick] (c3),
                          (w4) -- [boson, MyYellow, ultra thick] (c3),
                          (c3) -- [ultra thick, MyBlue] (f1), 
                          (f2) -- [ultra thick, MyBlue] (c3),  
                       };
                      \end{feynman}
                    \end{tikzpicture}}}  
                    + \cdots
    \end{aligned}
\end{equation}

\subsubsection*{Power-Law Divergences}\label{sec:power}

An important technical point is the presence 
of power-law divergences in our calculations. 
These divergences are unphysical 
and can be removed by local
counterterms. The power-law divergences
should be contrasted with logarithmic singularities,
which have consequences on large scales 
and capture the physical effect of short
modes that we have integrated out in the EFT. 

Let us first discuss the mass renormalization~\cite{Kol:2007rx}. 
Naively, there are infinite many 
diagrams renormalizing the point mass that
we need to compute even in the static case. 
However, we will see now that 
it is sufficient to ignore these diagrams and just
replace the  
bare point-particle mass $m$ with the physical
renormalized BH mass $M$ in all finial answers. 

The renormalized mass $M$ is defined as the dressed 0-point function on the worldline,
\begin{equation}
    (-iM) = \vcenter{\hbox{\begin{tikzpicture}[scale=0.7]
        \begin{feynman}
            \vertex (i) at (0,0);
            \vertex (e) at (0,3);
            \vertex[blob, scale=0.7] (f1) at (0, 1.5) {};

            \diagram*{
                (i) -- [double, double, thick] (f1),
                (f1) -- [double, double, thick] (e),
            };

        \end{feynman}
    \end{tikzpicture}}}
    = \vcenter{\hbox{\begin{tikzpicture}[scale=0.7]
        \begin{feynman}
            \vertex (i) at (0,0);
            \vertex (e) at (0,3);
            \vertex[dot, MyYellow] (f1) at (0, 1.5) {};

            \diagram*{
                (i) -- [double, double, thick] (e),
            };
        \end{feynman}
    \end{tikzpicture}}}
    \quad + \quad 
    \vcenter{\hbox{\begin{tikzpicture}[scale=0.7]
        \begin{feynman}
            \vertex (i) at (0,0);
            \vertex (e) at (0,3);
            \vertex[dot, MyYellow] (f1) at (0,2.5) {};
            \vertex[dot, MyYellow] (f2) at (0, 0.5) {};

            \diagram*{
                (i) -- [double, double, thick] (e),
                (f1) -- [boson, MyYellow, ultra thick, bend left = 50, edge label = $h_{00}$] (f2),
            };
        \end{feynman}   
    \end{tikzpicture}}}
    \quad + 
        \vcenter{\hbox{\begin{tikzpicture}[scale=0.7]
        \begin{feynman}
            \vertex (i) at (0,0);
            \vertex (e) at (0,3);
            \vertex[dot, MyYellow] (f1) at (0,2.5) {};
            \vertex[dot, MyYellow] (f2) at (0, 1.5) {};
            \vertex[dot, MyYellow] (f3) at (0, 0.5) {};

            \diagram*{
                (i) -- [double, double, thick] (e),
                (f1) -- [boson, MyYellow, ultra thick, bend left = 70, edge label = $h_{00}$] (f2),
                (f2) --[boson, MyYellow, ultra thick, bend left = 70, edge label = $h_{00}$] (f3),
            };
        \end{feynman}   
    \end{tikzpicture}}}
    \quad + \cdots
\end{equation}
Physically, this renormalized mass $M$ can be viewed as the BH asymptotic Arnowitt-Deser-Misner (ADM) mass measured at the spatial infinity. 
% Typically, we do not expect the BH ADM mass to be the same as the point particle bare mass, but to be the bare mass dressed with potential gravitons. 
Importantly, 
all loop corrections above lead to power-law divergences, 
and hence the presence of these diagrams does 
not change our power counting.
These divergences can be absorbed into 
local counterterms. Thus, for all practical 
applications we can just first perform all calculations with the point particle bare mass $m$ and then replace it with the physical renormalized mass $M$.
% An alternative prescription is to work in dimensional 
% regularization, where all power-law divergences
% disappear and hence there would be no difference 
% between $M$ and $m$. This is the approach that we will 
% employ in what follows.
 % For the future convenience, we will ignore such mass renormalization and keep on using the bare mass $m$ in the EFT calculation with the red dot in the diagram.

The second problematic aspect is the field strength renormalization of $h$. For example, if we assume a cutoff 
regularization, the following diagram 
would be linearly divergent:
\begin{equation}\label{EFT Building Block Diagram}
    \vcenter{\hbox{\begin{tikzpicture}[scale=0.7]
        \begin{feynman}
            \vertex (i) at (0, 0);
                  \vertex (e) at (0, 3);
                  \vertex[dot, MyYellow] (w1) at (0, 2.5) {};
                  \vertex[dot, MyYellow] (w2) at (0, 0.5) {};
                  \vertex (c1) at (1,1.5);
                  \vertex (c2) at (2.5,1.5);
            
                  \diagram*{
                    (i) -- [double,double, thick] (e), 
                    (w1) -- [boson, MyYellow, ultra thick] (c1),
                    (w2) -- [boson, MyYellow, ultra thick] (c1),
                    (c1) -- [boson, MyYellow, ultra thick, reversed momentum = {[arrow style = MyYellow] $\boldsymbol{k}$}] (c2),  
                  };
        \end{feynman}
    \end{tikzpicture}}}
    \sim \frac{-i}{\boldsymbol{k}^2} \Lambda + {\rm finite\, term} ~.
\end{equation}
In principle, the divergent part can be absorbed into the renormalized field strength $h$.
We need to keep the physical finite part though 
as it
produces a nonvanishing 
contribution to the amplitude.

An alternative option is to work in dimensional 
regularization, where all power-law divergences
are automatically set to zero. This means, in particular, that there would be no difference 
between $M$ and $m$. This is the approach that we will 
employ in what follows.

% Both the mass and field divergences can be absorbed by local counter terms without causing any physical consequences. 
% In dimensional regularization both these divergences disappear,
% which is why we stick to this prescription in what follows. 

\subsubsection*{Off-shell Amplitudes}\label{Off-shell Amplitudes}

It is useful to focus on a blob diagram inside 
the pyramid and ``amputate'' the worldline and  
external scalar legs. This reduced blob 
diagram can be treated now as an off-shell
amplitude. Let us estimate how this 
amplitude scales with external momentum 
in Fourier space. 
To that end we introduce 
$N_m^{\rm Pyr}$, $P_h^{\rm Pyr}$, $N_h^{\rm Pyr}$ and $V^{\rm Pyr}$ to denote the number of worldline vertices, bulk graviton propagators, external graviton legs, and the bulk vertices in a pyramid diagram, respectively. 

The point mass contribution 
is a delta function in position space. 
Hence, each point mass 
leg scales as $|\boldsymbol{k}|^{3}$, giving a total momentum
$|\boldsymbol{k}|^{3N_m^{\rm Pyr}}$.
Every graviton vertex $\sim \int d^3 x \d\d$ has two derivatives,
i.e., it scales
as $|\boldsymbol{k}|^2$, giving a total of 
$|\boldsymbol{k}|^{2 V^{\rm Pyr}}$.
All together, this gives a total number
of momenta in the numerator $3N_m^{\rm Pyr} + 2 V^{\rm Pyr}$.
The momenta in the denominator come from 
static propagators $\sim k^{-2}$. 
Then the cumulative number of the momenta in
the amplitude denominator from graviton propagators connecting point particles, scalar fields, and
bulk vertices is
$2 (N_m^{\rm Pyr} +  N_h^{\rm Pyr}+ P_h^{\rm Pyr})$. 
Note that the momenta flowing into our diagram from the connecting point
must satisfy momentum conservation.
The number of quantum loops in a pyramid diagram satisfies
\begin{equation}
    L^{\rm Pyr} = P_h^{\rm Pyr}  -V^{\rm Pyr}  + N_h^{\rm Pyr}\equiv 0\,.
\end{equation} 
Thus yields the total momentum scaling of an off-shell blob diagram 
\begin{equation}
        D = N_m^{\rm Pyr} - 3
\end{equation}

Let us study now off-shell scattering amplitudes shown in Eq.\eqref{Pyramid Structure}. Recall that they serve 
as building blocks for our EFT one-point function calculation. From the symmetry perspective, this amplitude should be ${\rm SO}(3)$ invariants. 
Suppose that the two external $\varphi$ fields have momentum $\boldsymbol{k_1}$ and $\boldsymbol{k_2}$. Then the amplitude of a generic pyramid diagram can only be a function of $|\boldsymbol{k_1}+\boldsymbol{k_2}|$ and $\boldsymbol{k_1} \cdot \boldsymbol{k_2}$. By using the momentum counting formula $D = N_{m}^{\rm Pyr} - 3 $ and taking into account two spatial derivatives acting on external $\varphi$ legs, we get the following estimate the 
total amplitude: 

\begin{equation}\label{1PI Amplitude}
    \vcenter{\hbox{\begin{tikzpicture}
        \begin{feynman}
          \vertex (i) at (0, 0);
          \vertex (e) at (0, 3);
          \vertex[dot, MyYellow] (w1) at (0, 2.5) {};
          \vertex[dot, MyYellow] (w2) at (0, 2.0) {};
          \vertex[dot, MyYellow] (w3) at (0, 1.0) {};
          \vertex[dot, MyYellow] (w4) at (0,0.5) {};
          \node at (0.5,1.6) {$\cdot$};
          \node at (0.5,1.4) {$\cdot$};
          \node at (0.5,1.5) {$\cdot$};
          \node at (2.7,1.5) {$\cdot$};
          \node at (2.7,1.6) {$\cdot$};
          \node at (2.7,1.4) {$\cdot$};
          \vertex[blob, scale =2] (c1) at (1.5, 1.5) {};
          \node[circle, fill=white, scale = 0.8] (c2) at (1.5,1.5) {static};
          \vertex[dot, black] (c3) at (3.5, 1.5) {};
          \vertex (f1) at (5,2.8) {};
          \vertex (f2) at (5,0.2) {}; 
    
          \diagram*{
            (c1) -- [boson, ultra thick, MyYellow, bend right=30] (c3),
            (c1) -- [boson, ultra thick, MyYellow, bend left=30] (c3),
            (i) -- [double,double, thick, edge label = $N_m^{\rm 1PI}$] (e), 
            (w1) -- [boson, MyYellow, ultra thick] (c1) -- [boson, MyYellow, ultra thick] (w2),  
            (w3) -- [boson, MyYellow, ultra thick] (c1) -- [boson, MyYellow, ultra thick] (w4),
            (c3) -- [ultra thick, MyBlue, reversed momentum ={[arrow style = MyBlue, arrow shorten = 0.25] $\boldsymbol{k_1}$}] (f1), 
            (f2) -- [ultra thick, MyBlue, momentum ={[arrow style = MyBlue, arrow shorten = 0.25] $\boldsymbol{k_2}$}] (c3),  
          };
        \end{feynman}
    \end{tikzpicture}}}
        \sim |\boldsymbol{k_1} + \boldsymbol{k_2}|^{N_m^{\rm Pyr}-3} \(\boldsymbol{k_1} \cdot \boldsymbol{k_2}\) ~.
\end{equation}

\subsection{EFT Diagramatic Recurrence Relation}\label{EFT Diagramatic Recurrence Relation}

As we discussed in Section \ref{Ladder and Pyramid}, 
the ladder structure hints on a relationship
between EFT PN diagrams of various orders. 
In other words, it suggests that higher order PN correction can be built from the lower order PN diagrams. 
Imagine that we want to compute the $n$PN correction to the 
scalar field profile $\delta\varphi^n$. 
It can be built from $(n-1)$PN, $(n-2)$PN, $\ldots$ 
by combining them with appropriate 
off-shell scattering amplitudes.

Suppose that we have a diagram made of two ladders.
Each ladder has $j_1$ and $j_2$ mass insertions, respectively. 
This diagram can be represented 
as a product of two diagrammatic 
elements.  
The first one is a single ladder corresponding to a one-point
function correction produced by $j_1$ mass insertions.
The second element is 
an off-shell amplitude at $j_2$PN order.
We have

\begin{equation}\label{General 2 Ladder Recurrence}
    \begin{aligned}
    \vcenter{\hbox{\begin{tikzpicture}[scale=0.7]
        \begin{feynman}
            \vertex (i) at (0, 0);
            \vertex (e) at (0, 4);
            \vertex (e1)at (0, 2);
            \vertex[dot, MyYellow] (w4) at (0, 3.6) {};
            \vertex[dot, MyYellow] (w5) at (0, 3.3) {};
            \vertex[dot, MyYellow] (w6) at (0, 2.6) {};
            \vertex[blob, scale =0.7] (c2) at ( 1, 3.1) {};
            \vertex[dot, MyYellow] (w7) at (0, 1.5) {};
            \vertex[dot, MyYellow] (w8) at (0, 1.2) {};
            \vertex[dot, MyYellow] (w9) at (0, 0.5) {};
            \vertex[blob, scale =0.7] (c3) at ( 1, 1) {};
            \vertex[dot, MyBlue, label= $\boldsymbol{x}$] (f1) at (4.0,3.6) {};
            \vertex[crossed dot, MyBlue] (fs) at (4.0,0.4) {};
            \vertex[dot, black] (d2) at (2.5, 3.1) {};
            \vertex[dot, black] (d3) at (2.5, 1) {};
            \node at (0.25,0.95) {$\cdot$};
            \node at (0.25,0.85) {$\cdot$};
            \node at (0.25,0.75) {$\cdot$};
            \node at (0.25,3.05) {$\cdot$};
            \node at (0.25,2.95) {$\cdot$};
            \node at (0.25,2.85) {$\cdot$};
            \node at (1.9,1.05) {$\cdot$};
            \node at (1.9,0.95) {$\cdot$};
            \node at (1.9,0.85) {$\cdot$};
            \node at (1.9,3.15) {$\cdot$};
            \node at (1.9,3.05) {$\cdot$};
            \node at (1.9,2.95) {$\cdot$};
      
            \diagram*{
              (i) -- [double,double, thick, edge label = $j_1$] (e1),
              (e1) -- [double, double, thick, edge label = $j_2$] (e),
              (w4) -- [boson, MyYellow, ultra thick] (c2) -- [boson, MyYellow, ultra thick] (w5),  
              (w6) -- [boson, MyYellow, ultra thick] (c2),
              (w7) -- [boson, MyYellow, ultra thick] (c3) -- [boson, MyYellow, ultra thick] (w8),  
              (w9) -- [boson, MyYellow, ultra thick] (c3),
              (c2) -- [boson, ultra thick, MyYellow, bend left=30] (d2),
              (c2) -- [boson, ultra thick, MyYellow, bend right=30] (d2),
              (c3) -- [boson, ultra thick, MyYellow, bend left=30] (d3),
              (c3) -- [boson, ultra thick, MyYellow, bend right=30] (d3),
              (d2) -- [ultra thick, MyBlue, reversed momentum ={[arrow style = MyBlue, arrow shorten = 0.25] $\boldsymbol{k}$}] (f1), 
              (fs) -- [ultra thick, MyBlue] (d3),  
              (d2) -- [ultra thick, MyBlue, momentum ={[arrow style = MyBlue, arrow shorten = 0.25] $\boldsymbol{k_1}$}] (d3),
            };
          \end{feynman}        
        
    \end{tikzpicture}}}
    &= 
    \int \frac{d^3 \boldsymbol{k_1}}{(2\pi)^3} \int \frac{d^3 \boldsymbol{k}}{(2\pi)^3} 
    \(\vcenter{\hbox{\begin{tikzpicture}[scale=0.7]
        \begin{feynman}
          \vertex (i) at (0, 0);
          \vertex (e) at (0, 3);
          \vertex[dot, MyYellow] (w1) at (0, 2.5) {};
          \vertex[dot, MyYellow] (w2) at (0, 2.0) {};
          \vertex[dot, MyYellow] (w3) at (0, 1.0) {};
          \vertex[dot, MyYellow] (w4) at (0,0.5) {};
          \node at (0.5,1.6) {$\cdot$};
          \node at (0.5,1.4) {$\cdot$};
          \node at (0.5,1.5) {$\cdot$};
          \node at (2.7,1.5) {$\cdot$};
          \node at (2.7,1.6) {$\cdot$};
          \node at (2.7,1.4) {$\cdot$};
          \vertex[blob, scale = 1.2] (c1) at (1.5, 1.5) {};
          \vertex[dot, black] (c3) at (3.5, 1.5) {};
          \vertex (f1) at (5,2.8) {};
          \vertex[crossed dot, MyBlue] (f2) at (4.8,0.2) {}; 
    
          \diagram*{
            (c1) -- [boson, ultra thick, MyYellow, bend right=30] (c3),
            (c1) -- [boson, ultra thick, MyYellow, bend left=30] (c3),
            (i) -- [double,double, thick, edge label = $j_1$] (e), 
            (w1) -- [boson, MyYellow, ultra thick] (c1) -- [boson, MyYellow, ultra thick] (w2),  
            (w3) -- [boson, MyYellow, ultra thick] (c1) -- [boson, MyYellow, ultra thick] (w4),
            (c3) -- [ultra thick, MyBlue, reversed momentum ={[arrow style = MyBlue, arrow shorten = 0.25] $\boldsymbol{k_1}$}] (f1), 
            (f2) -- [ultra thick, MyBlue] (c3),  
          };
        \end{feynman}
    \end{tikzpicture}}}
    \) \\
    & \quad \times 
    \(\vcenter{\hbox{\begin{tikzpicture}[scale=0.7]
        \begin{feynman}
          \vertex (i) at (0, 0);
          \vertex (e) at (0, 3);
          \vertex[dot, MyYellow] (w1) at (0, 2.5) {};
          \vertex[dot, MyYellow] (w2) at (0, 2.0) {};
          \vertex[dot, MyYellow] (w3) at (0, 1.0) {};
          \vertex[dot, MyYellow] (w4) at (0,0.5) {};
          \node at (0.5,1.6) {$\cdot$};
          \node at (0.5,1.4) {$\cdot$};
          \node at (0.5,1.5) {$\cdot$};
          \node at (2.7,1.5) {$\cdot$};
          \node at (2.7,1.6) {$\cdot$};
          \node at (2.7,1.4) {$\cdot$};
          \vertex[blob, scale = 1.2] (c1) at (1.5, 1.5) {};
          \vertex[dot, black] (c3) at (3.5, 1.5) {};
          \vertex (f1) at (5,2.8) {};
          \vertex (f2) at (5,0.2) {}; 
    
          \diagram*{
            (c1) -- [boson, ultra thick, MyYellow, bend right=30] (c3),
            (c1) -- [boson, ultra thick, MyYellow, bend left=30] (c3),
            (i) -- [double,double, thick, edge label = $j_2$] (e), 
            (w1) -- [boson, MyYellow, ultra thick] (c1) -- [boson, MyYellow, ultra thick] (w2),  
            (w3) -- [boson, MyYellow, ultra thick] (c1) -- [boson, MyYellow, ultra thick] (w4),
            (c3) -- [ultra thick, MyBlue, reversed momentum ={[arrow style = MyBlue, arrow shorten = 0.25] $\boldsymbol{k}$}] (f1), 
            (f2) -- [ultra thick, MyBlue, reversed momentum ={[arrow style = MyBlue, arrow shorten = 0.25] $\boldsymbol{k_1}$}] (c3),  
          };
        \end{feynman}
    \end{tikzpicture}}}
    \) \times \frac{-i}{\boldsymbol{k}^2} e^{i\boldsymbol{k} \cdot \boldsymbol{x}} ~.
    \end{aligned}
\end{equation}
% With this equation, by making use of Eq.~\eqref{1PI Amplitude}, 
It is convenient to use the inverse Laplacian $\partial^{-2}$ and the integration variable $\boldsymbol{k'}$ 
defined as $\boldsymbol{k} = \boldsymbol{k'}+\boldsymbol{k_1}$.
Then the above integral can be rewritten as
\begin{equation}
    \begin{aligned}
       \vcenter{\hbox{\begin{tikzpicture}[scale=0.7]
        \begin{feynman}
            \vertex (i) at (0, 0);
            \vertex (e) at (0, 4);
            \vertex (e1)at (0, 2);
            \vertex[dot, MyYellow] (w4) at (0, 3.6) {};
            \vertex[dot, MyYellow] (w5) at (0, 3.3) {};
            \vertex[dot, MyYellow] (w6) at (0, 2.6) {};
            \vertex[blob, scale =0.7] (c2) at ( 1, 3.1) {};
            \vertex[dot, MyYellow] (w7) at (0, 1.5) {};
            \vertex[dot, MyYellow] (w8) at (0, 1.2) {};
            \vertex[dot, MyYellow] (w9) at (0, 0.5) {};
            \vertex[blob, scale =0.7] (c3) at ( 1, 1) {};
            \vertex[dot, MyBlue, label= $\boldsymbol{x}$] (f1) at (4.0,3.6) {};
            \vertex[crossed dot, MyBlue] (fs) at (4.0,0.4) {};
            \vertex[dot, black] (d2) at (2.5, 3.1) {};
            \vertex[dot, black] (d3) at (2.5, 1) {};
            \node at (0.25,0.95) {$\cdot$};
            \node at (0.25,0.85) {$\cdot$};
            \node at (0.25,0.75) {$\cdot$};
            \node at (0.25,3.05) {$\cdot$};
            \node at (0.25,2.95) {$\cdot$};
            \node at (0.25,2.85) {$\cdot$};
            \node at (1.9,1.05) {$\cdot$};
            \node at (1.9,0.95) {$\cdot$};
            \node at (1.9,0.85) {$\cdot$};
            \node at (1.9,3.15) {$\cdot$};
            \node at (1.9,3.05) {$\cdot$};
            \node at (1.9,2.95) {$\cdot$};
      
            \diagram*{
              (i) -- [double,double, thick, edge label = $j_1$] (e1),
              (e1) -- [double, double, thick, edge label = $j_2$] (e),
              (w4) -- [boson, MyYellow, ultra thick] (c2) -- [boson, MyYellow, ultra thick] (w5),  
              (w6) -- [boson, MyYellow, ultra thick] (c2),
              (w7) -- [boson, MyYellow, ultra thick] (c3) -- [boson, MyYellow, ultra thick] (w8),  
              (w9) -- [boson, MyYellow, ultra thick] (c3),
              (c2) -- [boson, ultra thick, MyYellow, bend left=30] (d2),
              (c2) -- [boson, ultra thick, MyYellow, bend right=30] (d2),
              (c3) -- [boson, ultra thick, MyYellow, bend left=30] (d3),
              (c3) -- [boson, ultra thick, MyYellow, bend right=30] (d3),
              (d2) -- [ultra thick, MyBlue, reversed momentum ={[arrow style = MyBlue, arrow shorten = 0.25] $\boldsymbol{k}$}] (f1), 
              (fs) -- [ultra thick, MyBlue] (d3),  
              (d2) -- [ultra thick, MyBlue, momentum ={[arrow style = MyBlue, arrow shorten = 0.25] $\boldsymbol{k_1}$}] (d3),
            };
          \end{feynman}        
        
    \end{tikzpicture}}}
    & \sim m^{j_2} \partial^{-2}\Bigg[\(\int \frac{d^3 \boldsymbol{k_1}}{(2\pi)^3} 
    \vcenter{\hbox{\begin{tikzpicture}[scale=0.5]
        \begin{feynman}
          \vertex (i) at (0, 0);
          \vertex (e) at (0, 3);
          \vertex[dot, MyYellow] (w1) at (0, 2.5) {};
          \vertex[dot, MyYellow] (w2) at (0, 2.0) {};
          \vertex[dot, MyYellow] (w3) at (0, 1.0) {};
          \vertex[dot, MyYellow] (w4) at (0,0.5) {};
          \node at (0.5,1.6) {$\cdot$};
          \node at (0.5,1.4) {$\cdot$};
          \node at (0.5,1.5) {$\cdot$};
          \node at (2.7,1.5) {$\cdot$};
          \node at (2.7,1.6) {$\cdot$};
          \node at (2.7,1.4) {$\cdot$};
          \vertex[blob, scale = 1.0] (c1) at (1.5, 1.5) {};
          \vertex[dot, black] (c3) at (3.5, 1.5) {};
          \vertex (f1) at (5,2.8) {};
          \vertex[crossed dot, MyBlue] (f2) at (4.8,0.2) {}; 
    
          \diagram*{
            (c1) -- [boson, ultra thick, MyYellow, bend right=30] (c3),
            (c1) -- [boson, ultra thick, MyYellow, bend left=30] (c3),
            (i) -- [double,double, thick, edge label = $j_1$] (e), 
            (w1) -- [boson, MyYellow, ultra thick] (c1) -- [boson, MyYellow, ultra thick] (w2),  
            (w3) -- [boson, MyYellow, ultra thick] (c1) -- [boson, MyYellow, ultra thick] (w4),
            (c3) -- [ultra thick, MyBlue, reversed momentum ={[arrow style = MyBlue, arrow shorten = 0.25] $\boldsymbol{k_1}$}] (f1), 
            (f2) -- [ultra thick, MyBlue] (c3),  
          };
        \end{feynman}
    \end{tikzpicture}}}
    \boldsymbol{k_1} e^{i\boldsymbol{k_1}\cdot \boldsymbol{x}}\) \cdot \(\int \frac{d^3 \boldsymbol{k'}}{(2\pi)^3} \frac{\boldsymbol{k'}}{|\boldsymbol{k'}|^{3-j_2}} e^{i\boldsymbol{k'}\cdot \boldsymbol{x}}\) \\
    & \quad + \(\int \frac{d^3 \boldsymbol{k_1}}{(2\pi)^3} 
    \vcenter{\hbox{\begin{tikzpicture}[scale=0.5]
        \begin{feynman}
          \vertex (i) at (0, 0);
          \vertex (e) at (0, 3);
          \vertex[dot, MyYellow] (w1) at (0, 2.5) {};
          \vertex[dot, MyYellow] (w2) at (0, 2.0) {};
          \vertex[dot, MyYellow] (w3) at (0, 1.0) {};
          \vertex[dot, MyYellow] (w4) at (0,0.5) {};
          \node at (0.5,1.6) {$\cdot$};
          \node at (0.5,1.4) {$\cdot$};
          \node at (0.5,1.5) {$\cdot$};
          \node at (2.7,1.5) {$\cdot$};
          \node at (2.7,1.6) {$\cdot$};
          \node at (2.7,1.4) {$\cdot$};
          \vertex[blob, scale = 1.0] (c1) at (1.5, 1.5) {};
          \vertex[dot, black] (c3) at (3.5, 1.5) {};
          \vertex (f1) at (5,2.8) {};
          \vertex[crossed dot, MyBlue] (f2) at (4.8,0.2) {}; 
    
          \diagram*{
            (c1) -- [boson, ultra thick, MyYellow, bend right=30] (c3),
            (c1) -- [boson, ultra thick, MyYellow, bend left=30] (c3),
            (i) -- [double,double, thick, edge label = $j_1$] (e), 
            (w1) -- [boson, MyYellow, ultra thick] (c1) -- [boson, MyYellow, ultra thick] (w2),  
            (w3) -- [boson, MyYellow, ultra thick] (c1) -- [boson, MyYellow, ultra thick] (w4),
            (c3) -- [ultra thick, MyBlue, reversed momentum ={[arrow style = MyBlue, arrow shorten = 0.25] $\boldsymbol{k_1}$}] (f1), 
            (f2) -- [ultra thick, MyBlue] (c3),  
          };
        \end{feynman}
    \end{tikzpicture}}}
    \boldsymbol{k_1}^2 e^{i\boldsymbol{k_1}\cdot \boldsymbol{x}}\) \(\int \frac{d^3 \boldsymbol{k'}}{(2\pi)^3} \frac{1}{|\boldsymbol{k'}|^{3-j_2}} e^{i\boldsymbol{k'}\cdot \boldsymbol{x}}\) \Bigg] \\
    & = m^{j_2} \partial^{-2} \[ \(\frac{j_2 x^i}{r^{j_2+2}} \partial_i - \frac{1}{r^{j_2}} \partial^2 \)     \vcenter{\hbox{\begin{tikzpicture}[scale=0.5]
        \begin{feynman}
          \vertex (i) at (0, 0);
          \vertex (e) at (0, 3);
          \vertex[dot, MyYellow] (w1) at (0, 2.5) {};
          \vertex[dot, MyYellow] (w2) at (0, 2.0) {};
          \vertex[dot, MyYellow] (w3) at (0, 1.0) {};
          \vertex[dot, MyYellow] (w4) at (0,0.5) {};
          \node at (0.5,1.6) {$\cdot$};
          \node at (0.5,1.4) {$\cdot$};
          \node at (0.5,1.5) {$\cdot$};
          \node at (2.7,1.5) {$\cdot$};
          \node at (2.7,1.6) {$\cdot$};
          \node at (2.7,1.4) {$\cdot$};
          \vertex[blob, scale = 1.0] (c1) at (1.5, 1.5) {};
          \vertex[dot, black] (c3) at (3.5, 1.5) {};
          \vertex[dot, MyBlue, label= $\boldsymbol{x}$] (f1) at (5,2.8) {};
          \vertex[crossed dot, MyBlue] (f2) at (4.8,0.2) {}; 
    
          \diagram*{
            (c1) -- [boson, ultra thick, MyYellow, bend right=30] (c3),
            (c1) -- [boson, ultra thick, MyYellow, bend left=30] (c3),
            (i) -- [double,double, thick, edge label = $j_1$] (e), 
            (w1) -- [boson, MyYellow, ultra thick] (c1) -- [boson, MyYellow, ultra thick] (w2),  
            (w3) -- [boson, MyYellow, ultra thick] (c1) -- [boson, MyYellow, ultra thick] (w4),
            (c3) -- [ultra thick, MyBlue] (f1), 
            (f2) -- [ultra thick, MyBlue] (c3),  
          };
        \end{feynman}
    \end{tikzpicture}}}
    \]
    \end{aligned}
\end{equation}

Based on this factorization property,
all higher order PN corrections can be constructed
from lower order PN diagrams. 
Suppose that we want to compute 
an $n$PN correction $\delta\varphi^n(\boldsymbol{x})$.
It can be built from a sum of all $(n-j)$PN corrections $\delta\varphi^{n-j}(\boldsymbol{x})$ ($j=0,1\cdots n$) by combining them with the 
corresponding off-shell amplitudes

\begin{eBox}
\begin{equation}\label{General Diagramatic Recurrence Relation}
\begin{aligned}
    \delta\varphi^{n}(\boldsymbol{x}) &= \sum_{j=1}^{n}  \int \frac{d^3 \boldsymbol{k_1}}{(2\pi)^3} \int \frac{d^3 \boldsymbol{k}}{(2\pi)^3}\delta\varphi^{n-j}_{\boldsymbol{k_1}} \times
    \left( \sum_{\substack{\text{all} \\ \text{possible} \\ \text{diagrams}}}
    \vcenter{\hbox{\begin{tikzpicture}
        \begin{feynman}
          \vertex (i) at (0, 0);
          \vertex (e) at (0, 3);
          \vertex[dot, MyYellow] (w1) at (0, 2.5) {};
          \vertex[dot, MyYellow] (w2) at (0, 2.0) {};
          \vertex[dot, MyYellow] (w3) at (0, 1.0) {};
          \vertex[dot, MyYellow] (w4) at (0,0.5) {};
          \node at (0.5,1.6) {$\cdot$};
          \node at (0.5,1.4) {$\cdot$};
          \node at (0.5,1.5) {$\cdot$};
          \node at (2.7,1.5) {$\cdot$};
          \node at (2.7,1.6) {$\cdot$};
          \node at (2.7,1.4) {$\cdot$};
          \vertex[blob, scale =2] (c1) at (1.5, 1.5) {};
          \node[circle, fill=white, scale = 0.8] (c2) at (1.5,1.5) {static};
          \vertex[dot, black] (c3) at (3.5, 1.5) {};
          \vertex (f1) at (5,2.8) {};
          \vertex (f2) at (5,0.2) {}; 
    
          \diagram*{
            (c1) -- [boson, ultra thick, MyYellow, bend right=30] (c3),
            (c1) -- [boson, ultra thick, MyYellow, bend left=30] (c3),
            (i) -- [double,double, thick, edge label = $j$] (e), 
            (w1) -- [boson, MyYellow, ultra thick] (c1) -- [boson, MyYellow, ultra thick] (w2),  
            (w3) -- [boson, MyYellow, ultra thick] (c1) -- [boson, MyYellow, ultra thick] (w4),
            (c3) -- [ultra thick, MyBlue, reversed momentum ={[arrow style = MyBlue, arrow shorten = 0.25] $\boldsymbol{k}$}] (f1), 
            (f2) -- [ultra thick, MyBlue, reversed momentum ={[arrow style = MyBlue, arrow shorten = 0.25] $\boldsymbol{k_1}$}] (c3),  
          };
        \end{feynman}
    \end{tikzpicture}}}
    \right)
        \times \frac{-i}{\boldsymbol{k}^2} e^{i\boldsymbol{k}\cdot \boldsymbol{x}} \\
        &\sim \sum_{j=1}^{n} m^j \partial^{-2}\(\frac{j x^i}{r^{j+2}} \partial_i \delta\varphi^{n-j}(\boldsymbol{x}) - \frac{1}{r^j}\partial^2 \delta\varphi^{n-j}(\boldsymbol{x})\) ~,
\end{aligned}
\end{equation}
\end{eBox}
where inside the large brakets we sum over all possible off-shell amplitudes with $j$  worldline point mass vertices. 
Importantly, when the operator $\partial^{-2}$ 
hits corrections scaling as $r^{-\ell-3}$,
it
gives rise to logarithmic divergences,
\begin{equation}
    \partial^{-2} \(r^{\ell} \times r^{-2 \ell-3} Y_{\ell m}(\theta,\phi)\) \sim r^{\ell} \times r^{-2 \ell-1} \ln\({r}{\mu}\) Y_{\ell m}(\theta,\phi) ~,
\end{equation}
where $\mu$ is the renormalization (momentum) scale (sliding scale). 
As we can see, the logarithmic divergences 
are
generically present 
for each $2\ell+1$ PN order graph in an arbitrary gauge. 
But we should notice that only the sum over all graphs
is physical, which we denote by $c_{2\ell + 1}$:
\begin{equation}
    \sum_{\substack{\text{all} \\ \text{possible} \\ \text{diagrams}}}
    \vcenter{\hbox{\begin{tikzpicture}
        \begin{feynman}
          \vertex (i) at (0, 0);
          \vertex (e) at (0, 3);
          \vertex[dot, MyYellow] (w1) at (0, 2.5) {};
          \vertex[dot, MyYellow] (w2) at (0, 2.0) {};
          \vertex[dot, MyYellow] (w3) at (0, 0.5) {};
          \node at (0.5,1.1) {$\cdot$};
          \node at (0.5,1.3) {$\cdot$};
          \node at (0.5,1.5) {$\cdot$};
          \vertex[blob, scale =2] (c1) at ( 2, 1.5) {};
          \node[circle, fill=white, scale = 0.8] at (2,1.5) {static};
          \vertex[dot, MyBlue, label= $\boldsymbol{x}$] (f1) at (3.5,2.8) {};
          \vertex[crossed dot, MyBlue] (fs) at (3.5,0.2) {}; 
    
          \diagram*{
            (i) -- [double,double, thick, edge label = $2\ell + 1$] (e), 
            (w1) -- [boson, MyYellow, ultra thick] (c1) -- [boson, MyYellow, ultra thick] (w2),  
            (w3) -- [boson, MyYellow, ultra thick] (c1),
            (c1) -- [ultra thick, MyBlue] (f1), 
            (fs) -- [ultra thick, MyBlue, edge label' = $r^{\ell}$] (c1),  
          };
        \end{feynman}
    \end{tikzpicture}}}
    = c_{2\ell + 1} \mathcal{E}_{i_1\cdots i_\ell} x^{i_1} \cdots x^{i_\ell} \times \(\frac{m}{r}\)^{2\ell + 1} \times 
    \ln\({r}{\mu}\) ~,
\end{equation}
Physically, if $c_{2\ell + 1}\neq 0$, the Love numbers will exhibit a running behavior. Indeed, the total term of order $r^{-\ell-1}$ in the test field profile is given by
\begin{equation}
\begin{aligned}
    & \quad \sum_{\substack{\text{all} \\ \text{possible} \\ \text{diagrams}}}
        \vcenter{\hbox{\begin{tikzpicture}
        \begin{feynman}
          \vertex (i) at (0, 0);
          \vertex (e) at (0, 3);
          \vertex[dot, MyYellow] (w1) at (0, 2.5) {};
          \vertex[dot, MyYellow] (w2) at (0, 2.0) {};
          \vertex[dot, MyYellow] (w3) at (0, 0.5) {};
          \node at (0.5,1.1) {$\cdot$};
          \node at (0.5,1.3) {$\cdot$};
          \node at (0.5,1.5) {$\cdot$};
          \vertex[blob, scale =2] (c1) at ( 2, 1.5) {};
          \node[circle, fill=white, scale = 0.8] at (2,1.5) {static};
          \vertex[dot, MyBlue, label= $\boldsymbol{x}$] (f1) at (3.5,2.8) {};
          \vertex[crossed dot, MyBlue] (fs) at (3.5,0.2) {}; 
    
          \diagram*{
            (i) -- [double,double, thick, edge label = $2\ell + 1$] (e), 
            (w1) -- [boson, MyYellow, ultra thick] (c1) -- [boson, MyYellow, ultra thick] (w2),  
            (w3) -- [boson, MyYellow, ultra thick] (c1),
            (c1) -- [ultra thick, MyBlue] (f1), 
            (fs) -- [ultra thick, MyBlue, edge label' = $r^{\ell}$] (c1),  
          };
        \end{feynman}
    \end{tikzpicture}}}
    +
        \vcenter{\hbox{\begin{tikzpicture}[scale=0.7]
        \begin{feynman}
            \vertex (i) at (0,0);
            \vertex (e) at (0,3);
            \node[circle, draw=repGreen, fill = repGreen, scale=1, label=left:$\lambda_\ell(\mu)$] (w1) at (0, 1.5);
            \vertex[dot, MyBlue, label= $\boldsymbol{x}$] (f1) at (1.5,2.8) {};
            \vertex[crossed dot, MyBlue] (fs) at (1.5,0.2) {}; 

            \diagram*{
                (i) -- [double, double, thick] (w1),
                (w1) -- [double, double, thick] (e),
                (f1) -- [MyBlue, ultra thick] (w1),
                (fs) -- [MyBlue, ultra thick, edge label' = $r^\ell$] (w1)
            };
        \end{feynman}
    \end{tikzpicture}}}
     \\
     & =  \mathcal{E}_{i_1\cdots i_\ell} x^{i_1}\cdots x^{i_\ell} \times \(\frac{m}{r}\)^{2\ell + 1} \times \Bigg(c_{2\ell + 1} \ln\({r}{\mu}\) + m^{-2\ell-1}\frac{(2\ell-1)!!}{4\pi} \lambda_{\ell}(\mu) \Bigg) ~.
\end{aligned}
\end{equation}
Since the physical one-point function should not depend on the renormalization scale $\mu$, the Love number needs to flow 
with energy. 
The corresponding classical RG flow is given by 
\begin{equation}\label{eq:rg}
    \mu \frac{d}{d\mu} \lambda_\ell(\mu) = -c_{2\ell + 1} m^{2\ell+1}\frac{4\pi}{(2\ell - 1)!!} ~.
\end{equation}
% If $c_{2\ell + 1}=0$, the love number will not have the running behaviour. 
We see that the condition of the absence 
of RG running for Love numbers is $c_{2\ell + 1}=0$.
To determine $c_{2\ell + 1}$, we have two ways. The first one is by summing over all the related EFT diagrams, and the second one is by matching to the UV theory to see whether there is any logarithmic terms. 
In the following sections, we will first give some specific examples of the EFT diagrammatic computations of the spin-1 dipole,
and general spin-0 and spin-2 electric perturbations.
These calculations indeed imply a vanishing $c_{2\ell+1}$. 
After that  we will match the EFT and UV calculations,
which confirm that the Love numbers indeed vanish.

\section{PN Corrections to External Fields: Explicit Calculations}\label{sec:spec}

Let us demonstrate the general arguments 
given above on a concrete example. 
In this section we will explicitly 
compute $(2\ell+1)$PN corrections 
to the spin-0,1,2 test field profiles. 
Our main result will be that the 
coefficient
$c_{2\ell+1}$
vanishes for all types of perturbations.
This means that the $(2\ell+1)$PN
corrections to the source terms vanish identically
for Schwarzschild black holes, and hence 
Love numbers do not run. 
We prove that for a general orbital number $\ell$
in the case of spin-0 and spin-2
fields, and for the dipolar sector ($\ell=1$)
in the Maxwell field case.

\subsection{Consistent Gauge}

As a first step, we choose a convenient
coordinate system. 
Since we are interested in the non-relativistic regime,  
it is customary to use the 3+1 metric decomposition 
based on the Kaluza-Klein reduction formula~\cite{Kol:2011vg}\,,
\begin{equation}\label{KK Reduction}
    ds^2 = e^{2\phi}(dt- \mathcal{A}_i dx^i)^2 - e^{-2\phi} \gamma_{ij} dx^i dx^j ~,
\end{equation}
 where $\phi, \mathcal{A}_i, \gamma_{ij}$ is the Newtonian gravitational potential (dilaton field), 
  the gravitomagnetic vector, 
  and three-dimensional (3D) metric fields, respectively. 
  They satisfy $\gamma^{ij}\gamma_{ik} \equiv \delta^{i}_k$ and $\mathcal{A}^i \equiv \gamma^{ij} \mathcal{A}_j$. 
  The background field decomposition 
  in this setup can be written as 
\begin{equation}
    \gamma_{ij} \equiv \delta_{ij} + \sigma_{ij}  \,,
\end{equation}
where $\sigma_{ij}$ is the fluctuating part. 

As a second step, we fix the gauge. 
This is an important aspect since all EFT and GR 
calculations, in practice, have to be
carried out in a specific gauge (i.e., coordinate system). 
Ideally, one would want to match manifestly 
gauge-independent results, such as cross sections. 
In principle, one could also directly 
match one-point functions provided that they are computed in 
similar gauges. 
We formalize this statement by defining a notion of 
a ``consistent gauge.'' We call an EFT gauge consistent 
with the underlying GR background metric if the EFT calculation
of the off-shell graviton   one-point function  
reproduces the background 
metric perturbatively. 
The use of a consistent gauge ensures a correct matching 
between IR (EFT) and UV (GR) observables.

In practice, we choose a gauge that matches isotropic Schwarzschild coordinates.
Recall that in these coordinates the BH
solution takes the following form,
\begin{equation}\label{eq:schiso}
    ds^2 = \(\frac{1-M/2r}{1+M/2r}\)^2 dt^2 - \(1+ M/2r\)^4 (dr^2 + r^2 d\Omega^2 ) ~,
\end{equation}
where $M$ is the BH mass. From the EFT
point of view, a gauge consistent with Eq.~\eqref{eq:schiso} 
can be fixed by the following requirements: 
\begin{equation}
\label{eq:isogauge}
    \sigma_{ij} = \sigma \delta_{ij}, \quad \mathcal{A}_i =0 ~.
\end{equation}
In the static case, it is sufficient to consider 
off-shell potential modes for both external fields and GR degrees of freedom.
There are no propagating degrees of freedom. 
Now let us expand the action in field perturbations. We have the following:
\begin{itemize}
    \item Point particle term: 
    \begin{equation}
        S_{\rm pp} = \int dt \Bigg( -m - m\phi - \frac{m\phi^2}{2} - \cdots - \frac{m\phi^n}{n!}\Bigg) ~,        
    \end{equation} 

    \item Bulk graviton term:
    the EH action in the static limit takes the form
    \begin{equation}\label{Static Einstin-Hilbert Action}
        S_{\rm EH} = - \frac{1}{16 \pi} \int dt d^3 \boldsymbol{x} \sqrt{\gamma} \( {-R[\gamma]} +2 \gamma^{ij}\partial_i \phi  \partial_j \phi - \frac{1}{4} e^{4\phi} {\gamma^{ik}\gamma^{jl}\mathcal{F}_{ij}
        \mathcal{F}_{kl}}\) ~, 
    \end{equation}
    where $R[\gamma]$ is the 3D Ricci scalar of $\gamma_{ij}$, and $\mathcal{F}_{ij} = \partial_i \mathcal{A}_j - \partial_j \mathcal{A}_i$. Perturbations in the isotropic gauge take the form
    
    \begin{equation}
        \begin{aligned}
            S_{\rm EH}^{(2)} &= \frac{1}{8\pi} \int dt d^3 \boldsymbol{x} \(-\partial_i\phi \partial^i \phi + \frac{1}{4}\partial_i \sigma \partial^i \sigma\)  ~, \\
            S_{\rm EH}^{(3)} & = \frac{1}{8\pi} \int dt d^3 \boldsymbol{x} \(-\frac{3}{8} \sigma\partial_i \sigma \partial^i \sigma - \frac{1}{2} \sigma \partial_i \phi \partial^i \phi \) ~, \\
            & \cdots \cdots 
        \end{aligned}
    \end{equation}
    whilst the two-point correlation functions of $\phi$ and $\sigma$ are given by
\begin{equation}
    \langle\phi(\boldsymbol{x_1}) \phi(\boldsymbol{x_2}) \rangle = 
    \begin{tikzpicture}[baseline={([yshift=-2.0 ex]current bounding box.center)}]
        \begin{feynman}
            \vertex (a) at (0,0);
            \vertex (b) at (2,0);
            
            \diagram*{
                (a) -- [dashed, ultra thick, MyRed, edge label = $\phi$] (b), 
            };
        \end{feynman}
    \end{tikzpicture}
    = 4\pi \int \frac{d^3 \boldsymbol{k}}{(2\pi)^3} \frac{-i}{\boldsymbol{k}^2} e^{i\boldsymbol{k} \cdot (\boldsymbol{x_1}-\boldsymbol{x_2})} \propto \frac{1}{|\boldsymbol{x_1}-\boldsymbol{x_2}|} ~,
\end{equation}
\begin{equation}
    \langle\sigma(\boldsymbol{x_1}) \sigma(\boldsymbol{x_2}) \rangle = 
    \begin{tikzpicture}[baseline={([yshift=-1.5 ex]current bounding box.center)}]
        \begin{feynman}
            \vertex (a) at (0,0);
            \vertex (b) at (2,0);
            
            \diagram*{
                (a) -- [boson, MyYellow, ultra thick, edge label = $\sigma$] (b), 
            };
        \end{feynman}
    \end{tikzpicture}
    =-16\pi \int \frac{d^3 \boldsymbol{k}}{(2\pi)^3} \frac{-i}{\boldsymbol{k}^2} e^{i\boldsymbol{k} \cdot (\boldsymbol{x_1}-\boldsymbol{x_2})} \propto \frac{1}{|\boldsymbol{x_1}-\boldsymbol{x_2}|} ~.\\    
\end{equation}
\end{itemize}
Feynman rules for the above interactions are presented in Appendix \ref{Feynman Rule}.  For convenience, some of them are
presented below:
\begin{equation}
    \begin{aligned}
            \begin{tikzpicture}[baseline={([yshift=-0.5 ex]current bounding box.center)},scale=0.7]
                \begin{feynman}
                    \vertex (c1) at (0,0);
                    \vertex (c2) at (1.5, 0);
                    \vertex (c3) at (-1.5, 1);
                    \vertex (c4) at (-1.5, -1) ;

                    \diagram*{
                        (c1) -- [boson, MyYellow, ultra thick] (c2), 
                        (c1) -- [MyRed, dashed, ultra thick, reversed momentum'={[arrow style = MyRed, arrow shorten = 0.25] $\boldsymbol{k}$}] (c3),
                        (c1) -- [MyRed, dashed, ultra thick, reversed momentum={[arrow style = MyRed, arrow shorten = 0.25] $\boldsymbol{p}$}] (c4), 
                    };
                \end{feynman}
            \end{tikzpicture}
            &= i \frac{1}{8\pi} \boldsymbol{k} \cdot \boldsymbol{p} ~,
            \begin{tikzpicture}[baseline={([yshift=-0.5 ex]current bounding box.center)},scale=0.7]
                \begin{feynman}
                    \vertex (c1) at (0,0);
                    \vertex (c2) at (1.5, 0);
                    \vertex (c3) at (-1.5, 1);
                    \vertex (c4) at (-1.5, -1) ;

                    \diagram*{
                        (c1) -- [boson, MyYellow, ultra thick, reversed momentum={[arrow style = MyYellow, arrow shorten = 0.25] $\boldsymbol{k}_3$}] (c2), 
                        (c1) -- [boson, MyYellow, ultra thick, reversed momentum'={[arrow style = MyYellow, arrow shorten = 0.25] $\boldsymbol{k}_1$}] (c3),
                        (c1) -- [boson, MyYellow, ultra thick, reversed momentum={[arrow style = MyYellow, arrow shorten = 0.25] $\boldsymbol{k}_2$}] (c4), 
                    };
                \end{feynman}
            \end{tikzpicture}
            = i \frac{3}{32 \pi} (\boldsymbol{k_1} \cdot \boldsymbol{k_2} + \boldsymbol{k_2} \cdot \boldsymbol{k_3} + \boldsymbol{k_3} \cdot \boldsymbol{k_1}) 
    \end{aligned}
\end{equation}
% and the useful integral formula is included in Appendix \ref{Master Integrals}. 
% As a consistency check, we explicitly reproduce the Schwarzschild metric in isotropic gauge to $O\Big((m/r)^4\Big)$ in Appexdix \ref{Reproducing Schwarzschild Metric}.  
Now we can explicitly check that we reproduce the 
Schwarzschild metric in isotropic coordinates using our isotropic gauge;
see Appexdix \ref{Reproducing Schwarzschild Metric} for more detail.
Focusing on corrections up to $\mathcal{O}((m/r)^5)$, we get 
    \be 
    \phi (\boldsymbol{x}) ~~~~~= ~~~~~
        \vcenter{\hbox{\begin{tikzpicture}[scale=0.7]
        \begin{feynman}
            \vertex (i) at (0,0);
            \vertex (e) at (0,3);
            \vertex[dot, MyRed] (f1) at (0, 1.5) {};
            \vertex [dot, MyRed, label = $\boldsymbol{x}$] (c1) at (1.5,1.5) {};

            \diagram*{
                (i) -- [double, double, thick] (e),
                (f1) -- [dashed, MyRed, ultra thick] (c1),
            };
        \end{feynman}
    \end{tikzpicture}}}
 ~~~~~   + ~~~~~
    % = - \frac{m}{r} ~.
    \vcenter{\hbox{\begin{tikzpicture}[scale=0.7]
        \begin{feynman}
            \vertex (i) at (0, 0);
                  \vertex (e) at (0, 3);
                  \vertex[dot, MyRed] (w1) at (0, 2.2) {};
                  \vertex[dot, MyRed] (w2) at (0, 1.2) {};
                  \vertex[dot, MyRed] (w3) at (0, 0.5) {};
                  \vertex (c1) at (0.8,1.5);
                  \vertex (c2) at (2,1.5);
                  \vertex[dot, MyRed, label = $\boldsymbol{x}$] (c3) at (3.5,1.5) {} ;
            
                  \diagram*{
                    (i) -- [double,double, thick] (e), 
                    (w1) -- [dashed, MyRed, ultra thick] (c1),
                    (w2) -- [dashed, MyRed, ultra thick] (c1),
                    (c1) -- [boson, MyYellow, ultra thick] (c2),  
                    (w3) -- [dashed, MyRed, ultra thick] (c2),
                    (c2) -- [dashed, MyRed, ultra thick] (c3)
                  };
        \end{feynman}
    \end{tikzpicture}}}
 ~~~~   + ~~~~~... \,,
    \ee
    %%%%%% SIGMA 
        \be 
       \sigma (\boldsymbol{x}) ~  = ~ 
       \vcenter{\hbox{\begin{tikzpicture}[scale=0.7]
        \begin{feynman}
            \vertex (i) at (0, 0);
                  \vertex (e) at (0, 3);
                  \vertex[dot, MyRed] (w1) at (0, 2) {};
                  \vertex[dot, MyRed] (w2) at (0, 1) {};
                  \vertex (c1) at (1,1.5);
                  \vertex [dot, MyYellow, label = $\boldsymbol{x}$] (c2) at (2.5,1.5) {};
            
                  \diagram*{
                    (i) -- [double,double, thick] (e), 
                    (w1) -- [dashed, MyRed, ultra thick] (c1),
                    (w2) -- [dashed, MyRed, ultra thick] (c1),
                    (c1) -- [boson, MyYellow, ultra thick] (c2),  
                  };
        \end{feynman}
    \end{tikzpicture}}}
 ~  + ~ 
    % = - \frac{m}{r} ~.
\vcenter{\hbox{
            \begin{tikzpicture}[scale=0.7]
                \begin{feynman}
                  \vertex (i) at (0, 0);
                  \vertex (e) at (0, 3);
                  \vertex[dot, MyRed] (w1) at (0, 2.75) {};
                  \vertex[dot, MyRed] (w2) at (0, 2.00) {};
                  \vertex[dot, MyRed] (w3) at (0, 1.25) {};
                  \vertex[dot, MyRed] (w4) at (0, 0.5) {};
                  \vertex (c1) at (1, 2.375);
                  \vertex (c2) at (1, 0.875);
                  \vertex (c3) at (1.75, 1.5);
                  \vertex[dot, MyYellow, label = $\boldsymbol{x}$] (c4) at (3, 1.5) {};
            
                  \diagram*{
                    (i) -- [double,double, thick] (e), 
                    (w1) -- [dashed, MyRed, ultra thick] (c1),
                    (w2) -- [dashed, MyRed, ultra thick] (c1),
                    (c1) -- [boson, MyYellow, ultra thick] (c3),
                    (w3) -- [dashed, MyRed, ultra thick] (c2),
                    (w4) -- [dashed, MyRed, ultra thick] (c2),
                    (c2) -- [boson, MyYellow, ultra thick] (c3),
                    (c3) -- [boson, ultra thick, MyYellow] (c4), 
                  };
                \end{feynman}
            \end{tikzpicture}}}
             ~   + ~ 
             \vcenter{\hbox{
                \begin{tikzpicture}[scale=0.7]
                    \begin{feynman}
                      \vertex (i) at (0, 0);
                      \vertex (e) at (0, 3);
                      \vertex[dot, MyRed] (w1) at (0, 2.75) {};
                      \vertex[dot, MyRed] (w2) at (0, 2.00) {};
                      \vertex[dot, MyRed] (w3) at (0, 1.25) {};
                      \vertex[dot, MyRed] (w4) at (0, 0.5) {};
                      \vertex (c1) at (1, 2.375);
                      \vertex (c2) at (1, 0.875);
                      \vertex[dot, MyYellow, label = $\boldsymbol{x}$] (c4) at (2.5, 0.875) {};
                
                      \diagram*{
                        (i) -- [double,double, thick] (e), 
                        (w1) -- [dashed, MyRed, ultra thick] (c1),
                        (w2) -- [dashed, MyRed, ultra thick] (c1),
                        (c1) -- [boson, MyYellow, ultra thick] (c2),
                        (w3) -- [dashed, MyRed, ultra thick] (c2),
                        (w4) -- [dashed, MyRed, ultra thick] (c2),
                        (c2) -- [boson, MyYellow, ultra thick] (c4),
                      };
                    \end{feynman}
                \end{tikzpicture}}} 
 ~   + ~ 
                \vcenter{\hbox{
                    \begin{tikzpicture}[scale=0.7]
                        \begin{feynman}
                          \vertex (i) at (0, 0);
                          \vertex (e) at (0, 3);
                          \vertex[dot, MyRed] (w1) at (0, 2.75) {};
                          \vertex[dot, MyRed] (w2) at (0, 2.00) {};
                          \vertex[dot, MyRed] (w3) at (0, 1.25) {};
                          \vertex[dot, MyRed] (w4) at (0, 0.25) {};
                          \vertex (c1) at (1, 2.375);
                          \vertex (c2) at (1, 0.975);
                          \vertex[dot, MyYellow, label = $\boldsymbol{x}$] (c4) at (2.5, 0.35) {};
                          \vertex (c5) at (1, 0.35);
                    
                          \diagram*{
                            (i) -- [double,double, thick] (e), 
                            (w1) -- [dashed, MyRed, ultra thick] (c1),
                            (w2) -- [dashed, MyRed, ultra thick] (c1),
                            (c1) -- [boson, MyYellow, ultra thick] (c2),
                            (w3) -- [dashed, MyRed, ultra thick] (c2),
                            (w4) -- [dashed, MyRed, ultra thick] (c5),
                            (c5) -- [boson, MyYellow, ultra thick] (c4),
                            (c2) -- [dashed, MyRed, ultra thick] (c5),
                          };
                        \end{feynman}
                           \end{tikzpicture}}} 
 ~  + ~ ... \,.
    \ee
Plugging these expressions into Eq.~\eqref{KK Reduction}, we 
obtain 
\begin{align}
    g_{00} &= 1 - 2 \(\frac{m}{r}\) + 2 \(\frac{m}{r}\)^2 - \frac{3}{2} \(\frac{m}{r}\)^3 + \(\frac{m}{r}\)^4  + O\(\frac{m}{r}\)^5 ~, \\
    g_{ij} &= \( 1 + 2 \(\frac{m}{r}\) + \frac{3}{2} \(\frac{m}{r}\)^2 + \frac{1}{2} \(\frac{m}{r}\)^3 + \frac{1}{16} \(\frac{m}{r}\)^4 + O\(\frac{m}{r}\)^5 \) \delta_{ij}\,.
\end{align}
After identification $m\to M$, we see that this metric coincides with the Schwarzschild metric in isotropic gauge~\eqref{eq:schiso}
expanded up to $\mathcal{O}((m/r)^5)$. Thus, our isotropic KK gauge choices~\eqref{KK Reduction} and \eqref{eq:isogauge} are indeed consistent with the Schwarzschild metric.

% In order to match between the EFT and the full theory, we need to make sure that we are at the consistent gauge to compare results. The consistency here can be checked via off-shell graviton 1-point function to see if it correctly reproduces the Schwarzschild metric in certain coordinate. 

In addition, let us point out that the commonly used de Donder gauge in the EFT is consistent with the harmonic Schwarzschild coordinates~\cite{feynman1963quantum, duff1973quantum}
\begin{equation}
    ds^{2}=\frac{r-M}{r+M} dt^{2} - \frac{r+M}{r-M} dr^{2} - (r+M)^{2} d\Omega^2 ~.
\end{equation}

\subsection{Spin-0/2}\label{Spin-0 and Spin-2 Electric}

We analyze now spin-0 and the spin-2 electric-type perturbations. 
These two types of perturbations share the same structure in the isotropic gauge,
and hence it is natural to analyze them together. 
The bulk action for perturbations of a test scalar field $\Phi$ is given by
% For the spin-0 perturbations denoted by $\Phi$, in the static case, the effective action takes the form 
\begin{equation}
\label{eq:effPhi}
    S_{\Phi} = - \frac{1}{2} \int dt d^3 \boldsymbol{x} \sqrt{\gamma} \gamma^{ij} \partial_i \Phi \partial_j \Phi ~.
\end{equation}
The first important observation 
is that $\Phi$ only couples to the $\sigma$ in the isotropic gauge,
which significantly simplify our computations. 

As far as the static spin-2 electric perturbations are concerned, 
we perform the following decomposition of the dilaton:
\begin{equation}
     \phi = \phi_{\rm BH} + \frac{\delta \phi}{2 \sqrt{2} M_{\rm pl}} ~,
\end{equation}
where the Planck mass $M_{\rm pl}\equiv 1/\sqrt{32\pi}$ in our unit system with $G=1$. 
Physically, $\delta\phi$ is the perturbation of the Newtonian potential caused  by 
an external metric fluctuation, 
while $\phi_{\rm BH}$ is the background part that matches the Schwarzschild metric. 
Plugging this into the static Einstein-Hilbert action~\eqref{Static Einstin-Hilbert Action}, 
we obtain the following effective action for $\delta\phi$  
\begin{equation}
       S_{\delta\phi} = - \frac{1}{2} \int dt  d^3 \boldsymbol{x} \sqrt{\gamma} \(\gamma^{ij} \partial_i \delta\phi \partial_j \delta\phi\) ~,
\end{equation}
which is the same as the scalar field action~\eqref{eq:effPhi}.

Now let us move on to the finite size effects. 
In the spin-2 case they 
are controlled by the electric tidal field related to the Weyl tensor.
In the Newtonian limit, sufficient for
the extraction of the 
finite-size effects from the worldline action, 
it is straightforward to get
\begin{equation}\label{Spin-2 Electric Field}
    E_{ij} \equiv 2 \sqrt{2} M_{\rm pl}
    C_{0i0j} = -\(\partial_{i}\partial_j  - \frac{1}{3}\delta_{ij}\partial^2 \)\delta\phi ~,
\end{equation}
where $C_{0i0j}$ is the parity even component of the Weyl tensor.

% For the static spin-2 electric perturbations, in the Newtonian limit, it is straightforward to get
% \begin{equation}\label{Spin-2 Electric Field}
%     \mathcal{E}_{ij} \equiv C_{0i0j} = -\(\partial_{i}\partial_j  - \frac{1}{3}\delta_{ij}\partial^2 \)\delta\phi ~,
% \end{equation}
% where $C_{0i0j}$ is the parity even component of the Weyl tensor and $\delta\phi$ is the perturbation of the Newtonian potential. 
% Based on the static Einstein-Hilbert action in Eq.\eqref{Static Einstin-Hilbert Action}, it is already sufficient to consider the following decomposition
% \begin{equation}
%      \phi = \phi_{\rm BH} + \delta \phi ~,
% \end{equation}
% where $\phi_{\rm BH}$ is used for reproducing the Schwarzschild metric. The effective action for $\delta\phi$ then reads 
% \begin{equation}
%        S_{\delta\phi} = - \frac{1}{16 \pi} \int dt  d^3 \boldsymbol{x} \sqrt{\gamma} \(\gamma^{ij} \partial_i \delta\phi \partial_j \delta\phi\) ~,
% \end{equation}
% which only differs from the spin-0 by a normalization factor. 
Plugging this into~\eqref{eq:tidalgen} we find an expression 
for the static worldline action 
identical to
that of the spin-0 case,
\begin{align}
 S_{\rm s=0}^{\rm Love} = \frac{1}{2\ell!} \int dt \lambda_\ell^{s=0} \partial_{\langle i_1 \cdots i_\ell \rangle} \delta\phi \partial^{\langle i_1 \cdots i_\ell \rangle} \delta\phi \,,\quad\text{cf.}
 \quad
    % S_{\rm s=2}^{\rm love} &= \frac{1}{2\ell !} \int d\tau \lambda_\ell^{s=2} \(\partial_{\langle i_1 \cdots i_{\ell-2}}\mathcal{E}_{i_{\ell-1}i_{\ell}\rangle }\) \(\partial^{\langle i_1 \cdots i_{\ell-2}} \mathcal{E}^{i_{\ell-1}i_{\ell}\rangle}\) \,,
    S_{\rm s=2}^{\rm Love} = \frac{1}{2\ell!} \int dt \lambda_\ell^{s=0} \partial_{\langle i_1 \cdots i_\ell \rangle} \Phi \partial^{\langle i_1 \cdots i_\ell \rangle} \Phi \,.
\end{align}

Since the spin-0 and electric spin-2 sectors are described by identical actions, 
we focus on the spin-0 case in what follows.
Expanding in the number of fields and splitting each 
field component $\Phi$ into 
external source $\bar\Phi$ and response $\delta\Phi$,
we obtain, at quartic order,
% In the following, we will just focus on the spin-0 case, and the spin-2 electric perturbations will share the same story. 
% After performing the perturbative expansion, we get
\begin{equation}
        \begin{aligned}
            S_{\Phi}^{(2)} &= \int dt d^3\boldsymbol{x} \(- \frac{1}{2} (\partial_i \delta\Phi \partial^i \delta\Phi) - (\partial_i \bar\Phi \partial^i \delta\Phi) \) ~, \\
            S_{\Phi}^{(3)} &= \int dt d^3\boldsymbol{x} \(-\frac{1}{4} \sigma (\partial_i \delta\Phi \partial^i \delta\Phi) - \frac{1}{2} \sigma (\partial_i \bar \Phi \partial^i \delta\Phi) \) ~, \\
            S_{\Phi}^{(4)} & = \int dt d^3 \boldsymbol{x} \Bigg(\frac{1}{16} \sigma^2 \partial_i \delta\Phi \partial^i \delta\Phi + \frac{1}{8} \sigma^2 \partial_i \bar\Phi \partial^i \delta\Phi \Bigg) \,.
        \end{aligned}
\end{equation}
The corresponding two point function takes the form
\begin{equation}
    \langle\delta\Phi(\boldsymbol{x_1}) \delta\Phi(\boldsymbol{x_2}) \rangle = 
    \begin{tikzpicture}[baseline={([yshift=-2.0 ex]current bounding box.center)}]
        \begin{feynman}
            \vertex (a) at (0,0);
            \vertex (b) at (2,0);
            
            \diagram*{
                (a) -- [ultra thick, MyBlue, edge label = $\delta\Phi$] (b), 
            };
        \end{feynman}
    \end{tikzpicture}
    =\int \frac{d^3 \boldsymbol{k}}{(2\pi)^3} \frac{-i}{\boldsymbol{k}^2} e^{i\boldsymbol{k} \cdot (\boldsymbol{x_1}-\boldsymbol{x_2})} \propto \frac{1}{|\boldsymbol{x_1}-\boldsymbol{x_2}|}~.
\end{equation}
Feynman rules for propagators and the above interaction vertices are given in Appendix \ref{Feynman Rule}. The 
most important interaction vertices
are 
\begin{equation}
   \begin{aligned}
       \begin{tikzpicture}[baseline={([yshift=-0.5 ex]current bounding box.center)},scale=0.7]
                \begin{feynman}
                    \vertex (c1) at (0,0);
                    \vertex (c2) at (1.5, 0);
                    \vertex (c3) at (3, 1);
                    \vertex[crossed dot, MyBlue] (c4) at (3, -1) {};

                    \diagram*{
                        (c1) -- [boson, MyYellow, ultra thick, momentum={[arrow style = MyYellow, arrow shorten = 0.25] $\boldsymbol{p}$}] (c2), 
                        (c2) -- [MyBlue, ultra thick, reversed momentum={[arrow style = MyBlue, arrow shorten = 0.25] $\boldsymbol{k}$}] (c3),
                        (c2) -- [MyBlue, ultra thick, edge label = $r^{\ell}$] (c4), 
                    };
                \end{feynman}
            \end{tikzpicture}
         = - \frac{\ell}{2} k^i \mathcal{E}_{i i_2 \ldots i_\ell} (2\pi)^3 \(i^{\ell-1} \frac{d^{\ell-1}}{dk_{i_2} \cdots dk_{i_\ell}} \delta^{(3)}(\boldsymbol{k} + \boldsymbol{p})\) ~,
            \begin{tikzpicture}[baseline={([yshift=-0.5 ex]current bounding box.center)},scale=0.7]
                \begin{feynman}
                    \vertex (c1) at (0,0);
                    \vertex (c2) at (1.5, 0);
                    \vertex (c3) at (3, 1);
                    \vertex (c4) at (3, -1) ;

                    \diagram*{
                        (c1) -- [boson, MyYellow, ultra thick] (c2), 
                        (c2) -- [MyBlue, ultra thick, reversed momentum={[arrow style = MyBlue, arrow shorten = 0.25] $\boldsymbol{k}$}] (c3),
                        (c2) -- [MyBlue, ultra thick, reversed momentum'={[arrow style = MyBlue, arrow shorten = 0.25] $\boldsymbol{p}$}] (c4), 
                    };
                \end{feynman}
            \end{tikzpicture}
             = i \frac{1}{2} \boldsymbol{k} \cdot \boldsymbol{p} \,.
    \end{aligned}
\end{equation}

\subsection{Non-Renormaliztion of Love Numbers}

Crucially, 
one may notice that the 
absence of Love numbers'
running follow 
from the structure of 
metric perturbations 
in the isotropic Kaluza-Klein
gauge. Indeed, in this gauge 
the point mass $m$ couples to 
one $\phi$, at leading order.\footnote{This order will be sufficient
as diagrams with higher order interactions generate quantum loops and hence vanish in the classical 
limit.}
The Einstein-Hilbert   
action~\eqref{Static Einstin-Hilbert Action}, however, contains
two $\phi$'s; i.e., one can 
only have interactions 
such as $\phi^2 \sigma$, 
$\phi^2 \sigma^2$ etc.
(we ignore derivatives here as they are irrelevant for our discussion).
This means 
that it is only possible 
to draw 
a classical 
worldline diagram 
with an even number of 
$\phi$'s and hence an even number 
of point masses. 
% we notice that the static Einstein-Hilbert   
% action 
% Eq.\eqref{Static Einstin-Hilbert Action}
% is symmetric under the field sign flips $\phi\rightarrow-\phi$. 
% Since the external field only couples to $\sigma$, 
% whilst the worldline only couples to the $\phi$, 
% a diagram connecting the worldline with the external source
% can only have an even number of point-mass insertions.
% we conclude that there can only be even number of vertices on the worldline. 
Therefore, it is impossible to construct a diagram producing 
a $(2\ell+1)$PN order correction to the one-point function, as this 
diagram obviously 
requires an odd number of 
point mass insertions on
the worldline,\footnote{An equivalent argument was used by \cite{Kol:2011vg,unp1,unp2}, which argue for the absence of logs by clashing the  
$\phi\to -\phi$ symmetry of the 
Einstein-Hilbert action 
in the isotropic Kaluza-Klein gauge 
and the $\phi\to -\phi$ 
to $m\to -m$ symmetry 
of the leading order
point particle interaction.
} $2\ell+1$.
Hence, $c_{2\ell +1} = 0$. 
This means that Love numbers do not get 
renormalized by graviton corrections.
Recalling Eq.~
\eqref{eq:rg}, this also implies that 
Love numbers do not run. 
Since Love numbers are defined as 
gauge-invariant EFT Wilson coefficients,
the absence of their logarithmic running 
is a gauge-independent 
statement.

Note that the absence of logarithmic corrections
does not follow from EFT power counting rules. 
Indeed, as we have mentioned earlier, 
in an arbitrary gauge one generally obtains 
non-trivial $(2\ell+1)$PN corrections from individual 
diagrams to the one-point function. 
Every such diagram would naively imply a logarithmic running.
However, when summed together, these logs must 
cancel identically. 
From the EFT point of view, this cancellation 
is a fine-tuning, which is reminiscent of the apparent 
cancellation of loop corrections 
to the Higgs mass in the usual QFT.   

Note that logarithmic contributions to Love numbers, 
if present, can be found in both the UV (full GR) 
and IR (EFT) calculations; see e.g.,~\cite{Kol:2011vg}. 
In Ref.~\cite{Charalambous:2021mea} the absence of  
logarithmic running in four dimensions
was interpreted as a constraint 
imposed by the Love symmetry of GR, which 
is a UV symmetry from the EFT point of view. 
Since the EFT must be a consistent description 
of the UV theory, the logs should be absent in the EFT as well. 
In an arbitrary gauge this appears as a miraculous 
cancellation between different Feynman diagrams. 
The choice of the isotropic gauge makes this cancellation
manifest for spin-0 and spin-2 fields, 
but does not explain its origin.
In this sense we cannot claim that the nonrenormalization
of Love numbers is a consequence of some hidden 
structure of the GR action that is apparent in the isotropic gauge.
As an explicit example supporting this statement, 
we will compute the 
running of the 
spin-1 Love number corrections shorty. 
We will see that in this case the isotropic gauge 
itself does not forbid Love numbers to run; i.e.,
individual $(2\ell+1)$PN diagrams will contain 
logs as expected on general grounds.
However, these contributions will sum to zero,
implying the absence of running as enforced 
by the Love symmetry.

% At this point, we clearly see that the isotropic coordinate gauge is a good gauge such that there is no diagram contribute to the $2\ell+1$ PM order source correction from the first principle. Based on the this observation, it is reasonable to conclude that the logarithmic divergence is totally a gauge issue. If we are clever enough, it is possible to find a suitable gauge such that the divergence just does not exist.

\subsubsection*{Reconstruction of the Full one-Point Function}

% Based on such simplification, we now have the ability to study the gravitational nonlinear correction to any PM order by making use of the EFT diagramatic recurrence relation. 

Thanks to significant simplifications that take place in
the isotropic gauge, we can actually compute 
PN corrections to the test field profiles to 
\text{all} PN orders. 
Let us show this explicitly. 

The first important observation
is that all pyramid diagrams 
with more than two worldline point masses
are unphysical and hence must exactly cancel
with each other. 
Indeed,  each pyramid diagram with $N_m$
mass insertions 
scales as $|\boldsymbol{k_1} + \boldsymbol{k_2}|^{N_m^{\rm Pyr}-3} \(\boldsymbol{k_1} \cdot \boldsymbol{k_2}\) $
\eqref{1PI Amplitude}.
Physical amplitudes should peak 
at the momentum conserving configurations 
$\boldsymbol{k_1} + \boldsymbol{k_2} = 0$
because the momenta of external legs do not change drastically in a soft scattering process
characterized by $|\boldsymbol{k}|m \ll 1$.
In contrast to this physical expectation,  
the amplitudes do not peak within the momentum conserving region 
if $N_m^{\rm Pyr} \geq 4$.
For these diagrams they peak when the momentum transfer is large, instead. 
Since this behaviour is clearly unphysical, 
the corresponding amplitudes must cancel with each other at any given order in $m/r$. 
To illustrate this argument explicitly, we provide a concrete example for the $N_m^{\rm Pyr} = 4$ case. There are four possible diagrams in total: 
\begin{equation}
        \vcenter{\hbox{
            \begin{tikzpicture}[scale=0.7]
                \begin{feynman}
                  \vertex (i) at (0, 0);
                  \vertex (e) at (0, 3);
                  \vertex[dot, MyRed] (w1) at (0, 2.75) {};
                  \vertex[dot, MyRed] (w2) at (0, 2.00) {};
                  \vertex[dot, MyRed] (w3) at (0, 1.25) {};
                  \vertex[dot, MyRed] (w4) at (0, 0.5) {};
                  \vertex (c1) at (1, 2.375);
                  \vertex (c2) at (1, 0.875);
                  \vertex (c3) at (1.75, 1.5);
                  \vertex (c4) at (2.5, 1.5);
                  \vertex (f1) at (3.5,2.8) {};
                  \vertex (f2) at (3.5,0.2) {}; 
            
                  \diagram*{
                    (i) -- [double,double, thick] (e), 
                    (w1) -- [dashed, MyRed, ultra thick] (c1),
                    (w2) -- [dashed, MyRed, ultra thick] (c1),
                    (c1) -- [boson, MyYellow, ultra thick] (c3),
                    (w3) -- [dashed, MyRed, ultra thick] (c2),
                    (w4) -- [dashed, MyRed, ultra thick] (c2),
                    (c2) -- [boson, MyYellow, ultra thick] (c3),
                    (c4) -- [ultra thick, MyBlue, reversed momentum ={[arrow style = MyBlue, arrow shorten = 0.25] $\boldsymbol{k_1}$}] (f1), 
                    (f2) -- [ultra thick, MyBlue, momentum ={[arrow style = MyBlue, arrow shorten = 0.25] $\boldsymbol{k_2}$}] (c4), 
                    (c3) -- [boson, ultra thick, MyYellow] (c4), 
                  };
                \end{feynman}
            \end{tikzpicture}}} = (-i) \frac{1}{16} \pi^2 m^4 |\boldsymbol{k_1} + \boldsymbol{k_2}| (\boldsymbol{k_1} \cdot \boldsymbol{k_2}) ~,    
\end{equation}
\begin{equation}
                \vcenter{\hbox{
                \begin{tikzpicture}[scale=0.7]
                    \begin{feynman}
                      \vertex (i) at (0, 0);
                      \vertex (e) at (0, 3);
                      \vertex[dot, MyRed] (w1) at (0, 2.75) {};
                      \vertex[dot, MyRed] (w2) at (0, 2.00) {};
                      \vertex[dot, MyRed] (w3) at (0, 1.25) {};
                      \vertex[dot, MyRed] (w4) at (0, 0.5) {};
                      \vertex (c1) at (1, 2.375);
                      \vertex (c2) at (1, 0.875);
                      \vertex (c4) at (2.5, 0.875);
                      \vertex (f1) at (3.5,1.8) {};
                      \vertex (f2) at (3.5,0.2) {}; 
                
                      \diagram*{
                        (i) -- [double,double, thick] (e), 
                        (w1) -- [dashed, MyRed, ultra thick] (c1),
                        (w2) -- [dashed, MyRed, ultra thick] (c1),
                        (c1) -- [boson, MyYellow, ultra thick] (c2),
                        (w3) -- [dashed, MyRed, ultra thick] (c2),
                        (w4) -- [dashed, MyRed, ultra thick] (c2),
                        (c2) -- [boson, MyYellow, ultra thick] (c4),
                        (c4) -- [ultra thick, MyBlue, reversed momentum ={[arrow style = MyBlue, arrow shorten = 0.25] $\boldsymbol{k_1}$}] (f1), 
                        (f2) -- [ultra thick, MyBlue, momentum ={[arrow style = MyBlue, arrow shorten = 0.25] $\boldsymbol{k_2}$}] (c4),  
                      };
                    \end{feynman}
                \end{tikzpicture}}} 
                = i \frac{1}{96} \pi^2 m^4 |\boldsymbol{k_1} + \boldsymbol{k_2}| (\boldsymbol{k_1} \cdot \boldsymbol{k_2}) ~,
\end{equation}
\begin{equation}
                    \vcenter{\hbox{
                    \begin{tikzpicture}[scale=0.7]
                        \begin{feynman}
                          \vertex (i) at (0, 0);
                          \vertex (e) at (0, 3);
                          \vertex[dot, MyRed] (w1) at (0, 2.75) {};
                          \vertex[dot, MyRed] (w2) at (0, 2.00) {};
                          \vertex[dot, MyRed] (w3) at (0, 1.25) {};
                          \vertex[dot, MyRed] (w4) at (0, 0.25) {};
                          \vertex (c1) at (1, 2.375);
                          \vertex (c2) at (1, 0.975);
                          \vertex (c4) at (2.5, 0.35);
                          \vertex (c5) at (1, 0.35);
                          \vertex (f1) at (4,1.0) {};
                          \vertex (f2) at (4,-0.4) {};
                    
                          \diagram*{
                            (i) -- [double,double, thick] (e), 
                            (w1) -- [dashed, MyRed, ultra thick] (c1),
                            (w2) -- [dashed, MyRed, ultra thick] (c1),
                            (c1) -- [boson, MyYellow, ultra thick] (c2),
                            (w3) -- [dashed, MyRed, ultra thick] (c2),
                            (w4) -- [dashed, MyRed, ultra thick] (c5),
                            (c5) -- [boson, MyYellow, ultra thick] (c4),
                            (c2) -- [dashed, MyRed, ultra thick] (c5),
                            (c4) -- [ultra thick, MyBlue, reversed momentum ={[arrow style = MyBlue, arrow shorten = 0.25] $\boldsymbol{k_1}$}] (f1), 
                            (f2) -- [ultra thick, MyBlue, momentum ={[arrow style = MyBlue, arrow shorten = 0.25] $\boldsymbol{k_2}$}] (c4),  
                          };
                        \end{feynman}
                    \end{tikzpicture}}} 
                    = i \frac{1}{48} \pi^2 m^4 |\boldsymbol{k_1} + \boldsymbol{k_2}| (\boldsymbol{k_1} \cdot \boldsymbol{k_2}) ~,
\end{equation}
\begin{equation}
                    \vcenter{\hbox{
                    \begin{tikzpicture}[scale=0.7]
                        \begin{feynman}
                          \vertex (i) at (0, 0);
                          \vertex (e) at (0, 3);
                          \vertex[dot, MyRed] (w1) at (0, 2.75) {};
                          \vertex[dot, MyRed] (w2) at (0, 2.00) {};
                          \vertex[dot, MyRed] (w3) at (0, 1.25) {};
                          \vertex[dot, MyRed] (w4) at (0, 0.5) {};
                          \vertex (c1) at (1, 2.375);
                          \vertex (c2) at (1, 0.875);
                          \vertex (c3) at (1.75, 1.5);
                          \vertex (f1) at (3,2.8) {};
                          \vertex (f2) at (3,0.2) {}; 
            
                        \diagram*{
                          (i) -- [double,double, thick] (e), 
                          (w1) -- [dashed, MyRed, ultra thick] (c1),
                          (w2) -- [dashed, MyRed, ultra thick] (c1),
                          (c1) -- [boson, MyYellow, ultra thick] (c3),
                          (w3) -- [dashed, MyRed, ultra thick] (c2),
                          (w4) -- [dashed, MyRed, ultra thick] (c2),
                          (c2) -- [boson, MyYellow, ultra thick] (c3),
                          (c3) -- [ultra thick, MyBlue, reversed momentum ={[arrow style = MyBlue, arrow shorten = 0.25] $\boldsymbol{k_1}$}] (f1), 
                          (f2) -- [ultra thick, MyBlue, momentum ={[arrow style = MyBlue, arrow shorten = 0.25] $\boldsymbol{k_2}$}] (c3),  
                       };
                      \end{feynman}
                    \end{tikzpicture}}} 
                    = i \frac{1}{32} \pi^2 m^4 |\boldsymbol{k_1} + \boldsymbol{k_2}| (\boldsymbol{k_1} \cdot \boldsymbol{k_2}) ~.
\end{equation}
Obviously, these diagrams cancel when summed over. 
One can explicitly check that the same is true 
for any $N_m^{\rm 1PI} > 4$. 

We conclude that only the amplitudes with $N_m^{\rm 1PI} = 2$
have the expected physical behavior. These amplitudes 
have poles in $|\boldsymbol{k_1} + \boldsymbol{k_2}|$;
i.e., they indeed peak at the momentum conserving region. 
% In particular, at leading order
This means that, in particular, 
the only diagram at leading order that is relevant
for the pyramid structure is 
\begin{equation}\label{Amplitudes}
    \vcenter{\hbox{
        \begin{tikzpicture}[scale=0.7]
            \begin{feynman}
              \vertex (i) at (0, 0);
              \vertex (e) at (0, 3);
              \vertex[dot, MyRed] (w1) at (0, 2) {};
              \vertex[dot, MyRed] (w2) at (0, 1) {};
              \vertex (c1) at (1,1.5);
              \vertex (c2) at (2.5,1.5);
              \vertex (f1) at (3.5,2.8) {};
              \vertex (f2) at (3.5,0.2) {}; 
        
              \diagram*{
                (i) -- [double,double, thick] (e), 
                (w1) -- [dashed, MyRed, ultra thick] (c1),
                (w2) -- [dashed, MyRed, ultra thick] (c1),
                (c1) -- [boson, MyYellow, ultra thick] (c2),
                (c2) -- [ultra thick, MyBlue, reversed momentum = {[arrow style = MyBlue, arrow shorten = 0.25] $\boldsymbol{k_1}$}] (f1), 
                (f2) -- [ultra thick, MyBlue, momentum = {[arrow style = MyBlue, arrow shorten = 0.25] $\boldsymbol{k_2}$}] (c2),  
              };
            \end{feynman}
        \end{tikzpicture}}} 
        = (-i) \frac{1}{2} (\pi^2 m^2) \frac{\boldsymbol{k_1} \cdot \boldsymbol{k_2}}{|\boldsymbol{k_1} + \boldsymbol{k_2}|}  ~.
\end{equation}
Knowing this diagram, we can compute 
the scalar field profile at all PN orders. 
In analogy with Eq.\eqref{General 2 Ladder Recurrence}, 
we first establish the relation $4$PN and $2$PN corrections,
\begin{equation}
    \begin{aligned}
    \vcenter{\hbox{
    \begin{tikzpicture}[scale=0.7]
        \begin{feynman}
            \vertex (i) at (0, 0);
            \vertex (e) at (0, 3);
            \vertex[dot, MyRed] (w1) at (0, 2.75) {};
            \vertex[dot, MyRed] (w2) at (0, 2.00) {};
            \vertex[dot, MyRed] (w3) at (0, 1.25) {};
            \vertex[dot, MyRed] (w4) at (0, 0.5) {};
            \vertex (c1) at (1, 2.375);
            \vertex (c2) at (1, 0.875);
            \vertex (c3) at (2.5, 2.375);
            \vertex (c4) at (2.5, 0.875);
            \vertex[dot, MyBlue, label = $\boldsymbol{x}$] (f1) at (3.5,2.8) {};
            \vertex[crossed dot, MyBlue] (f2) at (3.5,0.2) {}; 
                
            \diagram*{
                (i) -- [double,double, thick] (e), 
                (w1) -- [dashed, MyRed, ultra thick] (c1),
                (w2) -- [dashed, MyRed, ultra thick] (c1),
                (c1) -- [boson, MyYellow, ultra thick] (c3),
                (w3) -- [dashed, MyRed, ultra thick] (c2),
                (w4) -- [dashed, MyRed, ultra thick] (c2),
                (c2) -- [boson, MyYellow, ultra thick] (c4),
                (c3) -- [ultra thick, MyBlue] (c4),
                (c3) -- [ultra thick, MyBlue] (f1), 
                (f2) -- [ultra thick, MyBlue] (c4),  
            };
        \end{feynman}
        \end{tikzpicture}}} 
        &= \int \frac{d^3 \boldsymbol{k_1}}{(2\pi)^3} \int \frac{d^3 \boldsymbol{k}}{(2\pi)^3}
        \vcenter{\hbox{\begin{tikzpicture}[scale=0.7]
            \begin{feynman}
              \vertex (i) at (0, 0);
              \vertex (e) at (0, 3);
              \vertex[dot, MyRed] (w1) at (0, 2) {};
              \vertex[dot, MyRed] (w2) at (0, 1) {};
              \vertex (c1) at (1,1.5);
              \vertex (c2) at (2.5,1.5);
              \vertex (f1) at (3.5,2.8) {};
              \vertex[crossed dot, MyBlue] (f2) at (3.3,0.2) {}; 
        
              \diagram*{
                (i) -- [double,double, thick] (e), 
                (w1) -- [dashed, MyRed, ultra thick] (c1),
                (w2) -- [dashed, MyRed, ultra thick] (c1),
                (c1) -- [boson, MyYellow, ultra thick] (c2),
                (c2) -- [ultra thick, MyBlue, reversed momentum = {[arrow style = MyBlue, arrow shorten = 0.25] $\boldsymbol{k_1}$}] (f1), 
                (f2) -- [ultra thick, MyBlue] (c2),  
              };
            \end{feynman}
        \end{tikzpicture}}} 
        \times 
        \vcenter{\hbox{\begin{tikzpicture}[scale=0.7]
            \begin{feynman}
              \vertex (i) at (0, 0);
              \vertex (e) at (0, 3);
              \vertex[dot, MyRed] (w1) at (0, 2) {};
              \vertex[dot, MyRed] (w2) at (0, 1) {};
              \vertex (c1) at (1,1.5);
              \vertex (c2) at (2.5,1.5);
              \vertex (f1) at (3.5,2.8) {};
              \vertex (f2) at (3.5,0.2) {}; 
        
              \diagram*{
                (i) -- [double,double, thick] (e), 
                (w1) -- [dashed, MyRed, ultra thick] (c1),
                (w2) -- [dashed, MyRed, ultra thick] (c1),
                (c1) -- [boson, MyYellow, ultra thick] (c2),
                (c2) -- [ultra thick, MyBlue, reversed momentum = {[arrow style = MyBlue, arrow shorten = 0.25] $\boldsymbol{k}$}] (f1), 
                (f2) -- [ultra thick, MyBlue, reversed momentum = {[arrow style = MyBlue, arrow shorten = 0.25] $\boldsymbol{k_1}$}] (c2),  
              };
            \end{feynman}
        \end{tikzpicture}}} 
        \times \frac{-i}{\boldsymbol{k}^2} e^{i\boldsymbol{k} \cdot \boldsymbol{x}} \\
        & = \(\frac{m}{2}\)^2 \partial^{-2} \[\(- \frac{2x^i}{r^4} \partial_i + \frac{1}{r^2} \partial^2\)         \vcenter{\hbox{\begin{tikzpicture}[scale=0.7]
            \begin{feynman}
              \vertex (i) at (0, 0);
              \vertex (e) at (0, 3);
              \vertex[dot, MyRed] (w1) at (0, 2) {};
              \vertex[dot, MyRed] (w2) at (0, 1) {};
              \vertex (c1) at (1,1.5);
              \vertex (c2) at (2.5,1.5);
              \vertex[dot, MyBlue, label = $x$] (f1) at (3.5,2.8) {};
              \vertex[crossed dot, MyBlue] (f2) at (3.5,0.2) {}; 
        
              \diagram*{
                (i) -- [double,double, thick] (e), 
                (w1) -- [dashed, MyRed, ultra thick] (c1),
                (w2) -- [dashed, MyRed, ultra thick] (c1),
                (c1) -- [boson, MyYellow, ultra thick] (c2),
                (c2) -- [ultra thick, MyBlue] (f1), 
                (f2) -- [ultra thick, MyBlue] (c2),  
              };
            \end{feynman}
        \end{tikzpicture}}} 
        \]
    \end{aligned}
\end{equation}
% \magenta{More generally, with Eq.\eqref{Amplitudes}, the $2(n+1)$ PM corrections can be built up from the $2n$ PM corrections}
% Now we make use of the diagramatic recurrence relation~Eq.~\eqref{General Diagramatic Recurrence Relation}, 
% that take the following form for a scalar field,
This expression can readily be generalized to an arbitrary ($2n$)PN order,
\begin{equation}\label{Diagramatic Recurrence Relation}
    \begin{aligned}
        \delta \Phi^{2(n+1)}(\boldsymbol{x}) &= \int \frac{d^3 \boldsymbol{k_1}}{(2\pi)^3} \int \frac{d^3 \boldsymbol{k}}{(2\pi)^3}\delta\Phi^{2n}_{\boldsymbol{k_1}} \times 
        \vcenter{\hbox{\begin{tikzpicture}[scale=0.7]
            \begin{feynman}
              \vertex (i) at (0, 0);
              \vertex (e) at (0, 3);
              \vertex[dot, MyRed] (w1) at (0, 2) {};
              \vertex[dot, MyRed] (w2) at (0, 1) {};
              \vertex (c1) at (1,1.5);
              \vertex (c2) at (2.5,1.5);
              \vertex (f1) at (3.5,2.8) {};
              \vertex (f2) at (3.5,0.2) {}; 
        
              \diagram*{
                (i) -- [double,double, thick] (e), 
                (w1) -- [dashed, MyRed, ultra thick] (c1),
                (w2) -- [dashed, MyRed, ultra thick] (c1),
                (c1) -- [boson, MyYellow, ultra thick] (c2),
                (c2) -- [ultra thick, MyBlue, reversed momentum = {[arrow style = MyBlue, arrow shorten = 0.25] $\boldsymbol{k}$}] (f1), 
                (f2) -- [ultra thick, MyBlue, reversed momentum = {[arrow style = MyBlue, arrow shorten = 0.25] $\boldsymbol{k_1}$}] (c2),  
              };
            \end{feynman}
        \end{tikzpicture}}} 
        \times \frac{-i}{\boldsymbol{k}^2} e^{i\boldsymbol{k}\cdot \boldsymbol{x}} \\
        & = \(\frac{m}{2}\)^2 \partial^{-2}\( - \frac{2 x^i}{r^4} \partial_{i} \delta\Phi^{2n}(\boldsymbol{x}) + \frac{1}{r^2}\partial^2 \delta\Phi^{2n}(\boldsymbol{x}) \) \,.
    \end{aligned}
\end{equation} 
Now let us assume an anzats
\begin{equation}
    \delta\Phi^{2n}(\boldsymbol{x}) = \mathcal{E}_{i_1\cdots i_\ell} x^{i_1} \cdots x^{i_\ell} \times c_{2n} \(\frac{m}{2r}\)^{2n} ~.
\end{equation} 
Then from Eq.\eqref{Diagramatic Recurrence Relation} we obtain 
\begin{equation}
    \delta\Phi^{2(n+1)}(\boldsymbol{x}) = \mathcal{E}_{i_1 \cdots i_\ell} x^{i_1}\cdots x^{i_\ell} \times \(\frac{(n-\ell)(2n+1)}{(n+1)(2n+1-2\ell)} c_{2n}\) \times \( \frac{m}{2r}\)^{2(n+1)} ~,
\end{equation}
implying the following recurrence relation for the 
PN coefficients $c_{2n}$:
\begin{equation}\label{Recurrence Relation}
    c_{2(n+1)} = \frac{(n-\ell)(2n+1)}{(n+1)(2n+1-2\ell)}c_{2n} ~.
\end{equation}
Note that the recurrence series truncates at $n=\ell$,
which means that the static one-point function (in the absence
of finite-size effects) is a polynomial in $r$.
The truncation of the EFT solution can be seen as a fine-tuning. 
Indeed, this means that diagrams of $(\ell+1)$PN order
and higher all cancel identically.
% \textcolor{blue}{Is this a fine tuning?}
% Now, we can write down the full GR nonlinear correction to all PM orders in the static limit and do the resummation 
Equation \eqref{Recurrence Relation} allows us to `resum'
PN corrections to all orders and obtain the full 
GR solution for the external field with a source 
boundary condition at spatial infinity, 
\begin{equation}
    \begin{aligned}
        \Phi(\boldsymbol{x}) & = \bar\Phi(\boldsymbol{x}) + \sum_{n=1}^{\infty} \delta\Phi^{2n}(\boldsymbol{x}) \\
        & = \mathcal{E}_{i_1 \cdots i_\ell} x^{i_1} \cdots x^{i_\ell}\( 1 + c_2 \( \frac{m}{2r}\)^2 + c_4 \(\frac{m}{2r}\)^4 + \cdots +  c_{2n} \(\frac{m}{2r}\)^{2n}  + \cdots \) \\
        & = \sum_{m=-\ell}^{\ell} \mathcal{E}_{\ell m } Y_{\ell m}(\theta,\phi) r^{\ell}  \( 1 + c_2 \( \frac{m}{2r}\)^2 + c_4 \(\frac{m}{2r}\)^4 + \cdots +  c_{2n} \(\frac{m}{2r}\)^{2n}  + \cdots \)  \,,
    \end{aligned}
\end{equation}
where 
the coefficients $c_{2n}$ satisfy~Eq.~\eqref{Recurrence Relation}.
One can easily identify this series with the Gauss hypergeometric
function, giving
 \be 
 \Phi(\boldsymbol{x}) = \sum_{m=-\ell}^\ell \mathcal{E}_{\ell m} Y_{\ell m}(\theta,\phi) r^{\ell} {}_2 F_1\(\frac{1}{2},-\ell,\frac{1}{2}-\ell,\(\frac{m}{2r}\)^2 \)
 \ee
 % From another perspective, it will be helpful to seek for the constraint equation that can provide us such power series behaviour and the recurrence relation. Based on Eq.\eqref{Recurrence Relation}, it is natural to assume that this constraint equation is the homogenous second order differential equation. After normalizing the second order derivative term, due to the spherical symmetry, we can choose the constant term to be $\ell(\ell+1) R_{\ell}(r) / r^2$. By dimensional analysis, there will be singular points appear in the denorminator of the first order term. 
It is straightforward to write down an equation that is solved by this function,
\begin{equation}\label{Radial KG equation EFT}
    R_{\ell}''(r) + \(\frac{2}{2r-m} + \frac{2}{m+2r}\) R_{\ell}'(r) - \frac{\ell (\ell+1)}{r^2} R_{\ell}(r) = 0 \,.
\end{equation} 
Upon identification $m=M$ 
we see that this equation 
exactly coincides with the radial part of the Klein-Gordon 
equation in Schwarzschild isotropic coordinates, see Appendix.~\ref{Teukolsky Equation in Schwarzschild BH} for more detail. 
The fact that we could completely reconstruct the Klein-Gordon
equation even when we have ignored the finite-size effects
suggests that Love numbers must be zero.

% this is exactly the static spin-0 Teukolsky equation in S when we replace the particle mass $m$ with BH mass $M$. This tells us one important aspect that the static spin-0 Teukolsky equation in Schwarzschild can be reproduced as the constraint equation for the recurrence relation in EFT diagrams, in which the BH horizon sits at one of the regular singular point.

\subsubsection*{Towards Reconstructing the spin-2 Teukolsky Equation}

% As a consistency check, in this section, we show that the spin-2 electric perturbation equation obtained from $\delta\phi$ can be successfully transferred to the well-known static spin-2 Teukolsky equation. 

As an additional consistency check, let us see
if we can reproduce the spin-2 Teukolsky master 
equation. To that end we need to convert our metric 
perturbations into the Newman-Penrose Weyl scalar.
In the isotropic coordinates, the background Schwarzschild metric  is given by~\eqref{eq:schiso}. The Kinnersley tetrads read~\cite{kinnersley1969type}
\begin{equation}
    \begin{aligned}
    l^\mu &= \(\frac{(M+2r)^2}{(M-2r)^2}, - \frac{4r^2}{M^2-4r^2},0,0\)~, \quad n^\mu = \(\frac{1}{2}, \frac{2(M-2r)r^2}{(M+2r)^3},0,0\) ~, \\
    m^\mu &= \(0,0,\frac{2\sqrt{2} r}{(M + 2r)^2}, \frac{i}{\sin\theta}\frac{2\sqrt{2}r}{(M+2r)^2}\) ,\quad \bar m^\mu = \(0,0,\frac{2\sqrt{2}r}{(M+2r)^2},-\frac{i}{\sin\theta}\frac{2\sqrt{2}r}{ (M+2r)^2}\) \,.
    \end{aligned}
\end{equation}
They satisfy 
$l^\mu n_\mu = 1, m^\mu \bar m_\mu = -1$. 
The Weyl scalar $\psi_0$ is defined as \cite{newman1962approach,newman1963errata}
\begin{equation}
    \psi_0 = - C_{\mu\nu\alpha\beta} l^\mu m^\nu l^\alpha m^\beta \,.
\end{equation}
% This potential is identical to a test 
% scalar field, and hence it also satisfies  
Using the Kaluza-Klein decomposition Eq.\eqref{KK Reduction}, 
let us choose $\sigma = \sigma_{BH}$, 
$\phi = \phi_{\rm BH} + \delta\phi/(2\sqrt{2}M_{\rm pl})$ where $\sigma_{\rm BH}$ and $\phi_{\rm BH}$ are determined by the background Schwarzschild metric, 
see Appendix \ref{Reproducing Schwarzschild Metric}. 
Let us obtain now a linear perturbation equation for $\psi_0$.
In the Newtonian limit, the perturbed Weyl scalar $\psi_0$ is sourced 
by the Newtonian potential $\delta\phi$.
Making use of the usual spin raising operator $\eth^s$ (provided in Appendix \ref{Spin-Weighted Spherical Harmonics}), 
we get the following expression in linear theory:
\begin{equation}\label{Teukolsky psi0}
    \psi_0 = 8\sqrt{2} M_{\rm pl} \frac{r^2}{(M^2 -4r^2)^2} \eth^1 \eth^0 \delta\phi \equiv \sum_{\ell=2}^\infty \sum_{m=-\ell}^{\ell}\mathcal{R}_\ell(r)\, {}_2Y_{\ell m}(\theta,\phi) ~,
\end{equation}
where ${}_2Y_{\ell m}(\theta,\phi)$ is the spin-2 spherical harmonics and $\mathcal{R}_{\ell}(r)$ is the radial part of $\psi_0$. 
Recall now that $\delta\phi$ has the same description as a 
test scalar in the PN EFT. Hence, it also satisfies 
the Klein-Gordon equation~\eqref{Radial KG equation EFT}.
Acting on this equation with the spin raising operators 
and substituting $\mathcal{R}_\ell = R_\ell \frac{r^2}{(M^2-4r^2)^2} $
as dictated by Eq.~\eqref{Teukolsky psi0}, we  
obtain 
 % The corresponding leading order equation satisfied by $\mathcal{R}_{\ell}(r)$ reads 
\begin{equation}\label{Radial Teukolsky EFT}
    \mathcal{R}''_\ell(r) + \(-\frac{10}{M-2r}-\frac{4}{r} + \frac{10}{M+2r}\) \mathcal{R}_{\ell}'(r) - \frac{-6+\ell+\ell^2}{r^2} \mathcal{R}_{\ell}(r) = 0~.
\end{equation}
% \textcolor{blue}{Point out that we match $\phi$ and its OK, because whiever means would work in the EFT}
We see that this equation 
reproduces the Teukolsky equation only up to subleading terms
$O(\mathcal{R}_{\ell} M/r)$. 
This is because the full relativistic Weyl scalar $\psi_0$ 
is not completely determined by the Newtonian potential $\phi$.
It also depends on the metric perturbations $\delta\gamma_{ij}$ 
and the gravito-magnetic field $\delta A_i$. 
We have not included these perturbations because they 
do not affect the extraction of Love numbers
from the one-point function matching. 
Indeed, for this purpose it is sufficient to use only
$\phi$. In principle, we could also 
perform matching at the level of the full Weyl scalar.
In this case we would need to include fluctuations 
of $\delta \gamma_{ij}$ and $\delta A_i$  components as well. 

% As long as we are interested in Love numbers, 
% it is sufficient to use only $\phi$, which
% determines the dynamics in the Newtonian limit. 
% For the interest of love numbers, it is sufficient to only focus on the asymptotic behaviour in the Newtonian limit, which indicates that the constraint equation satisfied by $\psi_0$ in the Newtonian limit can be deduced from the constraint equation of $\delta\phi$. When we include the $\delta \gamma_{ij}$ and $\delta A_i$ perturbations, in principle, we will be able to deduce the full Teukolsky equation.

% After neglecting the small $O(M/r)$ corrections \footnote{}, we get the spin-2 static radial Teukolsky equation \cite{teukolsky1972rotating,teukolsky1973perturbations,press1973perturbations,teukolsky1974perturbations} in isotropic coordinate derived in Appendix \ref{Teukolsky Equation in Schwarzschild BH}.

\subsection{Spin-1 Electric Dipole}\label{Spin-1 Electric Dipole Case}

Let us now focus on spin-1 perturbations. 
The action for a Maxwell field in a curved spacetime 
is given by 
\begin{equation}
    S_{\rm EM} = \int d^4 x\sqrt{-g} \(-\frac{1}{4} g^{\mu\alpha}g^{\nu\beta} \(\partial_\mu {A}_\nu - \partial_\nu {A}_\mu\)\(\partial_\alpha {A}_\beta - \partial_\beta {A}_\alpha\) \) ~.
\end{equation}
where ${A}_\mu$ is the vector potential. 
% In the static limit the electrostatic potential $\mathcal{A}_0$ and 
% the 3-vector $\mathcal{A}_i$ decouple from each other. 
The electric-magnetic duality of the Schwarzschild 
spacetime dictates that the spin-1 
electric and magnetic Love numbers coincide. 
Hence, 
it will be 
sufficient to consider only the electric field, 
which is fully determined by the Coulomb potential,
$E_i = -\partial_i A_0$.
% This can be intuitively understood from the definition $\mathcal{E}_i = -\partial_i \mathcal{A}_0 - \partial_t \mathcal{A}_i, \mathcal{B}_i = \epsilon_{ijk}\partial_j \mathcal{A}_k$. 
% In the static limit, the electric part is fully determined by the Coulomb potential while the magnetic part is determined by the vector potential. For this paper, we only look at the electric part by only turning on the $\mathcal{A}_0$ terms and turning off the $\mathcal{A}_i$ terms. 
Now we expand out the Maxwell action around the Minkowski spacetime 
using the isotropic gauge for gravitational 
perturbations, and separate the Coulomb potential 
into a source part $\bar A_0$ and the response part $A_0$.
We get the following interaction terms
up to the fourth order 
in the number of fields:
\begin{equation}
\label{eq:EMbulk}
        \begin{aligned}
            S_{{A}_0}^{(2)} &= \int dt d^3\boldsymbol{x} \( \frac{1}{2} (\partial_i {A}_0 \partial^i {A}_0) + (\partial_i \bar{{A}}_0 \partial^i {A}_0) \) ~, \\
            S_{{A}_0}^{(3)} &= \int dt d^3\boldsymbol{x} \(\frac{1}{4} \sigma (\partial_i {A}_0 \partial^i {A}_0) + \frac{1}{2} \sigma (\partial_i \bar{{A}}_0 \partial^i {A}_0) - \phi(\partial_i {A}_0 \partial^i {A}_0) - 2\phi (\partial_i \bar{{A}}_0 \partial^i {A}_0)\) \\
            S_{{A}_0}^{(4)} & = \int dt d^3 \boldsymbol{x} \Bigg(- \frac{1}{16} \sigma^2 (\partial_i {A}_0 \partial^i {A}_0) - \frac{1}{8} \sigma^2 (\partial_i \bar{{A}}_0 \partial^i \mathcal{A}_0) + \phi^2(\partial_i {A}_0 \partial^i \mathcal{A}_0) + 2 \phi^2 (\partial_i \bar{{A}}_0 \partial^i {A}_0) \\
            & \quad - \frac{1}{2} \phi\sigma(\partial_i {A}_0 \partial^i {A}_0 ) - \phi\sigma(\partial_i \bar{{A}}_0 \partial^i {A}_0)\Bigg) 
        \end{aligned}
\end{equation}
The static propagator reads
\begin{equation}
    \langle {A}_0 (\boldsymbol{x_1}) {A}_0(\boldsymbol{x_2}) \rangle     
    =
    \begin{tikzpicture}[baseline={([yshift=-2.0 ex]current bounding box.center)}]
        \begin{feynman}
            \vertex (a) at (0,0);
            \vertex (b) at (2,0);
            
            \diagram*{
                (a) -- [ultra thick, MyBlue, edge label = ${A}_0$] (b), 
            };
        \end{feynman}
    \end{tikzpicture}
    = (-1) \int \frac{d^3 \boldsymbol{k}}{(2\pi)^3} \frac{-i}{\boldsymbol{k}^2} e^{i\boldsymbol{k} \cdot (\boldsymbol{x_1}-\boldsymbol{x_2})} \propto \frac{1}{|\boldsymbol{x_1}-\boldsymbol{x_2}|}~.
\end{equation}
The Feynman rules for these vertices are included in Appendix~\ref{Feynman Rule}.
As shown in Section \ref{sec:fs}, the effective action for the spin-1 electric response 
has the form
\begin{equation}
    S_{\rm s=1}^{\rm Love} =  \frac{1}{2\ell !} \int d\tau \lambda_\ell^{\rm s=1} \(\partial_{\langle i_1 \cdots i_{\ell-1}}E_{i_{\ell}\rangle }\) \(\partial^{\langle i_1 \cdots i_{\ell-1}} E^{i_{\ell}\rangle}\)  = 
    \frac{1}{2\ell !} \int d\tau \lambda_\ell^{\rm s=1} \(\partial_{\langle i_1 \cdots i_{\ell}\rangle}A_0\) \(\partial^{\langle i_1 \cdots i_{\ell}\rangle} A_0\)\,,
\end{equation}
where in the last equation we used $E_i = -\partial_i A_0$.
We see that the finite-size action is determined by $A_0$
and it has the same complexity as the scalar field worldline 
action. The spin-1 case is, however, very different 
from the spin-0 and spin-2 in what the $A_0$ bulk
action~\eqref{eq:EMbulk} contains all possible 
interactions between $A_0$ and the gravitational scalars 
$\phi,\sigma$. Because of this reason, 
the Maxwell action generates new interaction terms 
at every new order in gravitational perturbations.
Thus, we could not find 
a systematic way to study PN corrections to $A_0$,
and restricted our analysis to the dipole-type ($\ell=1$) 
external perturbations. 

To determine $c_3$ we need to compute all diagrams up to 3PN order. 
This calculation is laborious, but straightforward. We
present it in detail in Appendix \ref{Spin-1 1-pt Function}.
The final result is 
    \begin{equation}
        \begin{aligned}
        & \quad 
                        \vcenter{\hbox{
                \begin{tikzpicture}[scale=0.8]
                    \begin{feynman}
                      \vertex (i) at (0, 0);
                      \vertex (e) at (0, 3);
                      \vertex[dot, MyRed] (w1) at (0, 2.5) {};
                      \vertex[dot, MyRed] (w2) at (0, 0.5) {};
                      \vertex[dot, MyRed] (w3) at (0,1.5) {};
                      \vertex (c1) at (1.5, 2.5);
                      \vertex (c2) at (1.5, 0.5);
                      \vertex (c3) at (1.5, 1.5);
                      \vertex[dot, MyBlue, label= $\boldsymbol{x}$] (f1) at (2.5,2.8) {};
                      \vertex[crossed dot, MyBlue] (fs) at (2.5,0.2) {};

                      \diagram*{
                        (i) -- [double,double, thick] (e), 
                        (w1) -- [dashed, MyRed, ultra thick] (c1),
                        (w2) -- [dashed, MyRed, ultra thick] (c2),
                        (w3) -- [dashed, MyRed, ultra thick] (c3),
                        (c1) -- [ultra thick, MyBlue] (f1), 
                        (fs) -- [ultra thick, MyBlue, edge label' = $r$] (c2), 
                        (c2) -- [ultra thick, MyBlue] (c1), 
                      };
                    \end{feynman}
                \end{tikzpicture}}} 
                +
                \vcenter{\hbox{
                    \begin{tikzpicture}[scale=0.7]
                        \begin{feynman}
                          \vertex (i) at (0, 0);
                          \vertex (e) at (0, 3);
                          \vertex[dot, MyRed] (w1) at (0, 1.5) {};
                          \vertex[dot, MyRed] (w2) at (0, 0.5) {};
                          \vertex[dot, MyRed] (w3) at (0, 2.5) {}; 
                          \vertex (c1) at (1,1.0);
                          \vertex (c2) at (2.5,1.0);
                          \vertex (c3) at (2.5, 2.5);
                          \vertex[dot, MyBlue, label= $\boldsymbol{x}$] (f1) at (3.5,2.8) {};
                          \vertex[crossed dot, MyBlue] (fs) at (3.5,0.2) {}; 
                    
                          \diagram*{
                            (i) -- [double,double, thick] (e), 
                            (w1) -- [dashed, MyRed, ultra thick] (c1),
                            (w2) -- [dashed, MyRed, ultra thick] (c1),
                            (w3) -- [dashed, MyRed, ultra thick] (c3),
                            (c1) -- [boson, MyYellow, ultra thick] (c2),
                            (c3) -- [ultra thick, MyBlue] (f1), 
                            (c3) -- [ultra thick, MyBlue] (c2),
                            (fs) -- [ultra thick, MyBlue, edge label' = $r$] (c2),  
                          };
                        \end{feynman}
                    \end{tikzpicture}}}
                    +
                    \vcenter{\hbox{
                        \begin{tikzpicture}[scale=0.7]
                            \begin{feynman}
                              \vertex (i) at (0, 0);
                              \vertex (e) at (0, 3);
                              \vertex[dot, MyRed] (w1) at (0, 1.5) {};
                              \vertex[dot, MyRed] (w2) at (0, 0.5) {};
                              \vertex[dot, MyRed] (w3) at (0, 2.5) {};
                              \vertex (c1) at (1.5,1.0);
                              \vertex (c2) at (1.5, 2.5);
                              \vertex[dot, MyBlue, label= $\boldsymbol{x}$] (f1) at (3.0,2.8) {};
                              \vertex[crossed dot, MyBlue] (fs) at (3.0,0.2) {}; 
                        
                              \diagram*{
                                (i) -- [double,double, thick] (e), 
                                (w1) -- [dashed, MyRed, ultra thick] (c1),
                                (w2) -- [dashed, MyRed, ultra thick] (c1),
                                (w3) -- [dashed, MyRed, ultra thick] (c2),
                                (c1) -- [ultra thick, MyBlue] (c2),
                                (c2) -- [ultra thick, MyBlue] (f1), 
                                (fs) -- [ultra thick, MyBlue, edge label' = $r$] (c1),  
                              };
                            \end{feynman}
                        \end{tikzpicture}}} 
                        +
                                    \vcenter{\hbox{
                \begin{tikzpicture}[scale=0.7]
                    \begin{feynman}
                      \vertex (i) at (0, 0);
                      \vertex (e) at (0, 3);
                      \vertex[dot, MyRed] (w1) at (0, 2.5) {};
                      \vertex[dot, MyRed] (w2) at (0, 1.5) {};
                      \vertex[dot, MyRed] (w3) at (0, 0.5) {};
                      \vertex (c1) at (1,2);
                      \vertex (c2) at (2.5,2);
                      \vertex (c3) at (2.5,0.5);
                      \vertex[dot, MyBlue, label= $\boldsymbol{x}$] (f1) at (3.5,2.8) {};
                      \vertex[crossed dot, MyBlue] (fs) at (3.5,0.2) {}; 
                
                      \diagram*{
                        (i) -- [double,double, thick] (e), 
                        (w1) -- [dashed, MyRed, ultra thick] (c1),
                        (w2) -- [dashed, MyRed, ultra thick] (c1),
                        (w3) -- [dashed, MyRed, ultra thick] (c3),
                        (c1) -- [boson, MyYellow, ultra thick] (c2),
                        (c2) -- [ultra thick, MyBlue] (c3),
                        (c2) -- [ultra thick, MyBlue] (f1), 
                        (fs) -- [ultra thick, MyBlue, edge label' = $r$] (c3),  
                      };
                    \end{feynman}
                \end{tikzpicture}}}
                +
                \vcenter{\hbox{
                    \begin{tikzpicture}[scale=0.7]
                        \begin{feynman}
                          \vertex (i) at (0, 0);
                          \vertex (e) at (0, 3);
                          \vertex[dot, MyRed] (w1) at (0, 2.5) {};
                          \vertex[dot, MyRed] (w2) at (0, 1.5) {};
                          \vertex[dot, MyRed] (w3) at (0, 0.5) {};
                          \vertex (c1) at (1.5,2);
                          \vertex (c3) at (1.5,0.5);
                          \vertex[dot, MyBlue, label= $\boldsymbol{x}$] (f1) at (3.0,2.8) {};
                          \vertex[crossed dot, MyBlue] (fs) at (3.0,0.2) {}; 
                    
                          \diagram*{
                            (i) -- [double,double, thick] (e), 
                            (w1) -- [dashed, MyRed, ultra thick] (c1),
                            (w2) -- [dashed, MyRed, ultra thick] (c1),
                            (w3) -- [dashed, MyRed, ultra thick] (c3),
                            (c1) -- [ultra thick, MyBlue] (c3),
                            (c1) -- [ultra thick, MyBlue] (f1), 
                            (fs) -- [ultra thick, MyBlue, edge label' = $r$] (c3),  
                          };
                        \end{feynman}
                    \end{tikzpicture}}}
                    \\
                    &+
                        \vcenter{\hbox{\begin{tikzpicture}[scale=0.7]
        \begin{feynman}
            \vertex (i) at (0, 0);
                  \vertex (e) at (0, 3);
                  \vertex[dot, MyRed] (w1) at (0, 2.2) {};
                  \vertex[dot, MyRed] (w2) at (0, 1.2) {};
                  \vertex[dot, MyRed] (w3) at (0, 0.5) {};
                  \vertex (c1) at (0.8,1.5);
                  \vertex (c2) at (2,1.5);
                  \vertex (c3) at (3.5,1.5);
                  \vertex[dot, MyBlue, label = $\boldsymbol{x}$] (f1) at (4.5,2.8) {};
                  \vertex[crossed dot, MyBlue] (fs) at (4.5,0.2) {};
            
                  \diagram*{
                    (i) -- [double,double, thick] (e), 
                    (w1) -- [dashed, MyRed, ultra thick] (c1),
                    (w2) -- [dashed, MyRed, ultra thick] (c1),
                    (c1) -- [boson, MyYellow, ultra thick] (c2),  
                    (w3) -- [dashed, MyRed, ultra thick] (c2),
                    (c2) -- [dashed, MyRed, ultra thick] (c3),
                    (c3) -- [ultra thick, MyBlue] (f1),
                    (fs) -- [ultra thick, MyBlue, edge label' = $r$] (c3),
                  };
        \end{feynman}
    \end{tikzpicture}}}
    +
                \vcenter{\hbox{\begin{tikzpicture}[scale=0.7]
        \begin{feynman}
            \vertex (i) at (0, 0);
                  \vertex (e) at (0, 3);
                  \vertex[dot, MyRed] (w1) at (0, 2.2) {};
                  \vertex[dot, MyRed] (w2) at (0, 1.2) {};
                  \vertex[dot, MyRed] (w3) at (0, 0.5) {};
                  \vertex (c1) at (0.8,1.5);
                  \vertex (c2) at (2,1.5);
                  \vertex[dot, MyBlue, label = $\boldsymbol{x}$] (f1) at (3.5,2.8) {};
                  \vertex[crossed dot, MyBlue] (fs) at (3.5,0.2) {};
            
                  \diagram*{
                    (i) -- [double,double, thick] (e), 
                    (w1) -- [dashed, MyRed, ultra thick] (c1),
                    (w2) -- [dashed, MyRed, ultra thick] (c1),
                    (c1) -- [boson, MyYellow, ultra thick] (c2),  
                    (w3) -- [dashed, MyRed, ultra thick] (c2),
                    (c2) -- [ultra thick, MyBlue] (f1),
                    (fs) -- [ultra thick, MyBlue, edge label' = $r$] (c2),
                  };
        \end{feynman}
    \end{tikzpicture}}}
    +
                            \vcenter{\hbox{
            \begin{tikzpicture}[scale=0.7]
                \begin{feynman}
                  \vertex (i) at (0, 0);
                  \vertex (e) at (0, 3);
                  \vertex[dot, MyRed] (w1) at (0, 2.5) {};
                  \vertex[dot, MyRed] (w2) at (0, 1.5) {};
                  \vertex[dot, MyRed] (w3) at (0,0.5) {};
                  \vertex (c1) at (1.5,1.5);
                  \vertex[dot, MyBlue, label = $\boldsymbol{x}$] (f1) at (3.0,2.8) {};
                  \vertex[crossed dot, MyBlue] (fs) at (3.0,0.2) {}; 
            
                  \diagram*{
                    (i) -- [double,double, thick] (e), 
                    (w1) -- [dashed, MyRed, ultra thick] (c1),
                    (w2) -- [dashed, MyRed, ultra thick] (c1),
                    (w3) -- [dashed, MyRed, ultra thick] (c1),
                    (c1) -- [ultra thick, MyBlue] (f1), 
                    (fs) -- [ultra thick, MyBlue, edge label' = $r$] (c1),  
                  };
                \end{feynman}
            \end{tikzpicture}}}
            =0 ~.
        \end{aligned}
\end{equation}
This tells us that for the spin-1 electric dipole case $c_{3}=0$, 
which means that the spin-1 dipole polarization coefficients 
do not flow under RG (see Eq.~\eqref{eq:rg}). 
Just like in the spin-0 and spin-2 examples, the 
absence of logarithmic running 
is fine-tuning from the worldline EFT perspective. 
Its origin can be traced to the Love symmetry in the UV~\cite{Charalambous:2021kcz}. 

Note that in contrast to the spin-0/2 cases,
the isotropic gauge does not seem to be 
particularly useful for the spin-1 perturbations.
As expected in the general case, here each PNEFT diagram 
carries a logarithm, which all cancel when all diagrams are summed
together. We believe that there should exist a gauge 
where the cancellation of PN corrections to the spin-1 one-point
functions is manifest 
to begin with. Since there are no logs in the UV theory, 
in such a gauge it should be possible to 
prove that $c_{2\ell+1}=0$ for the electromagnetic (EM) perturbations for 
general $\ell$. 
We leave an explicit construction of this gauge for future work.

% For generic spin-1 $\ell\in\mathbb{N}$ case, even though we do not have a good proof of vanishing $c_{2\ell+1}$ due to the complicated diagramatic structure in the EFT side, it is reasonable to make a conjecture that $c_{2\ell+1}=0$ holds for generic $\ell$. This argument will be verified without ambiguity when we match to the full theory.

\section{Black Hole Perturbation Theory}\label{sec:bhpt}

To determine the Love numbers we need to match 
the full EFT calculation, including finite-size effects, 
to the UV theory result. The UV theory for our problem 
is the black hole linear perturbation theory. 
We will see now that one can match the EFT and UV 
expressions without any need for an analytic continuation.
First, we will point out that the EFT calculation
of the PN corrections to the source term is equivalent to 
constructing the solution with the Frobenius method. 
Then we will obtain spin-0,1,2 solutions in BH perturbation theory. 
Remarkably, these always coincide with the EFT solutions
that describe external sources with PN corrections attached to them. 
Since the full solution is reproduced by the PN corrections alone, 
without any finite-size effects, the Love numbers
must vanish identically.

Note that for the spin-1 and spin-2 perturbations we will 
match directly the Coulomb potential and the dilaton field profiles, respectively. 
These choices 
are gauge-dependent, but the results of our matching are not,
as our EFT calculations are carried out 
in the gauges consistent with the background 
geometry and the gauge choice of the UV solution.

% Following the standard fashion, all the unknown coefficients in the EFT, in principle, can be determined by the matching with the UV theory. Here, our UV theory is the BH perturbation theory. We will show that the diagrams appear in the gravitational nonlinear corrections can also be determined by Frobenius method, which gives the same results as the EFT diagramatic calculations. The love number and the coefficient in front of the logarithmic divergence can be determined further by putting the regular boundary condition at the horizon without doing analytic continuation in $\ell$. For completeness, the traditional analytic continuation method is also mentioned to make comparison. But we should emphasize that there is actually no need to do the analytic continuation, and everything can be in the physical region $\ell \in \mathbb{N}$.

\subsection{EFT versus the Frobenius Method}\label{sec:froben}

The Frobenius method is a method to 
construct a power series solution to a differential equation.
We will see now that this power series exactly maps onto the 
PN diagrammatic method.
% The Frobenius method makes use of the power counting law of the Feynman diagrams shown in Eq.\eqref{Power Counting Law}. For the diagrams of source corrections, it follows a nice power series with certain coefficient in front of each term. Apart from the EFT diagrmatic method, there is another way to determine these coefficients by viewing static Teukolsky equation as a constraint equation. The key point here is that we do not need to put the regular boundary condition at the event horizon, but just use the power series ansatz and the static Teukolsky equation to find the recurrence relation.

Let us start with the spin-0 case.
For the spin-0 case, the corresponding Teukolsky equation is covariantly written as 
\begin{equation}
    \nabla_\mu \nabla^\mu \Phi(\boldsymbol{x},t) = 0 ~.
\end{equation}
Separating the variable as $\Phi(\boldsymbol{x},t) = \sum_{\ell m} R_{\ell}(r) Y_{\ell m}e^{-i \omega t}$, we obtain the following equation for the 
radial function in the static limit:
\begin{equation}\label{Static Radial Equation}
    R_{\ell}''(r) + \(\frac{2}{2r-M} + \frac{2}{M+2r}\) R_{\ell}'(r) - \frac{\ell (\ell+1)}{r^2} R_{\ell}(r) = 0 ~,
\end{equation}
which is exactly the same as the equation obatained
by the ``resummation'' of the EFT diagrams; see Eq.~\eqref{Radial KG equation EFT} with bare mass $m$ replaced by BH mass $M$. 
The power series ansatz to solve~\eqref{Static Radial Equation}
takes the form 
\begin{equation}\label{eq:ans}
    R_{\ell}(r) = r^\ell \(1 +c_1 \(\frac{M}{2r}\) + c_2 \(\frac{M}{2r}\)^2 + \cdots + c_{N_m}\(\frac{M}{2r}\)^{N_m} + \cdots \) ~.
\end{equation}
Note that this ansatz has an ambiguity for $N_m\geq 2\ell + 1$,
which corresponds to 
a freedom of adding a decaying solution $\sim r^{-\ell -1}$ at $r\to \infty$.
We will see in the next section
that the series actually truncates at $N_m=\ell$.
% This ansatz does not have ambiguity if we only focus on $N_m<2\ell + 1$, since there is no source and response mixing as mentioned in Section \ref{GR Nonlinear Correction BH Response}. 
Plugging~\eqref{eq:ans} into Eq.~\eqref{Static Radial Equation}, it is straightforward to find that recurrence relation 
\begin{equation}
    (n+1)(2n+1 - 2 \ell) c_{2(n+1)} = (2n+1)(n-\ell)c_{2n} ~,
\end{equation}
which coincides with the 
EFT diagramatic recurrence relation Eq.~\eqref{Recurrence Relation}. 
This relation can be solved iteratively,
\begin{equation}\label{Frobenius Coefficients}
    \begin{aligned}
        c_1 &= c_3 = \cdots = c_{2\ell-1}   =0 ~, \\
        c_2 &= \frac{\ell}{-1+2\ell} ~, c_4 = \frac{3(-1+\ell)\ell}{2(-3 +2\ell)(-1+2\ell)} ~, \cdots \\
        c_{2n} &= (-1)^n \frac{(2n-1)!!}{2^n n!}(-n+1+\ell)\cdots (-1+\ell)\ell \times \frac{\Gamma(\frac{1}{2}-\ell)}{\Gamma(\frac{2n+1}{2}-\ell)}~, ~ n\leq \ell ~.
    \end{aligned}
\end{equation}
Indeed, from the EFT diagrams we have:
% With this solution, it is clear that for the static case the non-vanishing diagram takes the form
\begin{equation}
    \begin{aligned}
        \vcenter{\hbox{\begin{tikzpicture}
        \begin{feynman}
          \vertex (i) at (0, 0);
          \vertex (e) at (0, 3);
          \vertex[dot, MyRed] (w1) at (0, 2.5) {};
          \vertex[dot, MyRed] (w2) at (0, 2.0) {};
          \vertex[dot, MyRed] (w3) at (0, 0.5) {};
          \node at (0.5,1.1) {$\cdot$};
          \node at (0.5,1.3) {$\cdot$};
          \node at (0.5,1.5) {$\cdot$};
          \vertex[blob, scale =2] (c1) at ( 2, 1.5) {};
          \node[circle, fill=white, scale = 0.8] at (2,1.5) {static};
          \vertex[dot, MyBlue, label= $\boldsymbol{x}$] (f1) at (3.5,2.8) {};
          \vertex[crossed dot, MyBlue] (fs) at (3.5,0.2) {}; 
    
          \diagram*{
            (i) -- [double,double, thick, edge label = $2n$] (e), 
            (w1) -- [dashed, MyRed, ultra thick] (c1) -- [dashed, MyRed, ultra thick] (w2),  
            (w3) -- [dashed, MyRed, ultra thick] (c1),
            (c1) -- [ultra thick, MyBlue] (f1), 
            (fs) -- [ultra thick, MyBlue, edge label' = $r^\ell$] (c1),  
          };
        \end{feynman}
    \end{tikzpicture}}}
    & = \vcenter{\hbox{
            \begin{tikzpicture}[scale=0.8]
                \begin{feynman}
                  \vertex (i) at (0, 0);
                  \vertex (e) at (0, 3);
                  \vertex[dot, MyRed] (w1) at (0, 2.75) {};
                  \vertex[dot, MyRed] (w2) at (0, 2.00) {};
                  \vertex[dot, MyRed] (w3) at (0, 1.25) {};
                  \vertex[dot, MyRed] (w4) at (0, 0.5) {};
                  \vertex (c1) at (1, 2.375);
                  \vertex (c2) at (1, 0.875);
                  \vertex (c3) at (2.5, 2.375);
                  \vertex (c4) at (2.5, 0.875);
                  \vertex[dot, MyBlue, label= $\boldsymbol{x}$] (f1) at (3.5,2.8) {};
                  \vertex[crossed dot, MyBlue] (fs) at (3.5,0.2) {}; 
                  \node at (1.75,1.5) {$\cdot$};
                  \node at (1.75,1.7) {$\cdot$};
                  \node at (1.75,1.3) {$\cdot$};

                  \diagram*{
                    (i) -- [double,double, thick, edge label = $2n$] (e), 
                    (w1) -- [dashed, MyRed, ultra thick] (c1),
                    (w2) -- [dashed, MyRed, ultra thick] (c1),
                    (c1) -- [boson, MyYellow, ultra thick] (c3),
                    (w3) -- [dashed, MyRed, ultra thick] (c2),
                    (w4) -- [dashed, MyRed, ultra thick] (c2),
                    (c2) -- [boson, MyYellow, ultra thick] (c4),
                    (c3) -- [ultra thick, MyBlue] (f1), 
                    (fs) -- [ultra thick, MyBlue, edge label' = $r^{\ell}$] (c4), 
                    (c3) -- [ultra thick, MyBlue] (c4), 
                  };
                \end{feynman}
            \end{tikzpicture}}} \\
            & =  r^\ell \times \((-1)^n \frac{(2n-1)!!}{2^n n!}(-n+1+\ell)\cdots (-1+\ell)\ell \frac{\Gamma(\frac{1}{2}-\ell)}{\Gamma(\frac{2n+1}{2}-\ell)}\) \\
            & \quad \times \(\frac{m}{2r}\)^{2n} \,,
    \end{aligned}
\end{equation}
where we took into account only the physical pyramid graphs. 
% where bare mass $m$ is replaced by BH mass $M$. 
% In a word, the EFT diagrams for the source corrections give the exact same results as the Frobenius method for spin-0 Teukoleky equation. 
The argument would be the same for static spin-2 electric perturbations captured by the dilaton field.

As far as the spin-1 perturbations are concerned, they satisfy the
the covariant Maxwell equation
\begin{equation}
    \nabla_\mu F^{\mu\nu} = 0 ~,
\end{equation}
where $F_{\mu\nu} = \partial_\mu A_\nu - \partial_\nu A_\mu$. In the static case, the equation for $A_0$ and $A_i$ decouple. Using $A_0(\boldsymbol{x},t) = \sum_{\ell m} R_{\ell}(r) Y_{\ell m}e^{-i \omega t}$ and taking the static limit, we find
\begin{equation}\label{spin-1 Electric Radial Equation}
    R''_\ell(r) + \(\frac{2}{M-2r} + \frac{6}{M+2r}\) R'_\ell (r) -\frac{\ell(\ell + 1)}{r^2}R_{\ell}(r) = 0 ~.
\end{equation}
Focusing on the dipole case, we can get the following coefficients
with the Frobenius method:
\begin{equation}
    c_1 = -2, \quad c_2 = 1 ~, \quad c_{n,~n>2}=0\,,
\end{equation}
which coincides with the explicit 
EFT diagrammatic computation presented in Appendix~\ref{Spin-1 1-pt Function}.

All in all, the upshot of this Section is that 
the EFT calculation of the PN corrections to the source
has a one-to-one mapping onto the Frobenius method
of solving field perturbations in the Schwarzschild background. 
The Frobenius method builds a solution in 
terms of a power series in $m/r$, 
attached to the growing source asymptotic $r^\ell$. 
In our case this solution happened to be 
a polynomial, which is also regular at the 
BH horizon. 
By the uniqueness theorem, 
another linearly independent solution
is singular at the horizon,
and hence it does not contribute to the
physical profile. In other words, 
the Frobenius 
solution is the full solution.
The structure of this solution
implies that the PN corrections to Love numbers
and Love numbers vanish altogether.

% This solution is, however, not 
% guaranteed, in general, 
% to converge around the BH horizon.
% This means that the Frobenius solution 
% is incomplete. 
% In order to obtain the full physical solution
% we need to find a solution regular at the BH horizon,
% to which we proceed now.

\subsection{UV Calculation}

% The physical full theory actually means we need to solve the static Teukolsky equation for $\ell \in \mathbb{N}$ within the regular boundary condition at the event horizon. In 4 dimensional Schwarzschild black hole, the correct matching procedure in the physical region $\ell \in \mathbb{N}$ between EFT and UV has seldom been down. Typically, one will either ignore the gravitational nonlinear corrections at $2\ell + 1$ PM order, or use the ad hoc analytic continuation to artificially separate the source and response.

\subsubsection*{Spin-1}

% For spin-1 electric perturbations, in order to compare with EFT, we need to derive the two linearly independent solutions based on expansions at infinity $r\rightarrow \infty$ and require the regularity at the horizon to fix the relative ratio.  To do this, it is convenient to introduce another variable $z=M/(2r)$, and rewrite Eq.\eqref{spin-1 Electric Radial Equation} into 
We seek a solution of the Maxwell equation~\eqref{spin-1 Electric Radial Equation} rewritten in a new variable $z=M/(2r)$,
\begin{equation}
    R_{\ell}''(z) + \(\frac{1}{1-z} + \frac{3}{1+z}\) R_{\ell}'(z) - \frac{\ell(1+\ell)}{z^2} R_{\ell}(z) =0 ~.
\end{equation}
The above equation can be recast into the standard form of the hypergeometric equation by redefining the field $R_\ell(z) = z^{-\ell} (1-z)^2 u_\ell (z)$ and introducing $x=z^2$,
\begin{equation}
    u_\ell ''(x) + \(\frac{3}{-1+x} + \frac{1-2\ell}{2x}\) u'_\ell (x) + \(- \frac{3(-1+\ell)}{2(-1+x)} + \frac{3(-1+\ell)}{2x}\) u_\ell (x) = 0~.
\end{equation}
This equation has two linearly independent solutions 
\begin{equation}
    u_\ell^1 (x) = {}_2 F_1 \(\frac{3}{2}, 1-\ell, \frac{1}{2} - \ell,x\) ~, \quad u_\ell^2 (x) = x^{\frac{2\ell + 1}{2}} {}_2 F_1 \(\frac{3}{2}, 2+\ell, \frac{3}{2} + \ell, x\) ~.
\end{equation}
For physical values $\ell \in \mathbb{N}$, only the first solution is regular at the horizon $x=1$. 
The full solution then is a polynomial of order $\ell$,
\begin{equation}\label{Full theory solution}
    \begin{aligned}
        R_\ell^{\rm full} (z) &= z^{-\ell}(1-z)^2 {}_2 F_1 \(\frac{3}{2}, 1-\ell,\frac{1}{2}-\ell,z^2\) \\
        & = z^{-\ell}(1-z)^2 \sum_{n=0}^{\ell-1} (-1)^n  \binom{\ell -1}{n}\frac{
        \left(\frac{3}{2}\right)_n}{\left(\frac{1}{2}-\ell\right)_n} z^{2n}\,.
    \end{aligned}
\end{equation}
This tells us that there are no logarithmic corrections at $2\ell + 1$ PN order
and the Love numbers vanish identically.  
Indeed, matching the one-point functions in the EFT and the full theory we get
\begin{equation}
    c_{2\ell + 1} \log\({r}{\mu}\) - m^{-2\ell-1}\frac{(2\ell-1)!!}{4\pi} \lambda_{\ell}^{s=1}(\mu) \equiv  0 ~, \forall r > 0 ~.
\end{equation}
Since the renormalized Love number is a coupling constant on the worldline, it could only depend on the renormalization scale $\mu$ and does not depend on $r$. 
This implies that 
\begin{equation}
    c_{2\ell + 1} = 0, \quad \lambda_\ell^{s=1} (\mu) = 0 ~.
\end{equation}
This confirms the computation shown in the specific examples in Section \ref{Spin-1 Electric Dipole Case} that the $2\ell + 1$ PN order gravitational nonlinear correction vanishes, and the Love number has no RG running behavior. Thus, the vanishing of the ($2\ell + 1$)PN term in the full theory tells us that the gravitational nonlinear correction $c_{2\ell + 1}$ and the Love number $\lambda_\ell$ vanish altogether.

\subsubsection*{Spin-0/2}

The argument is similar for the spin-0 and spin-2 perturbations.
Introducing $R(z) = z^{-\ell} u(z)$ and $x= z^2$, we rewrite Eq.~\eqref{Static Radial Equation} as
\begin{equation}
    u''(x) + \(\frac{1}{-1+x} + \frac{1-2\ell}{2x}\)u'(x) + \(\frac{\ell}{2-2x} + \frac{\ell}{2x}\) u(x) =0 ~.
\end{equation}
This equation admits two linearly independent solutions 
\begin{equation}\label{Scalar Linear Independent Solutions}
    u_1(x) = {}_2 F_1 \(\frac{1}{2},-\ell,\frac{1}{2}-\ell,x\), \quad u_2(x) = x^{\frac{2\ell + 1}{2}}{}_2 F\(\frac{1}{2}, 1 + \ell, \frac{3}{2} + \ell, x\)
\end{equation}
For $\ell \in \mathbb{N}$, only $u_1(x)$ is regular at event horizon $x=1$. Thus, the full theory solution is 
\begin{equation}
    R_{\ell}^{\rm full}(z) = z^{-\ell} {}_2 F_1\(\frac{1}{2}, -\ell, \frac{1}{2}-\ell,z^2\)  ~.
\end{equation}
This is a polynomial of order $\ell$, which fully coincides with 
the result of the EFT PN calculations. Matching this to the full 
EFT calculation including finite-size effects, we obtain
that the Love numbers vanish identically,
\be 
\lambda_{\ell}^{s=0}=0\,, \quad \lambda_{\ell}^{s=2}=0\,.
\ee

% This implies that the $2\ell + 1$ PM order corrections to the source completely vanish. 
% As a result, $c_{2\ell+1} = 0$, and the Love numbers vanish, $\lambda_{\ell}^{s=0}=0$, $\lambda_{\ell}^{s=2}=0$.

All in all, we have shown that the worldline EFT 
approach allows one to unambiguously separate 
the source and response components 
of external fields
in the response problem 
in full general 
relativity. In particular, we have shown that 
the ($2\ell + 1$)PN corrections to the source profiles vanish,
which implies the absence of RG running of Love numbers.  
The Love numbers vanish as well, so our final results 
can be represented in the following diagrammatic form:
% As a final remark, we emphasize that there is no ambiguity in the matching procedure due to the logarithmic behaviour of the gravitational nonlinear correction at the $2\ell + 1$ PM order. The love number and the gravitational nonlinear corrections can be determined separately, and they both vanishing identically for 4D Schwarzschild BH. All in all, we get our main conclusion without doing analytic continuation
\begin{eBox}
\begin{equation}\label{Diagramatic Intepretation of Vanishing Love}
     \sum_{\substack{\text{all} \\ \text{possible} \\ \text{diagrams}}}
        \vcenter{\hbox{\begin{tikzpicture}
        \begin{feynman}
          \vertex (i) at (0, 0);
          \vertex (e) at (0, 3);
          \vertex[dot, MyRed] (w1) at (0, 2.5) {};
          \vertex[dot, MyRed] (w2) at (0, 2.0) {};
          \vertex[dot, MyRed] (w3) at (0, 0.5) {};
          \node at (0.5,1.1) {$\cdot$};
          \node at (0.5,1.3) {$\cdot$};
          \node at (0.5,1.5) {$\cdot$};
          \vertex[blob, scale =2] (c1) at ( 2, 1.5) {};
          \node[circle, fill=white, scale = 0.8] at (2,1.5) {static};
          \vertex[dot, MyBlue, label= $\boldsymbol{x}$] (f1) at (3.5,2.8) {};
          \vertex[crossed dot, MyBlue] (fs) at (3.5,0.2) {}; 
    
          \diagram*{
            (i) -- [double,double, thick, edge label = $2\ell + 1$] (e), 
            (w1) -- [dashed, MyRed, ultra thick] (c1) -- [dashed, MyRed, ultra thick] (w2),  
            (w3) -- [dashed, MyRed, ultra thick] (c1),
            (c1) -- [ultra thick, MyBlue] (f1), 
            (fs) -- [ultra thick, MyBlue, edge label' = $r^{\ell}$] (c1),  
          };
        \end{feynman}
    \end{tikzpicture}}}
    =0 ~, \quad \quad 
        \vcenter{\hbox{\begin{tikzpicture}
        \begin{feynman}
            \vertex (i) at (0,0);
            \vertex (e) at (0,3);
            \node[circle, draw=repGreen, fill = repGreen, scale=1, label=left:$\lambda_\ell$] (w1) at (0, 1.5);
            \vertex[dot, MyBlue, label= $\boldsymbol{x}$] (f1) at (1.5,2.8) {};
            \vertex[crossed dot, MyBlue] (fs) at (1.5,0.2) {}; 

            \diagram*{
                (i) -- [double, double, thick] (w1),
                (w1) -- [double, double, thick] (e),
                (f1) -- [MyBlue, ultra thick] (w1),
                (fs) -- [MyBlue, ultra thick, edge label' = $r^{\ell}$] (w1)
            };
        \end{feynman}
    \end{tikzpicture}}}
    = 0 ~.
\end{equation}
\end{eBox}

\subsection{Analytic Continuation}\label{Analytic Continuation}

For completeness, let us match the Love numbers 
for the spin-0 case
using the analytic
continuation prescription.
This prescription amounts to treating multipole orbital numbers
as rational numbers, $\ell \in \mathbb{R}$. 
This approach is motivated by 
higher-dimensional black hole perturbation theory, 
where the relevant parameter in perturbation equations is $\ell/(d-3)$,
where $d$ is the number of spacetime dimensions~\cite{Kol:2011vg}.
% However, we stress that in the realistic case $d=4$ 
% this prescription is

For a generic $\ell$ neither of the two solutions $u_1(x)$ and $u_2(x)$ in Eq.\eqref{Scalar Linear Independent Solutions} 
is regular at the event horizon $x=1$. Instead, one obtains a regular solution
by combining the two, 
\begin{equation}
    \begin{aligned}
    u_\ell(x) &= \frac{\Gamma(\frac{1}{2} + \ell)}{\Gamma\(\ell + 1\) \Gamma\(\frac{1}{2}\)} {}_2F_1\(\frac{1}{2},-\ell,\frac{1}{2}-\ell,x\) + \frac{\Gamma\(-\frac{1}{2}-\ell\)}{\Gamma(\frac{1}{2})\Gamma\(-\ell\)}x^{\frac{2\ell + 1}{2}}{}_2 F\(\frac{1}{2}, 1 + \ell, \frac{3}{2} + \ell, x\) \\
    & = {}_2 F_1\(\frac{1}{2},-\ell,1,1-x\) ~.
    \end{aligned}
\end{equation}
The corresponding full theory solution then reads
\begin{equation}
    R_{\ell}^{\rm full}(z) = z^{-\ell} {}_2 F_1\(\frac{1}{2},-\ell,1,1-z^2\) ~.
\end{equation}
Now, we keep $\ell$ generic and do the asymptotic expansion of the full theory solution at the asymptotic infinity,
\begin{equation}
    \begin{aligned}
        \lim_{r\rightarrow \infty} R_{\ell}^{\rm full}(r) & \propto r^\ell \(1 +c_1 \(\frac{M}{2r}\) + c_2 \(\frac{M}{2r}\)^2 + \cdots + c_{N_m}\(\frac{M}{2r}\)^{N_m} + \cdots \) \\
        & \quad + r^{-\ell-1} \(- \(\frac{M}{2}\)^{2\ell + 1} \frac{\Gamma(\frac{1}{2}-\ell) \Gamma(1+\ell)}{\Gamma(-\ell) \Gamma(\frac{3}{2}+\ell)} + \cdots  \) ~,
    \end{aligned}
\end{equation}
where $c_1, c_2,\ldots$ are the coefficients given in Eq.~\eqref{Frobenius Coefficients}. 
For generic noninteger and non-half-integer $\ell$, the first series describes the PN correction to the source $r^{\ell}$, while the second series describes the BH response. Importantly, these two asymptotic series never mix. 
This is a celebrated success of the analytic continuation. 

% To get predictions for the physical $\ell \in \mathbb{N}$, we are now restricting from $\ell \in \mathbb{R}$ to $\ell \in \mathbb{N}$. 
Now we take the physical limit $\ell \in \mathbb{N}$.
The source series then exactly reduces to our results
obtained with the diagrammatic method.
The response part, however, vanishes 
since $\Gamma(-\ell)\rightarrow \infty$ when $\ell \in \mathbb{N}$.

% It is unambiguous that both EFT diagramatic method and Frobenius method clearly give the result in the first part which describes the gravitational nonlinear correction to the source. 
% The undetermined Wilson coefficient that describes the love number can be matched with the second response part. In Eq.\eqref{Analytic Continuation}, , then the love number vanishes.  

\section{Dissipation Numbers}\label{sec:dissN}

One can extract the dissipation number
by matching one-point functions
in the EFT and UV theories. 
In this section we perform this matching 
explicitly. Importantly, this procedure gives the same
dissipation numbers as a matching of absorption
cross sections. This is an valuable consistency
check of the EFT approach. 

To match the dissipation numbers,
we solve the time-dependent Teukolsky equation 
in the near zone approximation. Note that the near zone
approximation does not exactly correspond to a low-frequency
expansion~\cite{Charalambous:2021mea,Charalambous:2021kcz}. 
As a result, the near zone approximation does not 
correctly reproduce the time-dependent
conservative effects~\cite{Charalambous:2021mea}.
However, it is sufficient for the matching 
of the dissipation number, and hence it is adequate
for our purposes.

% we have mentioned that the dissipation number can also be matched through the 1-point function of the external scalar field through the in-in formalism. In this section, we complete the story by matching to the black hole perturbation theory.

\subsection{Near Horizon Teukolsky Equation}

The near horizon approximation is based on the fact that the
Teukolsky equation simplifies drastically 
in the regime 
\be
 \omega r \ll 1\,, \quad M\omega \ll 1\,.
\ee
Under these assumptions, the
spin-s Teukolsky equation can be truncated as (see Appendix \ref{Teukolsky Equation in Schwarzschild BH})
\be \label{Near Zone Teukolsky Isotropic Coordinate}
\begin{split}
        &R_{\ell}''(r) + \(- \frac{2+4s}{M-2r} - \frac{2s}{r} + \frac{2+4s}{M+2r}\) R_{\ell}'(r) \\
        &+ \(\frac{(s-\ell)(1+s+\ell)}{r^2} + \frac{128 M^3 \omega (-is+2M\omega)}{(4r^2-M^2)^2}\)R_{\ell}(r) =0 \,.
        \end{split}
\ee
Note that this equation enjoys an ${\rm SL}(2,\mathbb{R})$
near horizon symmetry~\cite{Charalambous:2021kcz}.

The  physical frequency dependent solution has the ingoing boundary condition at the black hole horizon~\cite{teukolsky1973perturbations,starobinskii1973amplification},
\begin{equation}
    R_{\ell}(r) = {\rm const} \times \(r - \frac{M}{2}\)^{-4iM \omega-2s} ~, \quad r\rightarrow \frac{M}{2} ~.
\end{equation}
Using the field redefinition $R_{\ell}(z) = z^{s-\ell}(1-z^2)^{-4iM\omega-2s}u_{\ell}(z)$ and introducing a new variable $x=z^2=M^2/(4r^2)$ 
we get
\begin{equation}
    \begin{aligned}
    & \quad u_{\ell}''(x) + \(\frac{1-2\ell}{2x}+\frac{1-2s-8iM\omega}{-1+x}\) u_{\ell}'(x) \\
    & + \(\frac{(s+\ell+4iM\omega)(-1+2s+8iM\omega)}{2(-1+x)} - \frac{(s+\ell + 4 iM\omega)(-1+2s +8 iM\omega)}{2x}\) u_{\ell}(x) = 0 ~.        
    \end{aligned}
\end{equation}
% The solution regular at the BH horizon is given by
% \begin{equation}\label{Finite Frequency Near Zone Solution}
%     \begin{aligned}
%     u_\ell(x) = {}_2 F_1\(\frac{1}{2}-s- 4 i M \omega, -s-\ell - 4 i M \omega, 1-2s-8i M \omega, 1-x\) ~,
%     \end{aligned}
% \end{equation}
This equation has two linearly independent solutions
\begin{equation}
    \begin{aligned}
    u_{\ell}^1(x) &= {}_2 F_1\(\frac{1}{2}-s-4iM\omega, -s-\ell - 4 iM\omega, \frac{1}{2}-\ell, x\)~,  \\
    u_{\ell}^2(x) &= x^{\frac{2\ell+1}{2}} {}_2 F_1\(\frac{1}{2}- s - 4i M\omega, 1- s+ \ell-4iM\omega, \frac{3}{2}+\ell,x\) ~.
    \end{aligned}
\end{equation}
For $\ell\in\mathbb{N},s\in \mathbb{N}$ and nonzero $\omega$, 
only the linear combination of these two solutions 
is regular at the horizon, 
\begin{equation}\label{Finite Frequency Near Zone Solution}
    \begin{aligned}
    u_\ell(x) &={}_2 F_1\(\frac{1}{2}-s- 4 i M \omega, -s-\ell - 4 i M \omega, 1-2s-8i M \omega, 1-x\) \\
  &=  \frac{\Gamma\(\frac{1}{2}+\ell\)\Gamma\(1-2s-8i M \omega\)}{\Gamma\(\frac{1}{2}-s-4i M \omega\) \Gamma\(1-s+\ell - 4i M \omega\)} {}_2 F_1\(\frac{1}{2}-s-4iM\omega, -s-\ell - 4 iM\omega, \frac{1}{2}-\ell, x\)  \\
    & \quad + \frac{\Gamma\(-\frac{1}{2}-\ell\)\Gamma\(1-2s-8i M \omega\)}{\Gamma\(\frac{1}{2}-s-4i M \omega\)\Gamma(-s-\ell - 4 i M \omega)} \\
    &\quad \times x^{\frac{2\ell+1}{2}} {}_2 F_1\(\frac{1}{2}-s - 4i M\omega, 1- s+ \ell-4iM\omega, \frac{3}{2}+\ell,x\) \,.
    \end{aligned}
\end{equation}
To get the dissipation number we need to read off
the coefficient in front of the $\sim r^{-s-\ell-1}$ term in 
the Taylor expansion of $R_\ell$. 
This is particularly easy with 
our variable choice $x=M^2/(4r^2)$. Indeed, 
 Taylor expansions of the 
prefactor $(1-z^2)^{-4iM\omega-2s}$
and the first hypergeometric function 
in the right-hand side of Eq.~\eqref{Finite Frequency Near Zone Solution}
produce only even powers of $r$, and hence they do not 
contribute to the $r^{2\ell+1}$ term.
This term stems only from the second 
term~\eqref{Finite Frequency Near Zone Solution} proportional to $x^{\frac{2\ell+1}{2}}$.
Therefore, 
Taylor expanding \eqref{Finite Frequency Near Zone Solution} at $r\to \infty$, we get
\be \label{Dissipative Response Coefficient}
\begin{split}
   R_{\ell}^{\text{near zone}}(r) 
   % &=  {}_2 F_1\(\frac{1}{2}-s-4iM\omega, -s-\ell - 4 iM\omega, \frac{1}{2}-\ell, x\)  \\
   %  +& \frac{  \Gamma\(1-s+\ell - 4i M \omega\)}{\Gamma\(\frac{1}{2}+\ell\)}
   %   \frac{\Gamma\(-\frac{1}{2}-\ell\)\Gamma\(1-2s-8i M \omega\)}{\Gamma\(\frac{1}{2}-s-4i M \omega\)} x^{\frac{2\ell+1}{2}} {}_2 F_1\(\frac{1}{2}-s - 4i M\omega, 1- s+ \ell-4iM\omega, \frac{3}{2}+\ell,x\) \\
   %  & 
   %  \\
    &\propto r^{\ell-s}\(1 + \cdots + r^{-2\ell-1} \(  i\frac{4 (-1)^{s} (\ell -s)! (\ell +s)!}{(2\ell+1)!!(2\ell-1)!!} M^{2(\ell+1)} \omega \) + \cdots \)  ~.
   \end{split}
\ee 
We stress that we have not used any
analytic continuation in $\ell$
to obtain the above formula.
With our choice of variables, it is obvious that
there is no source/response 
mixing in the full solution. 
Shortly we will give a simple EFT argument 
of why the source/response mixing at order $r_s\omega$ 
is absent 
in many popular gauges.

% But one can still use the analytic continuation trick, and the coefficient in front of $r^{\ell - s} \times r^{-2\ell - 1}$ will be
% \begin{equation}
%     - \frac{\Gamma\(\frac{1}{2} - \ell\) \Gamma\(1 - s + \ell - 4 i M\omega\)}{\Gamma\(\frac{3}{2} + \ell\)\Gamma(-s-\ell - 4 i M \omega)} \(\frac{M}{2}\)^{2\ell + 1} \simeq i\frac{4 (-1)^{s} (\ell -s)! (\ell +s)!}{(2\ell+1)!!(2\ell-1)!!} M^{2(\ell+1)} \omega + O\((M \omega)^2\) ~,
% \end{equation}
% which is the same as Eq.\eqref{Dissipative Response Coefficient}.

% Note that the coefficient in front of the 
% $r^{\ell-s}\times r^{-2\ell-1}$ term
% is the same in the Schwarzschild coordinates. 
% We will see shortly that this is the case 
% because it is proportional to the
% gauge-invariant dissipation number. 

% The coefficient in front of $r^{\ell-s}\times r^{-2\ell-1}$ is actually gauge invariant. To explicit show this, in Appendix \ref{Matching Dissipation Number in Schwarzschild Coordinate}, we compute this coefficient in the Schwarzschild coordinate and get the same result. 

\subsection{Dissipation Number Matching}

Now we can match the EFT one-point function correction
due to the dissipation number~\eqref{eq:disdiag}
and the UV result~\eqref{Dissipative Response Coefficient}.
Note that this matching is actually 
completely unambiguous in the isotropic gauge,
because the finite-frequency PN corrections 
are proportional to $\omega^2$. Thus, linear in $\omega$
terms unambiguously correspond to the finite-size 
dissipation 
contributions. 
In other words, the dissipation diagram cannot be canceled by any source corrections at the $2\ell+1$ order PN level
in the isotropic gauge. 

Now we can do the matching easily. 
In the spin-0 case, the Teukolsky variable $\psi^{[0]}=\Phi$, with the Feynman rules provided in Appendix \ref{Feynman Rule}, we can get the spin-0 dissipation number
\begin{equation}
    \lambda_{\ell(\omega)}^{s=0}= \frac{8 \pi (\ell!)^2}{(2\ell+1)!!((2\ell-1)!!)^2} M^{2\ell + 1} \,.
\end{equation}
In the spin-1 case, the Teukoleky variable $\psi^{[\pm 1]}$ is functions of the Maxwell-Newman-Penrose scalars $\Phi_0,\Phi_1$ and $\Phi_2$ \cite{newman1962approach,newman1963errata}
\begin{equation}
    \Phi_0 = F_{\mu\nu} l^\mu m^\nu, \quad \Phi_1 = \frac{1}{2} F_{\mu\nu}(l^\mu n^\nu + \bar m^\mu m^\nu), \quad \Phi_2 = F_{\mu\nu}\bar m^\mu n^\nu ~.
\end{equation}
In the isotropic coordinates, it is more convenient to work with the rescaled scalars 
\begin{equation}\label{Rescaled Maxwell-Newman-Penrose Scalars}
    \tilde\Phi_0 = \Phi_0~, \quad \tilde \Phi_1 = \frac{(M+2r)^4}{64M^2 r^2} \Phi_1 ~, \quad \tilde \Phi_2 = \frac{(M+2r)^4}{64M^2 r^2} \Phi_2 ~.
\end{equation}
The Teukolsky variables in this formalism take the form $\psi^{[1]}=\tilde\Phi_0$, $\psi^{[-1]}=\tilde\Phi_2$. To match the spin-1
dissipation number it is sufficient to use $\psi^{[1]}$, 
\begin{equation}\label{Teukolsky psi1}
    \psi^{[1]} = 2\sqrt{2}\frac{r}{(M- 2r)^2}\eth^0 {A}_0  ~,
\end{equation}
where $\eth^0$ is the spin raising operator defined in Appendix \ref{Spin-Weighted Spherical Harmonics}.
This simplification appears because the electric and magnetic fields are clearly separated in $\psi^{[1]}$, so 
we can set the magnetic source to zero.
In this case the real part of 
$\psi^{[1]}$ is sourced
by the electric field and 
hence entirely by $A_0$ (see 
Appendix~\ref{Comments on Maxwell-Newman-Penrose Phi_0} for more detail).
% \magenta{We demonstrate the sufficiency of such matching in Appendix \ref{Comments on Maxwell-Newman-Penrose Phi_0}}
%  \magenta{This is not correct for spin-1 Maxwell-Weyl scalar, and only true for spin-2 Weyl scalar. Let's just comment that at the quasi-static level, if we only apply the electric source, the magnetic response will not be leading at $r^{-\ell - 1}$.}
% \textcolor{blue}{The real part of $\Phi_1$
% is the electric field. Do you mean that here you use $\Phi_0$ for which it's not the case?
% You can check Eq 4.12 of  2102.08917.
% Maybe it's best to match $\Phi_1$ here because 
% the eclectic and magnetic fields decouple in it- ?
% Let's add more details on this matching here.}\magenta{Yes, for $\Phi_1$ it decouples, but for $\Psi_0$ it does not. But for $\Psi_0$, the Teukolsky equation factorizes, for $\Psi_1$ it does not.}\textbf{}
With the Feynman rules provided in Appendix \ref{Feynman Rule}, we get the spin-1 electric dissipation number 
\begin{equation}
    \lambda_{\ell(\omega)}^{s=1} = \frac{8\pi (\ell-1)!(\ell+1)!}{(2\ell+1)!!((2\ell-1)!!)^2} M^{2\ell + 1}  \,.
\end{equation}
In the spin-2 electric case, the Weyl scalar $\psi^{[2]} = \psi_0$ is 
sourced, at the leading order, by the perturbation of the Newtonian potential $\delta\phi$, see~\eqref{Teukolsky psi0}.
Since we focus 
on the parity 
even sector, the magnetic contributions
to the Weyl scalar
can be ignored. 
The responses 
from other parity-even
metric fluctuations 
only appear at higher orders in the distance expansion,
and hence can be ignored
when calculating the Weyl
scalar in the EFT.
Hence, 
matching the EFT and the GR expressions we get
\begin{equation}\label{eq:s2diss}
    \lambda_{\ell(\omega)}^{s=2} = \frac{8\pi (\ell-2)!(\ell+2)!}{(2\ell+1)!!((2\ell-1)!!)^2} M^{2\ell + 1} \,.
\end{equation}
All together, these results can be combined in a master formula
for a generic spin-$s$ field,
\begin{eBox}
    \begin{equation}\label{Dissipation Number Master Formula}
  \lambda_{1~(s)}^{\text{non-loc.}} = 
  \frac{8 \pi (\ell-s)! (\ell+s)!}{(2\ell+1)!!((2\ell-1)!!)^2\ell! 2^{2\ell+1}} r_s^{2\ell+1} \,,
    \end{equation}
\end{eBox}
where we used~\eqref{eq:redef} and expressed
the result in terms of the Schwarzschild radius $r_s=2M$.
This expression coincides with the 
dissipation numbers extracted from the absorption 
cross sections for $\ell=s$; see Eq.~\eqref{eq:absmatch}.

\subsection{On Cancellations of Dissipative Response in Advanced Coordinates}
In Section \ref{EFT Diagramatic Recurrence Relation}
 we have shown 
 that individual static PN diagrams 
 always produce logarithmic divergences at the ($2\ell+1$)PN order.
 These logs then cancel after all diagrams are summed over.  
 The situation is different at $\mathcal{O}(\omega r_s)$,
 where gravitational nonlinear corrections 
are absent in most gauges. 
Indeed, in Schwarzschild, isotropic, and harmonic coordinates,
the Schwarzschild metric is diagonal in time, 
which means that any finite frequency interaction 
is quadratic in frequency, 
and has the typical form $g^{00}\partial_t\Phi \partial_t\Phi \propto \omega^2$. 
Thus, at $O(r_s\omega)$, there is no source/response mixing,
and hence there is no ambiguity in the matching 
procedure. 

However, the situation is different in 
the advanced (Eddington–Finkelstein) coordinates, which have 
a non-vanishing off-diagonal term $g_{0i}$ \cite{Charalambous:2021mea}. 
Therefore, in these coordinates we always have the interactions $g^{0i}\partial_t\Phi\partial_i\Phi \propto i\omega$, which can cancel the dissipative response. 
In the diagrammatic language, the imaginary part of the full theory solution at $2\ell+1.5$ PN order in the 
advanced coordinates has two distinctive 
contributions now,
\begin{equation}\label{Diagramatic Intepretation of Vanishing Love}
     \sum_{\substack{\text{all} \\ \text{possible} \\ \text{diagrams}}}
        \vcenter{\hbox{\begin{tikzpicture}
        \begin{feynman}
          \vertex (i) at (0, 0);
          \vertex (e) at (0, 3);
          \vertex[dot, MyRed] (w1) at (0, 2.5) {};
          \vertex[dot, MyRed] (w2) at (0, 2.0) {};
          \vertex[dot, MyRed] (w3) at (0, 0.5) {};
          \node at (0.5,1.1) {$\cdot$};
          \node at (0.5,1.3) {$\cdot$};
          \node at (0.5,1.5) {$\cdot$};
          \vertex[blob, scale =2] (c1) at ( 2, 1.5) {};
          \node[circle, fill=white, scale = 1.0] at (2,1.5) {$i\omega$};
          \vertex[dot, MyBlue, label= $\boldsymbol{x}$] (f1) at (3.5,2.8) {};
          \vertex[crossed dot, MyBlue] (fs) at (3.5,0.2) {}; 
    
          \diagram*{
            (i) -- [double,double, thick, edge label = $2\ell + 1$] (e), 
            (w1) -- [dashed, MyRed, ultra thick] (c1) -- [dashed, MyRed, ultra thick] (w2),  
            (w3) -- [dashed, MyRed, ultra thick] (c1),
            (c1) -- [ultra thick, MyBlue] (f1), 
            (fs) -- [ultra thick, MyBlue, edge label' = $r^{\ell}$] (c1),  
          };
        \end{feynman}
    \end{tikzpicture}}}
    + 
             \vcenter{\hbox{\begin{tikzpicture}[scale=0.7]
        \begin{feynman}
            \vertex (i) at (0,0);
            \vertex (e) at (0,3);
            \node[circle, draw=Orange, fill = Orange, scale=0.5, label=left:$Q^a$] (w1) at (0, 1.0);
            \node[circle, draw=Orange, fill = Orange, scale=0.5, label=left:$Q^b$] (w2) at (0, 2.0);
            \vertex[dot, MyBlue, label= $x$] (f1) at (1.5,2.8) {};
            \vertex[crossed dot, MyBlue] (fs) at (1.5,0.2) {}; 

            \diagram*{
                (i) -- [double, double, thick] (w1),
                (w1) -- [dashed, double, double, thick] (w2),
                (w2) -- [double, double, thick] (e),
                (f1) -- [MyBlue, ultra thick] (w2),
                (fs) -- [MyBlue, ultra thick, edge label' = $r^\ell$] (w1)
            };
        \end{feynman}
    \end{tikzpicture}}}
    \sim i r^\ell \times 
    \frac{1}{r^{2\ell+1}} 
    \(A m^{2\ell+2}\omega + \lambda_{\ell(\omega)} (r_s \omega) \)\,,
\end{equation}
where $A$ is an order-one
numerical 
coefficient. 
As argued in \cite{Charalambous:2021mea}, these two different
diagrams must exactly cancel each other 
out
in order to reproduce the full theory calculation in the advanced coordinates.
In contrast to the Love number vanishing, this particular 
cancellation 
does not represent fine-tuning, as the 
actual physical dissipative response does not vanish,
and can be easily extracted in other gauges. 
The fact that the cancellation between the $(2\ell+1.5)$PN
graviton corrections and the dissipative response happens 
only in the advanced coordinates suggests that this cancellation
is merely a gauge artifact. Indeed, the physical 
dissipation is not zero and has a scaling 
consistent with the Wilsonian naturalness principle; cf.~\eqref{Dissipation Number Master Formula}.

% But we should make it clear that this cancellation is not a fine-tuning problem, it is still a gauge issue, just like the cancellation of all the logarithmic divergences of $2\ell+1$ PM order source corrections. If we are clever enough to choose a nice coordinate, such as the isotropic coordinate here, these source diagrams just do not exist. So, there is no need of the ad hoc analytic continuation to match the dissipation number.

An interesting implication of this argument 
is that the source/response mixing should generally be 
present for Kerr BHs. Indeed, the Kerr metric 
has nonvanishing off-diagonal components in any coordinate system. 
This is the reason why the Kerr Love response coefficients' calculation
is obscured by the source/response mixing~\cite{Charalambous:2021mea}. Although 
the analytic continuation prescription 
allows one to correctly 
extract
the response coefficients, 
we believe that any robust analysis 
should be based on using the EFT.
We leave a detailed EFT calculation 
of the Kerr response coefficients for future work.

% Even though there are a lot of arguments given in \cite{charalambous2021vanishing} about the validity of analytic continuation in $\ell$, we still believe that it will be important to do everything in physical region to confirm the results.

\section{Conclusions and Outlook}\label{sec:concl}

We have computed the EFT one-point functions 
of static scalar, photon, and graviton perturbations of
four-dimensional Schwarzschild black holes. 
We developed a diagrammatic expansion that computes 
post-Newtonian corrections to the external field sources.  
Using the isotropic Kaluza-Klein gauge, we have explicitly
resummed the EFT PN diagrams 
to an arbitrary PN order
in the case of spin-0 and spin-2 dilaton fluctuations. 
These results are valid for any multipolar index $\ell$.
For the Maxwell field we have obtained explicit results 
for the $\ell=1$ case, i.e., at the 3PN order. 

Comparing our results with the full
BH perturbation theory calculations we have found 
that the static PN one-point functions
explicitly reproduce the full general relativity results
without having to include any finite-size effects. 
This implies that Love numbers vanish identically. 

In the second part of our paper we have matched the 
Schwarzschild BH
dissipation numbers. Using the in-in 
approach we have computed the dissipative corrections
to the one-point functions and extracted 
the dissipation numbers by comparing our EFT field profiles
to the general relativity results. 
Our expressions
for the dissipation numbers exactly coincide with the 
results of matching in a gauge-invariant manner
obtained by comparing 
cross sections for the absorption of massless particles 
by black holes
versus the absorption cross sections in the point-particle 
theory. 

We have also obtained some important results clarifying 
the EFT description of black holes. 
At the conceptual level,
we have shown how the EFT resolves the so-called 
source/response ambiguity.
Using the EFT we can extract the finite-size effects without having 
to use the so-called analytic continuation prescription. 
The second important result is that 
the individual EFT diagrams generically produce logarithmic corrections
to Love numbers. These corrections, however, cancel when all diagrams 
of the ($2\ell+1$)PN order are summed over.
We interpret this apparent fine-tuning as a manifestation of the 
Love symmetry of BH perturbations~\cite{Charalambous:2021kcz}.
The third important result is a consistent definition 
of dissipation numbers in the EFT, and their explicit relation 
to the absorption cross sections. 

We have also obtained some new results at the technical level,
which will be useful in future studies of BH within the worldline EFT 
formalism. First, we have defined a notion of a consistent gauge that allows for unambiguous matching. Second, we have set up an 
EFT diagrammatic expansion for external probes 
and studied its topological properties.
Our analysis facilitates the resummation 
of Feynman diagrams at high PN orders. 
Key to this resummation
is the diagrammatic recurrence relation between 
PN diagrams of different orders.
This recurrence relation is, essentially,
the diagrammatic 
version of the recurrence 
relation that appears in the Frobenius solution to the Teukolsky equation. Thus, 
one can use it to systematically resum the EFT diagrams 
and in this way to 
recover the entire 
static 
solution to the Teukolsky equation
that includes all 
necessary relativistic 
corrections.

Our study can be extended in multiple ways. 
First, it will be important to include BH spin 
and generalize our EFT PN expansion to the case of 
Kerr black holes along the lines of~\cite{Porto:2005ac,Porto:2007qi,Goldberger:2020fot}. 
Although the Kerr BH Love numbers 
were shown to vanish~\cite{Chia:2020yla,Charalambous:2021mea},
these results relied on the analytic continuation prescription,
whose validity is not completely warranted.
Another project in this research direction would be 
an explicit matching of dissipation numbers for Kerr black holes,
which do not vanish even for static external perturbations. 
The second potential line of research is to better understand 
the nature of fine-tuning in the EFT PN expansion. 
This fine-tuning includes the cancellation of logarithmic 
corrections to static Love numbers. In four dimensions this 
can be explained as a result of the Love symmetry, which 
enforces the polynomial structure of one-point functions. 
It would be interesting to study higher-dimensional
BHs in the EFT framework and see how the Love symmetry manifests 
itself there. 
Finally, it will be important to extract the Love numbers 
in a fully gauge-invariant manner by comparing the elastic 
scattering 
cross sections of external fields off the BH geometry against the EFT
on-shell scattering amplitudes (see~\cite{Ivanov:2022qqt} for
recent progress). 
We leave these research directions for future work.

\paragraph*{Acknowledgements}

We are grateful to 
Panos Charalambous, 
Horng Sheng Chia, 
Sergei Dubovsky, 
Gregor K\"{a}lin, 
Barak Kol,
Rafael A. Porto,
Michael Smolkin,
and
Matias Zaldarriaga
for their comments on the draft 
and enlightening discussions.
We also thank Mengyang Zhang for useful discussions.
Z.Z. thanks the long term hospitality of the Institute for Advanced Study.

\appendix

\section{Feynman Rules}\label{Feynman Rule}

In this section we present Feynman rules for the EFT diagramatic computation. 
We work in the isotropic gauge of the background gravitational field.
Since the spin-2 electric perturbations share the same structure with the spin-0 perturbations, we only provide the related Feynman rules for the spin-0 case.

\textbf{Gravitational Sector}:
\begin{itemize}
    \item Propagators:
    \begin{equation}
        \begin{aligned}
            \begin{tikzpicture}[baseline={([yshift=-2.0 ex]current bounding box.center)}]
                \begin{feynman}
                    \vertex (a) at (0,0);
                    \vertex (b) at (3,0);
                    
                    \diagram*{
                        (a) -- [dashed, ultra thick, MyRed, edge label = $\phi$] (b), 
                    };
                \end{feynman}
            \end{tikzpicture}
            &= 4 \pi G \frac{-i}{\boldsymbol{k}^2} ~, \quad 
            \begin{tikzpicture}[baseline={([yshift=-1.5 ex]current bounding box.center)}]
                \begin{feynman}
                    \vertex (a) at (0,0);
                    \vertex (b) at (3,0);
                    
                    \diagram*{
                        (a) -- [boson, MyYellow, ultra thick, edge label = $\sigma$] (b), 
                    };
                \end{feynman}
            \end{tikzpicture}
            = -16 \pi G \frac{-i}{\boldsymbol{k}^2}
        \end{aligned}
    \end{equation}

    \item Static vertices:
    
    Worldline vertex:
    \begin{equation}
        \begin{tikzpicture}[baseline={([yshift=-0.4 ex]current bounding box.center)}]
            \begin{feynman}
                \vertex (i) at (0,0);
                \vertex (e) at (0,1);
                \vertex[dot, MyRed] (f1) at (0,0.5) {};
                \vertex (c1) at (1,0.5);

                \diagram*{
                    (i) -- [double, double, thick] (e),
                    (f1) -- [dashed, MyRed, ultra thick, edge label = $\phi$] (c1),
                };
            \end{feynman}
        \end{tikzpicture}
        = - im ~, \quad \quad 
    \end{equation} 

    Bulk vertices:
    \begin{equation}
        \begin{aligned}
            \begin{tikzpicture}[baseline={([yshift=-0.5 ex]current bounding box.center)}]
                \begin{feynman}
                    \vertex (c1) at (0,0);
                    \vertex (c2) at (1.5, 0);
                    \vertex (c3) at (-1.5, 1);
                    \vertex (c4) at (-1.5, -1) ;

                    \diagram*{
                        (c1) -- [boson, MyYellow, ultra thick] (c2), 
                        (c1) -- [MyRed, dashed, ultra thick, reversed momentum'={[arrow style = MyRed, arrow shorten = 0.25] $\boldsymbol{k}$}] (c3),
                        (c1) -- [MyRed, dashed, ultra thick, reversed momentum={[arrow style = MyRed, arrow shorten = 0.25] $\boldsymbol{p}$}] (c4), 
                    };
                \end{feynman}
            \end{tikzpicture}
            &= i \frac{1}{8\pi} \boldsymbol{k} \cdot \boldsymbol{p} ~,
            \begin{tikzpicture}[baseline={([yshift=-0.5 ex]current bounding box.center)}]
                \begin{feynman}
                    \vertex (c1) at (0,0);
                    \vertex (c2) at (1.5, 0);
                    \vertex (c3) at (-1.5, 1);
                    \vertex (c4) at (-1.5, -1) ;

                    \diagram*{
                        (c1) -- [boson, MyYellow, ultra thick, reversed momentum={[arrow style = MyYellow, arrow shorten = 0.25] $\boldsymbol{k}_3$}] (c2), 
                        (c1) -- [boson, MyYellow, ultra thick, reversed momentum'={[arrow style = MyYellow, arrow shorten = 0.25] $\boldsymbol{k}_1$}] (c3),
                        (c1) -- [boson, MyYellow, ultra thick, reversed momentum={[arrow style = MyYellow, arrow shorten = 0.25] $\boldsymbol{k}_2$}] (c4), 
                    };
                \end{feynman}
            \end{tikzpicture}
            = i \frac{3}{32 \pi} (\boldsymbol{k_1} \cdot \boldsymbol{k_2} + \boldsymbol{k_2} \cdot \boldsymbol{k_3} + \boldsymbol{k_3} \cdot \boldsymbol{k_1}) \\
            &\cdots \cdots
        \end{aligned}
    \end{equation}
\end{itemize}

~\\
\textbf{Spin-0 Perturbations}:

\begin{itemize}
    \item Propagator:
    \begin{equation}
        \begin{tikzpicture}[baseline={([yshift=-2.0 ex]current bounding box.center)}]
                \begin{feynman}
                    \vertex (a) at (0,0);
                    \vertex (b) at (3,0);
                    
                    \diagram*{
                        (a) -- [ultra thick, MyBlue, edge label = $\delta\Phi$] (b), 
                    };
                \end{feynman}
            \end{tikzpicture}
             = \frac{-i}{\boldsymbol{k}^2} ~.   
    \end{equation}
  
    \item  Source vertices:
    \begin{equation}
        \begin{aligned}
            \begin{tikzpicture}[baseline={([yshift=-0.8 ex]current bounding box.center)}]
                \begin{feynman}
                    \vertex[crossed dot, MyBlue] (cs) at (0,0) {};
                    \vertex (c1) at (1.5,0) ;
                    
                    \diagram*{
                        (c1) -- [MyBlue, ultra thick, momentum'={[arrow style = MyBlue, arrow shorten=0.25]$\boldsymbol{k}$}, edge label = $\bar\Phi$] (cs),
                    };
                \end{feynman}
            \end{tikzpicture}
            &= \bar\Phi_{\boldsymbol{k}}, \quad \quad \\
            \begin{tikzpicture}[baseline={([yshift=-0.5 ex]current bounding box.center)}]
                \begin{feynman}
                    \vertex (c1) at (0,0);
                    \vertex (c2) at (1.5, 0);
                    \vertex (c3) at (3, 1);
                    \vertex[crossed dot, MyBlue] (c4) at (3, -1) {};

                    \diagram*{
                        (c1) -- [boson, MyYellow, ultra thick, momentum={[arrow style = MyYellow, arrow shorten = 0.25] $\boldsymbol{p}$}] (c2), 
                        (c2) -- [MyBlue, ultra thick, reversed momentum={[arrow style = MyBlue, arrow shorten = 0.25] $\boldsymbol{k}$}] (c3),
                        (c2) -- [MyBlue, ultra thick, edge label = $r^{\ell}$] (c4), 
                    };
                \end{feynman}
            \end{tikzpicture}
            & = - \frac{\ell}{2} k^i \mathcal{E}_{i i_2 \ldots i_\ell} (2\pi)^3 \(i^{\ell-1} \frac{d^{\ell-1}}{dk_{i_2} \cdots dk_{i_\ell}} \delta^{(3)}(\boldsymbol{k} + \boldsymbol{p})\) \\
            \cdots \cdots
        \end{aligned}
    \end{equation}
    
    \item Bulk vertices:
    \begin{equation}
        \begin{aligned}
                \begin{tikzpicture}[baseline={([yshift=-0.5 ex]current bounding box.center)}]
                \begin{feynman}
                    \vertex (c1) at (0,0);
                    \vertex (c2) at (1.5, 0);
                    \vertex (c3) at (3, 1);
                    \vertex (c4) at (3, -1) ;

                    \diagram*{
                        (c1) -- [boson, MyYellow, ultra thick] (c2), 
                        (c2) -- [MyBlue, ultra thick, reversed momentum={[arrow style = MyBlue, arrow shorten = 0.25] $\boldsymbol{k}$}] (c3),
                        (c2) -- [MyBlue, ultra thick, reversed momentum'={[arrow style = MyBlue, arrow shorten = 0.25] $\boldsymbol{p}$}] (c4), 
                    };
                \end{feynman}
            \end{tikzpicture}
             = i \frac{1}{2} \boldsymbol{k} \cdot \boldsymbol{p} ~\cdots \cdots
        \end{aligned}
    \end{equation}
    
    \item Love number vertex:
    \begin{equation}
            \vcenter{\hbox{\begin{tikzpicture}[scale=0.7]
        \begin{feynman}
            \vertex (i) at (0,0);
            \vertex (e) at (0,3);
            \node[circle, draw=repGreen, fill = repGreen, scale=1, label=left:$\lambda_\ell^{s=0}$] (w1) at (0, 1.5);
            \vertex[dot, MyBlue, label= $\boldsymbol{x}$] (f1) at (1.5,2.8) {};
            \vertex[crossed dot, MyBlue] (fs) at (1.5,0.2) {}; 

            \diagram*{
                (i) -- [double, double, thick] (w1),
                (w1) -- [double, double, thick] (e),
                (f1) -- [MyBlue, ultra thick] (w1),
                (fs) -- [MyBlue, ultra thick, edge label' = $r^\ell$] (w1)
            };
        \end{feynman}
    \end{tikzpicture}}}
    =   \frac{(2\ell - 1) !!}{4\pi} \lambda_{\ell}^{s=0} \mathcal{E}_{i_1\ldots i_\ell} x^{i_1} \cdots x^{i_\ell} \frac{1}{r^{2\ell + 1}} ~.
    \end{equation}
    
    \item Dissipation nonlocal vertex:
    \begin{equation}
            \vcenter{\hbox{\begin{tikzpicture}[scale=0.7]
        \begin{feynman}
            \vertex (i) at (0,0);
            \vertex (e) at (0,3);
            \node[circle, draw=Orange, fill = Orange, scale=0.5, label=left:$Q^a$] (w1) at (0, 1.0);
            \node[circle, draw=Orange, fill = Orange, scale=0.5, label=left:$Q^b$] (w2) at (0, 2.0);
            \vertex[dot, MyBlue, label= $x$] (f1) at (1.5,2.8) {};
            \vertex[crossed dot, MyBlue] (fs) at (1.5,0.2) {}; 

            \diagram*{
                (i) -- [double, double, thick] (w1),
                (w1) -- [dashed, double, double, thick] (w2),
                (w2) -- [double, double, thick] (e),
                (f1) -- [MyBlue, ultra thick] (w2),
                (fs) -- [MyBlue, ultra thick, edge label' = $r^\ell$] (w1)
            };
        \end{feynman}
    \end{tikzpicture}}}
    = i (r_s \omega) \frac{(2\ell-1)!!}{4\pi} \lambda_{\ell (\omega)}^{s=0} \mathcal{E}_{i_1\cdots i_\ell} x^{i_1}\cdots x^{i_\ell} \frac{1}{r^{2\ell+1}} e^{-i\omega t} ~.
    \end{equation}
\end{itemize}

~\\
\textbf{Spin-1 Electric Perturbations}:

\begin{itemize}

    \item Propagator:
    
    \begin{equation}
        \begin{tikzpicture}[baseline={([yshift=-2.0 ex]current bounding box.center)}]
                \begin{feynman}
                    \vertex (a) at (0,0);
                    \vertex (b) at (3,0);
                    
                    \diagram*{
                        (a) -- [ultra thick, MyBlue, edge label = $\mathcal{A}_0$] (b), 
                    };
                \end{feynman}
            \end{tikzpicture}        
        = (-1) \frac{-i}{\boldsymbol{k}^2} ~. 
    \end{equation}
    
    \item Source vertices:
    \begin{equation}
        \begin{aligned}
            \begin{tikzpicture}[baseline={([yshift=-0.8 ex]current bounding box.center)}]
                \begin{feynman}
                    \vertex[crossed dot, MyBlue] (cs) at (0,0) {};
                    \vertex (c1) at (1.5,0) ;
                    
                    \diagram*{
                        (c1) -- [MyBlue, ultra thick, momentum'={[arrow style = MyBlue, arrow shorten=0.25]$\boldsymbol{k}$}, edge label = $\bar{\mathcal{A}}_0$] (cs),
                    };
                \end{feynman}
            \end{tikzpicture}
            &= {\bar{\mathcal{A}}_0}{}_{\boldsymbol{k}}, \quad \quad \\
            \begin{tikzpicture}[baseline={([yshift=-0.5 ex]current bounding box.center)}]
                \begin{feynman}
                    \vertex (c1) at (0,0);
                    \vertex (c2) at (1.5, 0);
                    \vertex (c3) at (3, 1);
                    \vertex[crossed dot, MyBlue] (c4) at (3, -1) {};

                    \diagram*{
                        (c1) -- [boson, MyYellow, ultra thick, momentum={[arrow style = MyYellow, arrow shorten = 0.25] $\boldsymbol{p}$}] (c2), 
                        (c2) -- [MyBlue, ultra thick, reversed momentum={[arrow style = MyBlue, arrow shorten = 0.25] $\boldsymbol{k}$}] (c3),
                        (c2) -- [MyBlue, ultra thick, edge label = $r^{\ell}$] (c4), 
                    };
                \end{feynman}
            \end{tikzpicture}
            & = \frac{\ell}{2} k^i \mathcal{E}_{i i_2 \ldots i_\ell} (2\pi)^3 \(i^{\ell-1} \frac{d^{\ell-1}}{dk_{i_2} \cdots dk_{i_\ell}} \delta^{(3)}(\boldsymbol{k} + \boldsymbol{p})\) \\
            \begin{tikzpicture}[baseline={([yshift=-0.5 ex]current bounding box.center)}]
                \begin{feynman}
                    \vertex (c1) at (0,0);
                    \vertex (c2) at (1.5, 0);
                    \vertex (c3) at (3, 1);
                    \vertex[crossed dot, MyBlue] (c4) at (3, -1) {};

                    \diagram*{
                        (c1) -- [dashed, MyRed, ultra thick, momentum={[arrow style = MyRed, arrow shorten = 0.25] $\boldsymbol{p}$}] (c2), 
                        (c2) -- [MyBlue, ultra thick, reversed momentum={[arrow style = MyBlue, arrow shorten = 0.25] $\boldsymbol{k}$}] (c3),
                        (c2) -- [MyBlue, ultra thick, edge label = $r^{\ell}$] (c4), 
                    };
                \end{feynman}
            \end{tikzpicture}        
            & = - 2 \ell k^i \mathcal{E}_{ii_2 \cdots i_\ell} (2\pi)^3 \(i^{\ell-1} \frac{d^{\ell-1}}{dk_{i_2} \cdots dk_{i_\ell}} \delta^{(3)}(\boldsymbol{k}+\boldsymbol{p}) \) \\
            \cdots \cdots
        \end{aligned}
    \end{equation}
    
    \item Bulk vertices:
    \begin{equation}
        \begin{aligned}
                \begin{tikzpicture}[baseline={([yshift=-0.5 ex]current bounding box.center)}]
                \begin{feynman}
                    \vertex (c1) at (0,0);
                    \vertex (c2) at (1.5, 0);
                    \vertex (c3) at (3, 1);
                    \vertex (c4) at (3, -1) ;

                    \diagram*{
                        (c1) -- [boson, MyYellow, ultra thick] (c2), 
                        (c2) -- [MyBlue, ultra thick, reversed momentum={[arrow style = MyBlue, arrow shorten = 0.25] $\boldsymbol{k}$}] (c3),
                        (c2) -- [MyBlue, ultra thick, reversed momentum'={[arrow style = MyBlue, arrow shorten = 0.25] $\boldsymbol{p}$}] (c4), 
                    };
                \end{feynman}
            \end{tikzpicture}
             = - i \frac{1}{2} \boldsymbol{k} \cdot \boldsymbol{p} ~, \quad 
                \begin{tikzpicture}[baseline={([yshift=-0.5 ex]current bounding box.center)}]
                \begin{feynman}
                    \vertex (c1) at (0,0);
                    \vertex (c2) at (1.5, 0);
                    \vertex (c3) at (3, 1);
                    \vertex (c4) at (3, -1) ;

                    \diagram*{
                        (c1) -- [dashed, MyRed, ultra thick] (c2), 
                        (c2) -- [MyBlue, ultra thick, reversed momentum={[arrow style = MyBlue, arrow shorten = 0.25] $\boldsymbol{k}$}] (c3),
                        (c2) -- [MyBlue, ultra thick, reversed momentum'={[arrow style = MyBlue, arrow shorten = 0.25] $\boldsymbol{p}$}] (c4), 
                    };
                \end{feynman}
            \end{tikzpicture}
            = i 2 \boldsymbol{k} \cdot \boldsymbol{p} ~ \cdots \cdots
        \end{aligned}
    \end{equation}
    
    \item Love number vertex:
    \begin{equation}
            \vcenter{\hbox{\begin{tikzpicture}[scale=0.7]
        \begin{feynman}
            \vertex (i) at (0,0);
            \vertex (e) at (0,3);
            \node[circle, draw=repGreen, fill = repGreen, scale=1, label=left:$\lambda_\ell^{s=1}$] (w1) at (0, 1.5);
            \vertex[dot, MyBlue, label= $\boldsymbol{x}$] (f1) at (1.5,2.8) {};
            \vertex[crossed dot, MyBlue] (fs) at (1.5,0.2) {}; 

            \diagram*{
                (i) -- [double, double, thick] (w1),
                (w1) -- [double, double, thick] (e),
                (f1) -- [MyBlue, ultra thick] (w1),
                (fs) -- [MyBlue, ultra thick, edge label' = $r^\ell$] (w1)
            };
        \end{feynman}
    \end{tikzpicture}}}
    =   - \frac{(2\ell - 1) !!}{4\pi} \lambda_{\ell}^{s=1} \mathcal{E}_{i_1\ldots i_\ell} x^{i_1} \cdots x^{i_\ell} \frac{1}{r^{2\ell + 1}} ~.
    \end{equation}
    
    \item Dissipation non-local vertex:
    \begin{equation}
            \vcenter{\hbox{\begin{tikzpicture}[scale=0.7]
        \begin{feynman}
            \vertex (i) at (0,0);
            \vertex (e) at (0,3);
            \node[circle, draw=Orange, fill = Orange, scale=0.5, label=left:$Q^a$] (w1) at (0, 1.0);
            \node[circle, draw=Orange, fill = Orange, scale=0.5, label=left:$Q^b$] (w2) at (0, 2.0);
            \vertex[dot, MyBlue, label= $x$] (f1) at (1.5,2.8) {};
            \vertex[crossed dot, MyBlue] (fs) at (1.5,0.2) {}; 

            \diagram*{
                (i) -- [double, double, thick] (w1),
                (w1) -- [dashed, double, double, thick] (w2),
                (w2) -- [double, double, thick] (e),
                (f1) -- [MyBlue, ultra thick] (w2),
                (fs) -- [MyBlue, ultra thick, edge label' = $r^\ell$] (w1)
            };
        \end{feynman}
    \end{tikzpicture}}}
    = - i (r_s \omega) \frac{(2\ell-1)!!}{4\pi} \lambda_{\ell (\omega)}^{s=1} \mathcal{E}_{i_1\cdots i_\ell} x^{i_1}\cdots x^{i_\ell} \frac{1}{r^{2\ell+1}} e^{-i\omega t} ~.
    \end{equation}
\end{itemize}

\section{Useful Mathematical Relations}\label{Useful Mathematical Relations}

\subsection{(Spin-Weighted) Spherical Harmonics}\label{Spin-Weighted Spherical Harmonics}
\textbf{(Scalar) Spherical Harmonics}: We use the following definition of the (scalar) spherical harmonics: 
\begin{equation}
    Y_{\ell m}(\theta, \phi)=\frac{(-1)^{\ell+\frac{|m|+m}{2}}}{2^{\ell} \ell !}\left[\frac{2 \ell+1}{4 \pi} \frac{(\ell-|m|) !}{(\ell+|m|) !}\right]^{1 / 2} e^{i m \phi}(\sin \theta)^{|m|}\left(\frac{d}{d \cos \theta}\right)^{\ell+|m|}(\sin \theta)^{2 \ell} ~,
\end{equation}
with the parameter range $\ell \geq 0$, $-\ell \leq m \leq \ell$ and $\ell,m$ are all integers. These harmonics obey the following relations: 
\begin{equation}
    \Delta_{\mathbb{S}^{2}} Y_{\ell m}=-\ell(\ell+1) Y_{\ell m}, \quad Y_{\ell m}^{*}(\theta,\phi)=(-1)^{m} Y_{\ell(-m)}(\theta,\phi), \quad \int_{\mathbb{S}^{2}} d \Omega Y_{\ell m} Y_{\ell^{\prime} m^{\prime}}^{*}=\delta_{\ell \ell^{\prime}} \delta_{m m^{\prime}} ~,
\end{equation}
where $\Delta_{\mathbb{S}^{2}}$ is the two-sphere Laplacian.

With the spherical harmonics, we can represent the scalar field contracted between STF tensor $\mathcal{E}_L$ and $\ell$ copies of normal vector $n^{L}$ with the spherical harmonic basis
\begin{equation}
    n^{L}(\theta, \varphi) \mathcal{E}_{L}=\sum_{m=-\ell}^{\ell} \mathcal{E}_{\ell m} Y_{\ell m}(\theta, \varphi), \quad \text { where } \quad \mathcal{E}_{\ell m}=\mathcal{E}_{L} \int_{\mathbb{S}^{2}} n^{L} {Y}_{\ell m}^* \mathrm{~d} \Omega ~.
\end{equation}
This is equivalent to say 
\begin{equation}
    \mathcal{E}_{i_1 \cdots i_\ell} x^{i_1} \cdots x^{i\ell} = \sum_{m=-\ell}^{\ell}\mathcal{E}_{\ell m} r^\ell Y_{\ell m}(\theta,\phi) \,.
\end{equation}

\textbf{Spin-Weighted Spherical Harmonics}: To study the spin-1 and spin-2 perturbations, it is useful to review basic properties of spin-weighted spherical harmonics. We introduce the spin raising and lowering operator 
\begin{equation}
    \eth^s \equiv-\left(\partial_{\theta}+\frac{i}{\sin \theta} \partial_{\phi}-s \frac{\cos \theta}{\sin \theta}\right), \quad \bar{\eth}^s \equiv-\left(\partial_{\theta}-\frac{i}{\sin \theta} \partial_{\phi}+s \frac{\cos \theta}{\sin \theta}\right) ~,
\end{equation}
with raising and lowering operation 
\begin{equation}
    \begin{aligned}
        \eth^s \({}_s Y_{\ell m}\) = + \sqrt{(\ell-s)(\ell + s +1)} {}_{s+1}Y_{\ell m} ~, \\
        \bar{\eth}^s \({}_s Y_{\ell m}\) = -\sqrt{(\ell + s)(\ell - s +1)} {}_{s-1} Y_{\ell m} ~,
    \end{aligned}
\end{equation}
where $\ell \geq |s|$. Here, ${}_s Y_{\ell m }$ is defined as the spin-s spherical harmonics. These harmonics obey 
\begin{equation}
    { }_{s} Y_{\ell m}^{*}(\mathbf{n})=(-1)^{m+s}{ }_{-s} Y_{\ell(-m)}(\mathbf{n}), \quad \int_{\mathbb{S}^{2}} d \Omega{ }_{s} Y_{\ell m s} Y_{\ell^{\prime} m^{\prime}}^{*}=\delta_{\ell \ell^{\prime}} \delta_{m m^{\prime}} ~.
\end{equation}

\subsection{Master Integrals}\label{Master Integrals}
Dimensional regularization is widely used in the EFT diagrammatic computation. In this appendix, we summarize the useful d-dimensional momentum integration and Fourier transformation formula \cite{Kol:2009mj}.
The momentum integration formula
\begin{align}
    J &=\int \frac{d^{d} \boldsymbol{q}}{(2 \pi)^{d}} \frac{1}{\left(\boldsymbol{q}^{2}\right)^{\alpha}\left[(\boldsymbol{q}+\boldsymbol{k})^{2}\right]^{\beta}}=\frac{\left(\boldsymbol{k}^{2}\right)^{d / 2-\alpha-\beta}}{(4 \pi)^{d / 2}} \frac{\Gamma(\alpha+\beta-d / 2)}{\Gamma(\alpha) \Gamma(\beta)} \frac{\Gamma(d / 2-\alpha) \Gamma(d / 2-\beta)}{\Gamma(d-\alpha-\beta)} \\
    J_{i} &=\int \frac{d^{d} \boldsymbol{q}}{(2 \pi)^{d}} \frac{q_{i}}{\left(\boldsymbol{q}^{2}\right)^{\alpha}\left[(\boldsymbol{q} + \boldsymbol{k})^{2}\right]^{\beta}}= - \frac{d / 2-\alpha}{d-\alpha-\beta} J k_{i} \\
    J_{i j} & =\int \frac{d^{d} \boldsymbol{q}}{(2 \pi)^{d}}  \frac{q_{i} q_{j}}{\left(\boldsymbol{q}^{2}\right)^{\alpha}\left[(\boldsymbol{q} + \boldsymbol{k})^{2}\right]^{\beta}} \nonumber \\
    &=\frac{1}{(4 \pi)^{d / 2}} \frac{\Gamma(\alpha+\beta-d / 2-1)}{\Gamma(\alpha) \Gamma(\beta)} \frac{\Gamma(d / 2-\alpha+1) \Gamma(d / 2-\beta)}{\Gamma(d-\alpha-\beta+2)} \nonumber \\
    & \quad \times\left[(d / 2-\alpha+1)(\alpha+\beta-d / 2-1) k_{i} k_{j}+(d / 2-\beta) \frac{\boldsymbol{k}^{2}}{2} \delta_{i j}\right]\left(\boldsymbol{k}^{2}\right)^{d / 2-\alpha-\beta}  \\
    J_{i j k} & =\int \frac{d^{d} \boldsymbol{q}}{(2 \pi)^{d}} \frac{q_{i} q_{j} q_{k}}{\left(\boldsymbol{q}^{2}\right)^{\alpha}\left[(\boldsymbol{q} + \boldsymbol{k})^{2}\right]^{\beta}} \nonumber \\
    &=\frac{\left(\boldsymbol{k}^{2}\right)^{d / 2-\alpha-\beta}}{(4 \pi)^{d / 2}} \frac{\Gamma(\alpha+\beta-d / 2-1)}{\Gamma(\alpha) \Gamma(\beta)} \frac{\Gamma(d / 2-\alpha+2) \Gamma(d / 2-\beta)}{\Gamma(d-\alpha-\beta+3)} \nonumber \\
    & \quad \times\left[- (d / 2-\alpha+2)(\alpha+\beta-d / 2-1) k_{i} k_{j} k_{k}+(d / 2-\beta) \frac{\boldsymbol{k}^{2}}{2}\left(\delta_{i j} k_{k}+\delta_{j k} k_{i}+\delta_{i k} k_{j}\right)\right]  \\
    J_{i j k l} &=\int \frac{d^{d} \boldsymbol{q}}{(2 \pi)^{d}} \frac{q_{i} q_{j} q_{k} q_{l}}{\left(\boldsymbol{q}^{2}\right)^{\alpha}\left[(\boldsymbol{q}+\boldsymbol{k})^{2}\right]^{\beta}}=\frac{\left(\boldsymbol{k}^{2}\right)^{d / 2-\alpha-\beta}}{(4 \pi)^{d / 2}} \frac{\Gamma(d / 2-\alpha+2) \Gamma(d / 2-\beta)}{\Gamma(d-\alpha-\beta+4)} \nonumber \\
    & \quad \times \frac{\Gamma(\alpha+\beta-d / 2-2)}{\Gamma(\alpha) \Gamma(\beta)}\Bigg[(d / 2-\beta)(d / 2-\beta+1)\left(\delta_{i j} \delta_{k l}+\delta_{i k} \delta_{j l}+\delta_{i l} \delta_{j k}\right) \frac{\left(\boldsymbol{k}^{2}\right)^{2}}{4} \nonumber \\
    & \quad +\left(\delta_{i j} k_{k} k_{l}+\delta_{i k} k_{j} k_{l}+\delta_{i l} k_{j} k_{k}+\delta_{j k} k_{i} k_{l}+\delta_{j l} k_{i} k_{k}+\delta_{k l} k_{i} k_{j}\right) \frac{\boldsymbol{k}^{2}}{2} \nonumber \\
    & \quad \times(\alpha+\beta-d / 2-2)(d / 2-\alpha+2)(d / 2-\beta)  \nonumber \\
    & \quad +k_{i} k_{j} k_{k} k_{l}(\alpha+\beta-d / 2-2)(\alpha+\beta-d / 2-1)(d / 2-\alpha+2)(d / 2-\alpha+3) \Bigg]
\end{align} 

Fourier transformation formula:
\begin{align}
    \int \frac{d^{d} \boldsymbol{k}}{(2 \pi)^{d}} \frac{e^{i \boldsymbol{k} \cdot \boldsymbol{r}}}{\left(\boldsymbol{k}^{2}\right)^{\alpha}} & =\frac{1}{(4 \pi)^{d / 2}} \frac{\Gamma(d / 2-\alpha)}{\Gamma(\alpha)}\left(\frac{\boldsymbol{r}^{2}}{4}\right)^{\alpha-d / 2} \\
    \int \frac{d^{d} \boldsymbol{k}}{(2 \pi)^{d}} \frac{\boldsymbol{k}_{i}}{\left(\boldsymbol{k}^{2}\right)^{\alpha}} e^{i \boldsymbol{k} \cdot \boldsymbol{r}} & =i x_{i} \frac{\Gamma(d / 2-\alpha+1)}{2(4 \pi)^{d / 2} \Gamma(\alpha)}\left(\frac{\boldsymbol{r}^{2}}{4}\right)^{\alpha-d / 2-1} \\
    \int \frac{d^{d} \boldsymbol{k}}{(2 \pi)^{d}} \frac{\boldsymbol{k}_{i} \boldsymbol{k}_{j}}{\left(\boldsymbol{k}^{2}\right)^{\alpha}} e^{i \boldsymbol{k} \cdot \boldsymbol{r}} & =\frac{\Gamma(d / 2-\alpha+1)}{(4 \pi)^{d / 2} \Gamma(\alpha)}\left(\frac{\delta_{i j}}{2}+(\alpha-d / 2-1) \frac{x_{i} x_{j}}{\boldsymbol{r}^{2}}\right)\left(\frac{\boldsymbol{r}^{2}}{4}\right)^{\alpha-d / 2-1} \\
    \int \frac{d^{d} \boldsymbol{k}}{(2 \pi)^{d}} \frac{\boldsymbol{k}_{i} \boldsymbol{k}_{j} \boldsymbol{k}_{l}}{\left(\boldsymbol{k}^{2}\right)^{\alpha}} e^{i \boldsymbol{k} \cdot \boldsymbol{r}} &= \frac{i \Gamma(d / 2-\alpha+2)}{16(4 \pi)^{d / 2} \Gamma(\alpha)}\left(\frac{\boldsymbol{r}^{2}}{4}\right)^{\alpha-d / 2-3} \nonumber \\
    & \quad \times\left[\boldsymbol{r}^{2}\left(\delta_{i l} x_{j}+\delta_{j l} x_{i}+\delta_{i j} x_{l}\right)-(d-2 \alpha+4) x_{i} x_{j} x_{l}\right] \\
    \int \frac{d^{d} \boldsymbol{k}}{(2 \pi)^{d}} \frac{\boldsymbol{k}_{i} \boldsymbol{k}_{j} \boldsymbol{k}_{l} \boldsymbol{k}_{m}}{\left(\boldsymbol{k}^{2}\right)^{\alpha}} e^{i \boldsymbol{k} \cdot \boldsymbol{r}} & =\frac{\Gamma(d / 2-\alpha+3)}{32(4 \pi)^{d / 2} \Gamma(\alpha)}\left(\frac{\boldsymbol{r}^{2}}{4}\right)^{\alpha-d / 2-4}\Bigg[(d-2 \alpha+6) x_{i} x_{j} x_{l} x_{m} \nonumber \\
    & \quad  -\boldsymbol{r}^{2}\left(\delta_{i m} x_{j} x_{l}+\delta_{j m} x_{i} x_{l}+\delta_{l m} x_{i} x_{j}+\delta_{i l} x_{m} x_{j}+\delta_{j l} x_{i} x_{m}+\delta_{i j} x_{l} x_{m}\right) \nonumber \\
    & \quad +\frac{\left(\boldsymbol{r}^{2}\right)^{2}}{(d-2 \alpha+4)}\left(\delta_{i l} \delta_{j m}+\delta_{j l} \delta_{i m}+\delta_{i j} \delta_{l m}\right)\Bigg]
\end{align}

\section{Reproducing Schwarzschild Metric at $O\Big((m/r)^4\Big)$}\label{Reproducing Schwarzschild Metric}
As a consistency check, in this Appendix we show that the graviton one-point function in the EFT can actually reproduce the Schwarzschild metric perturbatively. We have:
\begin{itemize}
    \item $O\Big((m/r)\Big)$:
    \begin{equation}
         \vcenter{\hbox{\begin{tikzpicture}[scale=0.7]
        \begin{feynman}
            \vertex (i) at (0,0);
            \vertex (e) at (0,3);
            \vertex[dot, MyRed] (f1) at (0, 1.5) {};
            \vertex [dot, MyRed, label = $\boldsymbol{x}$] (c1) at (1.5,1.5) {};

            \diagram*{
                (i) -- [double, double, thick] (e),
                (f1) -- [dashed, MyRed, ultra thick] (c1),
            };
        \end{feynman}
    \end{tikzpicture}}}
    = - \frac{m}{r} ~.
    \end{equation}
    
    \item $O\Big((m/r)^2\Big)$:
    \begin{equation}
            \vcenter{\hbox{\begin{tikzpicture}[scale=0.7]
        \begin{feynman}
            \vertex (i) at (0, 0);
                  \vertex (e) at (0, 3);
                  \vertex[dot, MyRed] (w1) at (0, 2) {};
                  \vertex[dot, MyRed] (w2) at (0, 1) {};
                  \vertex (c1) at (1,1.5);
                  \vertex [dot, MyYellow, label = $\boldsymbol{x}$] (c2) at (2.5,1.5) {};
            
                  \diagram*{
                    (i) -- [double,double, thick] (e), 
                    (w1) -- [dashed, MyRed, ultra thick] (c1),
                    (w2) -- [dashed, MyRed, ultra thick] (c1),
                    (c1) -- [boson, MyYellow, ultra thick] (c2),  
                  };
        \end{feynman}
    \end{tikzpicture}}}
    = - \frac{1}{2} \(\frac{m}{r}\)^2 ~.
    \end{equation}
    
    \ $O((m/r)^3)$:
    % \itme $O\Big((m/r)^3\Big)$:
    \begin{equation}
                    \vcenter{\hbox{\begin{tikzpicture}[scale=0.7]
        \begin{feynman}
            \vertex (i) at (0, 0);
                  \vertex (e) at (0, 3);
                  \vertex[dot, MyRed] (w1) at (0, 2.2) {};
                  \vertex[dot, MyRed] (w2) at (0, 1.2) {};
                  \vertex[dot, MyRed] (w3) at (0, 0.5) {};
                  \vertex (c1) at (0.8,1.5);
                  \vertex (c2) at (2,1.5);
                  \vertex[dot, MyRed, label = $\boldsymbol{x}$] (c3) at (3.5,1.5) {} ;
            
                  \diagram*{
                    (i) -- [double,double, thick] (e), 
                    (w1) -- [dashed, MyRed, ultra thick] (c1),
                    (w2) -- [dashed, MyRed, ultra thick] (c1),
                    (c1) -- [boson, MyYellow, ultra thick] (c2),  
                    (w3) -- [dashed, MyRed, ultra thick] (c2),
                    (c2) -- [dashed, MyRed, ultra thick] (c3)
                  };
        \end{feynman}
    \end{tikzpicture}}}
    = - \frac{1}{12} \(\frac{m}{r}\)^3 ~.
    \end{equation}
    
    \item $O\Big((m/r)^4\Big)$:
    \begin{equation}
                \vcenter{\hbox{
            \begin{tikzpicture}[scale=0.7]
                \begin{feynman}
                  \vertex (i) at (0, 0);
                  \vertex (e) at (0, 3);
                  \vertex[dot, MyRed] (w1) at (0, 2.75) {};
                  \vertex[dot, MyRed] (w2) at (0, 2.00) {};
                  \vertex[dot, MyRed] (w3) at (0, 1.25) {};
                  \vertex[dot, MyRed] (w4) at (0, 0.5) {};
                  \vertex (c1) at (1, 2.375);
                  \vertex (c2) at (1, 0.875);
                  \vertex (c3) at (1.75, 1.5);
                  \vertex[dot, MyYellow, label = $\boldsymbol{x}$] (c4) at (3, 1.5) {};
            
                  \diagram*{
                    (i) -- [double,double, thick] (e), 
                    (w1) -- [dashed, MyRed, ultra thick] (c1),
                    (w2) -- [dashed, MyRed, ultra thick] (c1),
                    (c1) -- [boson, MyYellow, ultra thick] (c3),
                    (w3) -- [dashed, MyRed, ultra thick] (c2),
                    (w4) -- [dashed, MyRed, ultra thick] (c2),
                    (c2) -- [boson, MyYellow, ultra thick] (c3),
                    (c3) -- [boson, ultra thick, MyYellow] (c4), 
                  };
                \end{feynman}
            \end{tikzpicture}}} = \frac{1}{8} \(\frac{m}{r}\)^4 ~.
    \end{equation}
    
    \begin{equation}
                       \vcenter{\hbox{
                \begin{tikzpicture}[scale=0.7]
                    \begin{feynman}
                      \vertex (i) at (0, 0);
                      \vertex (e) at (0, 3);
                      \vertex[dot, MyRed] (w1) at (0, 2.75) {};
                      \vertex[dot, MyRed] (w2) at (0, 2.00) {};
                      \vertex[dot, MyRed] (w3) at (0, 1.25) {};
                      \vertex[dot, MyRed] (w4) at (0, 0.5) {};
                      \vertex (c1) at (1, 2.375);
                      \vertex (c2) at (1, 0.875);
                      \vertex[dot, MyYellow, label = $\boldsymbol{x}$] (c4) at (2.5, 0.875) {};
                
                      \diagram*{
                        (i) -- [double,double, thick] (e), 
                        (w1) -- [dashed, MyRed, ultra thick] (c1),
                        (w2) -- [dashed, MyRed, ultra thick] (c1),
                        (c1) -- [boson, MyYellow, ultra thick] (c2),
                        (w3) -- [dashed, MyRed, ultra thick] (c2),
                        (w4) -- [dashed, MyRed, ultra thick] (c2),
                        (c2) -- [boson, MyYellow, ultra thick] (c4),
                      };
                    \end{feynman}
                \end{tikzpicture}}} 
                = - \frac{1}{48} \(\frac{m}{r}\)^4 ~.
    \end{equation}
    
    \begin{equation}
                    \vcenter{\hbox{
                    \begin{tikzpicture}[scale=0.7]
                        \begin{feynman}
                          \vertex (i) at (0, 0);
                          \vertex (e) at (0, 3);
                          \vertex[dot, MyRed] (w1) at (0, 2.75) {};
                          \vertex[dot, MyRed] (w2) at (0, 2.00) {};
                          \vertex[dot, MyRed] (w3) at (0, 1.25) {};
                          \vertex[dot, MyRed] (w4) at (0, 0.25) {};
                          \vertex (c1) at (1, 2.375);
                          \vertex (c2) at (1, 0.975);
                          \vertex[dot, MyYellow, label = $\boldsymbol{x}$] (c4) at (2.5, 0.35) {};
                          \vertex (c5) at (1, 0.35);
                    
                          \diagram*{
                            (i) -- [double,double, thick] (e), 
                            (w1) -- [dashed, MyRed, ultra thick] (c1),
                            (w2) -- [dashed, MyRed, ultra thick] (c1),
                            (c1) -- [boson, MyYellow, ultra thick] (c2),
                            (w3) -- [dashed, MyRed, ultra thick] (c2),
                            (w4) -- [dashed, MyRed, ultra thick] (c5),
                            (c5) -- [boson, MyYellow, ultra thick] (c4),
                            (c2) -- [dashed, MyRed, ultra thick] (c5),
                          };
                        \end{feynman}
                    \end{tikzpicture}}}  
                = - \frac{1}{24} \(\frac{m}{r}\)^2 ~.
    \end{equation}
\end{itemize}

After combining all these results, we get the one-point function 
\begin{equation}
    \phi(\boldsymbol{x}) = - \frac{m}{r} - \frac{1}{12} \(\frac{m}{r}\)^3 ~, \quad \sigma(\boldsymbol{x}) = - \frac{1}{2} \(\frac{m}{r}\)^2 + \frac{1}{16} \(\frac{m}{r}\)^4 ~. 
\end{equation}
Substituting this into Eq.\eqref{KK Reduction}, we perturbatively derive the metric of Schwarzschild BH in isotropic gauge, 
\begin{align}
    g_{00} &= 1 - 2 \(\frac{m}{r}\) + 2 \(\frac{m}{r}\)^2 - \frac{3}{2} \(\frac{m}{r}\)^3 + \(\frac{m}{r}\)^4  + O\(\frac{m}{r}\)^5 ~, \\
    g_{ij} &= \( 1 + 2 \(\frac{m}{r}\) + \frac{3}{2} \(\frac{m}{r}\)^2 + \frac{1}{2} \(\frac{m}{r}\)^3 + \frac{1}{16} \(\frac{m}{r}\)^4 + O\(\frac{m}{r}\)^5 \) \delta_{ij}
\end{align}

\section{One-point Function of Spin-1 Electric Dipole}\label{Spin-1 1-pt Function}
In this appendix, we provide the explicit one-point function computation for the spin-1 electric dipole case.

\begin{itemize}
    \item $O(m/r)$:
    \begin{equation}
        \begin{aligned}
        \vcenter{\hbox{\begin{tikzpicture}[scale=0.7]
            \begin{feynman}
                \vertex (i) at (0,0);
                \vertex (e) at (0,3);
                \vertex[dot, MyRed] (w1) at (0, 1.5) {};
                \vertex [dot, MyRed] (c1) at (1.5,1.5);
                \vertex[dot, MyBlue, label = $\boldsymbol{x}$] (f1) at (3,2.8) {};
                \vertex[crossed dot, MyBlue] (fs) at (3,0.2) {};
    
                \diagram*{
                    (i) -- [double, double, thick] (e),
                    (w1) -- [dashed, MyRed, ultra thick] (c1),
                    (f1) -- [MyBlue, ultra thick] (c1),
                    (fs) -- [MyBlue, ultra thick] (c1)
                };
            \end{feynman}
        \end{tikzpicture}}}
        = \mathcal{E}_i x^i \(\frac{m}{2r}\) \times (-2)     
        \end{aligned}
    \end{equation}

    \item $O\((m/r)^2\)$:
    \begin{equation}
        \vcenter{\hbox{
            \begin{tikzpicture}[scale=0.7]
                \begin{feynman}
                  \vertex (i) at (0, 0);
                  \vertex (e) at (0, 3);
                  \vertex[dot, MyRed] (w1) at (0, 2) {};
                  \vertex[dot, MyRed] (w2) at (0, 1) {};
                  \vertex (c1) at (1,1.5);
                  \vertex (c2) at (2.5,1.5);
                  \vertex[dot, MyBlue, label= $\boldsymbol{x}$] (f1) at (3.5,2.8) {};
                  \vertex[crossed dot, MyBlue] (fs) at (3.5,0.2) {}; 
            
                  \diagram*{
                    (i) -- [double,double, thick] (e), 
                    (w1) -- [dashed, MyRed, ultra thick] (c1),
                    (w2) -- [dashed, MyRed, ultra thick] (c1),
                    (c1) -- [boson, MyYellow, ultra thick] (c2),
                    (c2) -- [ultra thick, MyBlue] (f1), 
                    (fs) -- [ultra thick, MyBlue, edge label' = $r$] (c2),  
                  };
                \end{feynman}
            \end{tikzpicture}}} = \mathcal{E}_i x^i \(\frac{m}{2r}\)^2 ~.
    \end{equation}
    
    \begin{equation}
        \begin{aligned}
        \vcenter{\hbox{
            \begin{tikzpicture}[scale=0.7]
                \begin{feynman}
                  \vertex (i) at (0, 0);
                  \vertex (e) at (0, 3);
                  \vertex[dot, MyRed] (w1) at (0, 2) {};
                  \vertex[dot, MyRed] (w2) at (0, 1) {};
                  \vertex (c1) at (1.5,1.5);
                  \vertex[dot, MyBlue, label= $\boldsymbol{x}$] (f1) at (3.0,2.8) {};
                  \vertex[crossed dot, MyBlue] (fs) at (3.0,0.2) {}; 
            
                  \diagram*{
                    (i) -- [double,double, thick] (e), 
                    (w1) -- [dashed, MyRed, ultra thick] (c1),
                    (w2) -- [dashed, MyRed, ultra thick] (c1),
                    (c1) -- [ultra thick, MyBlue] (f1), 
                    (fs) -- [ultra thick, MyBlue, edge label' = $r$] (c1),  
                  };
                \end{feynman}
            \end{tikzpicture}}} 
            =\mathcal{E}_i x^i \(\frac{m}{2r}\)^2\times (-8) ~.
        \end{aligned}
    \end{equation}

    \begin{equation}
        \begin{aligned}
            \vcenter{\hbox{
                \begin{tikzpicture}[scale=0.8]
                    \begin{feynman}
                      \vertex (i) at (0, 0);
                      \vertex (e) at (0, 3);
                      \vertex[dot, MyRed] (w1) at (0, 2.375) {};
                      \vertex[dot, MyRed] (w2) at (0, 0.875) {};
                      \vertex (c1) at (1.5, 2.375);
                      \vertex (c2) at (1.5, 0.875);
                      \vertex[dot, MyBlue, label= $\boldsymbol{x}$] (f1) at (2.5,2.8) {};
                      \vertex[crossed dot, MyBlue] (fs) at (2.5,0.2) {};

                      \diagram*{
                        (i) -- [double,double, thick] (e), 
                        (w1) -- [dashed, MyRed, ultra thick] (c1),
                        (w2) -- [dashed, MyRed, ultra thick] (c2),
                        (c1) -- [ultra thick, MyBlue] (f1), 
                        (fs) -- [ultra thick, MyBlue, edge label' = $r$] (c2), 
                        (c2) -- [ultra thick, MyBlue] (c1), 
                      };
                    \end{feynman}
                \end{tikzpicture}}}  
                = \mathcal{E}_i x^i \(\frac{m}{2r}\)^2 \times 8 ~.          
        \end{aligned}
    \end{equation}

    \item $O\((m/r)^3\)$:
    \begin{equation}
        \begin{aligned}
            \vcenter{\hbox{
                \begin{tikzpicture}[scale=0.8]
                    \begin{feynman}
                      \vertex (i) at (0, 0);
                      \vertex (e) at (0, 3);
                      \vertex[dot, MyRed] (w1) at (0, 2.5) {};
                      \vertex[dot, MyRed] (w2) at (0, 0.5) {};
                      \vertex[dot, MyRed] (w3) at (0,1.5) {};
                      \vertex (c1) at (1.5, 2.5);
                      \vertex (c2) at (1.5, 0.5);
                      \vertex (c3) at (1.5, 1.5);
                      \vertex[dot, MyBlue, label= $\boldsymbol{x}$] (f1) at (2.5,2.8) {};
                      \vertex[crossed dot, MyBlue] (fs) at (2.5,0.2) {};

                      \diagram*{
                        (i) -- [double,double, thick] (e), 
                        (w1) -- [dashed, MyRed, ultra thick] (c1),
                        (w2) -- [dashed, MyRed, ultra thick] (c2),
                        (w3) -- [dashed, MyRed, ultra thick] (c3),
                        (c1) -- [ultra thick, MyBlue] (f1), 
                        (fs) -- [ultra thick, MyBlue, edge label' = $r$] (c2), 
                        (c2) -- [ultra thick, MyBlue] (c1), 
                      };
                    \end{feynman}
                \end{tikzpicture}}} 
                +
                \vcenter{\hbox{
                    \begin{tikzpicture}[scale=0.7]
                        \begin{feynman}
                          \vertex (i) at (0, 0);
                          \vertex (e) at (0, 3);
                          \vertex[dot, MyRed] (w1) at (0, 1.5) {};
                          \vertex[dot, MyRed] (w2) at (0, 0.5) {};
                          \vertex[dot, MyRed] (w3) at (0, 2.5) {}; 
                          \vertex (c1) at (1,1.0);
                          \vertex (c2) at (2.5,1.0);
                          \vertex (c3) at (2.5, 2.5);
                          \vertex[dot, MyBlue, label= $\boldsymbol{x}$] (f1) at (3.5,2.8) {};
                          \vertex[crossed dot, MyBlue] (fs) at (3.5,0.2) {}; 
                    
                          \diagram*{
                            (i) -- [double,double, thick] (e), 
                            (w1) -- [dashed, MyRed, ultra thick] (c1),
                            (w2) -- [dashed, MyRed, ultra thick] (c1),
                            (w3) -- [dashed, MyRed, ultra thick] (c3),
                            (c1) -- [boson, MyYellow, ultra thick] (c2),
                            (c3) -- [ultra thick, MyBlue] (f1), 
                            (c3) -- [ultra thick, MyBlue] (c2),
                            (fs) -- [ultra thick, MyBlue, edge label' = $r$] (c2),  
                          };
                        \end{feynman}
                    \end{tikzpicture}}}
                    +
                    \vcenter{\hbox{
                        \begin{tikzpicture}[scale=0.7]
                            \begin{feynman}
                              \vertex (i) at (0, 0);
                              \vertex (e) at (0, 3);
                              \vertex[dot, MyRed] (w1) at (0, 1.5) {};
                              \vertex[dot, MyRed] (w2) at (0, 0.5) {};
                              \vertex[dot, MyRed] (w3) at (0, 2.5) {};
                              \vertex (c1) at (1.5,1.0);
                              \vertex (c2) at (1.5, 2.5);
                              \vertex[dot, MyBlue, label= $\boldsymbol{x}$] (f1) at (3.0,2.8) {};
                              \vertex[crossed dot, MyBlue] (fs) at (3.0,0.2) {}; 
                        
                              \diagram*{
                                (i) -- [double,double, thick] (e), 
                                (w1) -- [dashed, MyRed, ultra thick] (c1),
                                (w2) -- [dashed, MyRed, ultra thick] (c1),
                                (w3) -- [dashed, MyRed, ultra thick] (c2),
                                (c1) -- [ultra thick, MyBlue] (c2),
                                (c2) -- [ultra thick, MyBlue] (f1), 
                                (fs) -- [ultra thick, MyBlue, edge label' = $r$] (c1),  
                              };
                            \end{feynman}
                        \end{tikzpicture}}} 
                        = - \mathcal{E}_i x^i \(\frac{m}{2r}\)^3 \log\(r\mu\) \times \frac{4}{3} ~.
        \end{aligned}
    \end{equation}

    \begin{equation}
        \begin{aligned}
            \vcenter{\hbox{
                \begin{tikzpicture}[scale=0.7]
                    \begin{feynman}
                      \vertex (i) at (0, 0);
                      \vertex (e) at (0, 3);
                      \vertex[dot, MyRed] (w1) at (0, 2.5) {};
                      \vertex[dot, MyRed] (w2) at (0, 1.5) {};
                      \vertex[dot, MyRed] (w3) at (0, 0.5) {};
                      \vertex (c1) at (1,2);
                      \vertex (c2) at (2.5,2);
                      \vertex (c3) at (2.5,0.5);
                      \vertex[dot, MyBlue, label= $\boldsymbol{x}$] (f1) at (3.5,2.8) {};
                      \vertex[crossed dot, MyBlue] (fs) at (3.5,0.2) {}; 
                
                      \diagram*{
                        (i) -- [double,double, thick] (e), 
                        (w1) -- [dashed, MyRed, ultra thick] (c1),
                        (w2) -- [dashed, MyRed, ultra thick] (c1),
                        (w3) -- [dashed, MyRed, ultra thick] (c3),
                        (c1) -- [boson, MyYellow, ultra thick] (c2),
                        (c2) -- [ultra thick, MyBlue] (c3),
                        (c2) -- [ultra thick, MyBlue] (f1), 
                        (fs) -- [ultra thick, MyBlue, edge label' = $r$] (c3),  
                      };
                    \end{feynman}
                \end{tikzpicture}}}
                +
                \vcenter{\hbox{
                    \begin{tikzpicture}[scale=0.7]
                        \begin{feynman}
                          \vertex (i) at (0, 0);
                          \vertex (e) at (0, 3);
                          \vertex[dot, MyRed] (w1) at (0, 2.5) {};
                          \vertex[dot, MyRed] (w2) at (0, 1.5) {};
                          \vertex[dot, MyRed] (w3) at (0, 0.5) {};
                          \vertex (c1) at (1.5,2);
                          \vertex (c3) at (1.5,0.5);
                          \vertex[dot, MyBlue, label= $\boldsymbol{x}$] (f1) at (3.0,2.8) {};
                          \vertex[crossed dot, MyBlue] (fs) at (3.0,0.2) {}; 
                    
                          \diagram*{
                            (i) -- [double,double, thick] (e), 
                            (w1) -- [dashed, MyRed, ultra thick] (c1),
                            (w2) -- [dashed, MyRed, ultra thick] (c1),
                            (w3) -- [dashed, MyRed, ultra thick] (c3),
                            (c1) -- [ultra thick, MyBlue] (c3),
                            (c1) -- [ultra thick, MyBlue] (f1), 
                            (fs) -- [ultra thick, MyBlue, edge label' = $r$] (c3),  
                          };
                        \end{feynman}
                    \end{tikzpicture}}}
                    = \mathcal{E}_i x^i \(\frac{m}{2r}\)^{3} \log\(r\mu\) \times \frac{28}{3} ~.    
        \end{aligned}
    \end{equation}
    
    \begin{equation}
    \vcenter{\hbox{\begin{tikzpicture}[scale=0.7]
        \begin{feynman}
            \vertex (i) at (0, 0);
                  \vertex (e) at (0, 3);
                  \vertex[dot, MyRed] (w1) at (0, 2.2) {};
                  \vertex[dot, MyRed] (w2) at (0, 1.2) {};
                  \vertex[dot, MyRed] (w3) at (0, 0.5) {};
                  \vertex (c1) at (0.8,1.5);
                  \vertex (c2) at (2,1.5);
                  \vertex (c3) at (3.5,1.5);
                  \vertex[dot, MyBlue, label = $\boldsymbol{x}$] (f1) at (4.5,2.8) {};
                  \vertex[crossed dot, MyBlue] (fs) at (4.5,0.2) {};
            
                  \diagram*{
                    (i) -- [double,double, thick] (e), 
                    (w1) -- [dashed, MyRed, ultra thick] (c1),
                    (w2) -- [dashed, MyRed, ultra thick] (c1),
                    (c1) -- [boson, MyYellow, ultra thick] (c2),  
                    (w3) -- [dashed, MyRed, ultra thick] (c2),
                    (c2) -- [dashed, MyRed, ultra thick] (c3),
                    (c3) -- [ultra thick, MyBlue] (f1),
                    (fs) -- [ultra thick, MyBlue] (c3),
                  };
        \end{feynman}
    \end{tikzpicture}}}
    = - \mathcal{E}_i x^i \(\frac{m}{2r}\)^3 \log\(r\mu\) \times \frac{4}{3} ~.
    \end{equation}

    \begin{equation}
            \vcenter{\hbox{\begin{tikzpicture}[scale=0.7]
        \begin{feynman}
            \vertex (i) at (0, 0);
                  \vertex (e) at (0, 3);
                  \vertex[dot, MyRed] (w1) at (0, 2.2) {};
                  \vertex[dot, MyRed] (w2) at (0, 1.2) {};
                  \vertex[dot, MyRed] (w3) at (0, 0.5) {};
                  \vertex (c1) at (0.8,1.5);
                  \vertex (c2) at (2,1.5);
                  \vertex[dot, MyBlue, label = $\boldsymbol{x}$] (f1) at (3.5,2.8) {};
                  \vertex[crossed dot, MyBlue] (fs) at (3.5,0.2) {};
            
                  \diagram*{
                    (i) -- [double,double, thick] (e), 
                    (w1) -- [dashed, MyRed, ultra thick] (c1),
                    (w2) -- [dashed, MyRed, ultra thick] (c1),
                    (c1) -- [boson, MyYellow, ultra thick] (c2),  
                    (w3) -- [dashed, MyRed, ultra thick] (c2),
                    (c2) -- [ultra thick, MyBlue] (f1),
                    (fs) -- [ultra thick, MyBlue] (c2),
                  };
        \end{feynman}
    \end{tikzpicture}}}
     = \mathcal{E}_i x^i \(\frac{m}{2r}\)^{3} \log\(r\mu\) \times 4 ~. 
    \end{equation}
    
    \begin{equation}
                        \vcenter{\hbox{
            \begin{tikzpicture}[scale=0.7]
                \begin{feynman}
                  \vertex (i) at (0, 0);
                  \vertex (e) at (0, 3);
                  \vertex[dot, MyRed] (w1) at (0, 2.5) {};
                  \vertex[dot, MyRed] (w2) at (0, 1.5) {};
                  \vertex[dot, MyRed] (w3) at (0,0.5) {};
                  \vertex (c1) at (1.5,1.5);
                  \vertex[dot, MyBlue, label = $\boldsymbol{x}$] (f1) at (3.0,2.8) {};
                  \vertex[crossed dot, MyBlue] (fs) at (3.0,0.2) {}; 
            
                  \diagram*{
                    (i) -- [double,double, thick] (e), 
                    (w1) -- [dashed, MyRed, ultra thick] (c1),
                    (w2) -- [dashed, MyRed, ultra thick] (c1),
                    (w3) -- [dashed, MyRed, ultra thick] (c1),
                    (c1) -- [ultra thick, MyBlue] (f1), 
                    (fs) -- [ultra thick, MyBlue] (c1),  
                  };
                \end{feynman}
            \end{tikzpicture}}}
           = - \mathcal{E}_i x^i \(\frac{m}{2r}\)^3\log\(r\mu\) \times \frac{32}{3} ~.
    \end{equation}
\end{itemize}
    
\section{Teukolsky Equation in Schwarzschild BH}\label{Teukolsky Equation in Schwarzschild BH}

\subsection{Equation In Different Coordinates}

We derive now the Teukolsky master equation for radial functions in an isotropic coordinate. The full spin-s Teukolsky equation in the Boyer–Lindquist coordinate of Kerr BH for a generic spin-s field reads~\cite{teukolsky1972rotating,teukolsky1973perturbations,press1973perturbations,teukolsky1974perturbations}
\begin{equation}
    \begin{aligned}
        &{\left[\frac{\left(r^{2}+a^{2}\right)^{2}}{\Delta}-a^{2} \sin ^{2} \theta\right] \frac{\partial^{2} \psi^{[s]}}{\partial t^{2}}+\frac{4 M a r}{\Delta} \frac{\partial^{2} \psi^{[s]}}{\partial t \partial \phi}+\left[\frac{a^{2}}{\Delta}-\frac{1}{\sin ^{2} \theta}\right] \frac{\partial^{2} \psi^{[s]}}{\partial \phi^{2}}} \\
        &-\Delta^{-s} \frac{\partial}{\partial r}\left(\Delta^{s+1} \frac{\partial \psi^{[s]}}{\partial r}\right)-\frac{1}{\sin \theta} \frac{\partial}{\partial \theta}\left(\sin \theta \frac{\partial \psi^{[s]}}{\partial \theta}\right)-2 s\left[\frac{a(r-M)}{\Delta}+\frac{i \cos \theta}{\sin ^{2} \theta}\right] \frac{\partial \psi^{[s]}}{\partial \phi} \\
        &-2 s\left[\frac{M\left(r^{2}-a^{2}\right)}{\Delta}-r-i a \cos \theta\right] \frac{\partial \psi^{[s]}}{\partial t}+\left(s^{2} \cot ^{2} \theta-s\right) \psi^{[s]} = 0 ~,
    \end{aligned}
\end{equation} 
where $\Delta = (r-r_+)(r-r_-)$, $r_+ = M + \sqrt{M^2 - a^2}$, and $r_- = M - \sqrt{M^2 - a^2}$. In the Schwarzschild limit $a=0$, and in the static limit $\partial_t \rightarrow 0$, we use the variable separation ansatz $\psi^{[s]} = \sum_{\ell m} \mathcal{R}_\ell (r) {}_s Y_{\ell m}(\theta,\phi)$ and get the radial equation in the Schwarzschild coordinates:
\begin{equation}
    \mathcal{R}''_{\ell}(r) + \(\frac{1+s}{r} + \frac{1+s}{-2M + r}\)\mathcal{R}'_{\ell}(r) +\( -\frac{(s-\ell)(1+s+\ell)}{2Mr} + \frac{(s-\ell)(1+s+\ell)}{2M(-2M+ r)}\) \mathcal{R}_\ell(r) =0 ~.
\end{equation}
Transforming from the Schwarzschild coordinates to the isotropic coordinates, we get the version of this equation in the isotropic coordinates,
\begin{equation}
    \mathcal{R}''_{\ell}(r) + \(- \frac{2+4s}{M-2r} - \frac{2s}{r} + \frac{2+4s}{M+2r}\)\mathcal{R}'_{\ell}(r) + \frac{(s-\ell)(1+s+\ell)}{r^2} \mathcal{R}_\ell(r) = 0~.
\end{equation}

For the finite frequency perturbations in the Schwarzschild background we 
use the variable separation ansatz $\psi^{[s]} = \sum_{\ell m}\mathcal{R}_\ell(r) \, {}_s Y_{\ell m}(\theta,\phi) e^{-i\omega t}$, and the corresponding radial equation in the Schwarzschild coordinates reads 
\begin{equation}
    \begin{aligned}
        &r(r-2M)\mathcal{R}_{\ell}''(r) + 2(r-M)(1+s)\mathcal{R}_{\ell}'(r) + \Bigg(s(1+s)-\ell(1+\ell) \\
        & \quad +4isr\omega + \frac{-2is\omega r^2(r-M)+r^4 \omega^2}{r(r-2M)}\Bigg)\mathcal{R}_{\ell}(r) =0~.
    \end{aligned}
\end{equation}
In the near zone region $\omega r \ll 1$, the above equation simplifies: 
\begin{equation}\label{Near Zone Teukolsky Schwarzschild Coordiante}
    r(r-2M)\mathcal{R}_{\ell}''(r) + 2(r-M)(1+s)\mathcal{R}_{\ell}'(r) + \Bigg(s(1+s)-\ell(1+\ell) + \frac{-8is M^3 \omega+16 M^4 \omega^2}{r(r-2M)}\Bigg)\mathcal{R}_{\ell}(r) =0~.
\end{equation}
In the isotropic coordinates, the finite frequency radial equation takes
the following form:  
\begin{equation}\label{Finite Frequency Teukoleky}
    \mathcal{R}_{\ell}''(r) + \(- \frac{2+4s}{M-2r} - \frac{2s}{r} + \frac{2+4s}{M+2r}\) \mathcal{R}_{\ell}'(r) + \(\frac{(s-\ell)(1+s+\ell)}{r^2} + \frac{F(\omega,r)}{(4r^2-M^2)^2}\)\mathcal{R}_{\ell}(r) =0~,
\end{equation}
where
\begin{equation}
    F(\omega, r) = \frac{(M+2r)^4 (8is\omega r(M^2 - 8Mr + 4r^2) + (M+2r)^4 \omega^2)}{16 r^4} ~.
\end{equation}
In the near zone region, we use the approximation $F(\omega,M/2)\approx 128 M^3 \omega (-is+2M\omega)$.

\subsection{Matching Dissipation Number In Schwarzschild Coordinate}\label{Matching Dissipation Number in Schwarzschild Coordinate}
Based on the near zone spin-s Teukolsky equation in Schwarzschild coordinate Eq.\eqref{Near Zone Teukolsky Schwarzschild Coordiante}, we introduce the following raising and lowering operator to explicitly see the ${\rm SL}(2,\mathbb{R})$ symmetry:
\begin{equation}
    \begin{aligned}
        L_1 & =  \exp\(\frac{t}{4M}\) \( \Delta^{1/2} \partial_r - 4 M (r-M) \Delta^{-1/2} \partial_t + 2 (r- M) s \Delta^{-1/2} \) ~, \\
        L_{-1} &= - \exp\(-\frac{t}{4M}\) \( \Delta^{1/2} \partial_r + 4 M (r-M) \Delta^{-1/2} \partial_t \) ~,\\
        L_0 &= -4  M \partial_t + s ~,
    \end{aligned}
\end{equation}
where $\Delta = r(r-2M)$. These operators obey the ${\rm SL}(2,\mathbb{R})$ commutation relations
\begin{equation}
    [L_0, L_{\pm 1}] = \mp L_{\pm 1} ~, \quad [L_1,L_{-1}] = 2 L_0 ~.
\end{equation}
The near zone equation Eq.\eqref{Near Zone Teukolsky Schwarzschild Coordiante} can be written as 
\begin{equation}
    \mathcal{C}_2 \psi^{[s]} = \ell(\ell + 1)\psi^{[s]} ~, 
\end{equation}
where $\mathcal{C}_2$ is the quadratic Casimir operator. The ingoing boundary condition at the event horizon in these coordinates takes the form
\begin{equation}
    R_{\ell}(r) = {\rm const} \times \(r - 2M\)^{-2iM \omega-s} ~, \quad r\rightarrow 2M ~.
\end{equation}
It is useful to introduce the near zone variable $z=(r-2M)/2M$ and rewrite Eq.\eqref{Near Zone Teukolsky Schwarzschild Coordiante} as 
\begin{equation}
    z(1+z)R_{\ell}''(z)+ (1+s)(1+2z)R_{\ell}'(z) + \((s-\ell)(1+s+\ell) - \frac{-2is M\omega + 4 (M\omega)^2}{z(1+z)}\) R_{\ell}(z) = 0~.
\end{equation}
The solution that satisfies the ingoing boundary condition is
\begin{equation}
    R_{\ell}^{\rm full}(z) = z^{-s/2}(1+z)^{-s/2} P_\ell^{s+4iM\omega}(1+2z) ~,
\end{equation}
where $P_n^m(z)$ is the associated Legendre function. We perform the asymptotic expansion of $R_{\ell}^{\rm full}$ and find that the coefficient in front of the $r^{\ell-s} \times r^{-2\ell-1}$ term is the same as Eq.~\eqref{Dissipative Response Coefficient}. This will lead to the same dissipation number as in Eq.~\eqref{Dissipation Number Master Formula}.

\subsection{Comments on Maxwell-Newman-Penrose $\tilde{\Phi}_0$}\label{Comments on Maxwell-Newman-Penrose Phi_0}
In this appendix, we show it is sufficient 
to consider $A_0$
for the spin-1 dissipation number matching. The Newman-Penrose-Maxwell 
scalar $\tilde{\Phi}_0$ in Eq.\eqref{Rescaled Maxwell-Newman-Penrose Scalars} takes the following
form in the isotropic coordinates:
\begin{equation}\label{Full Expression of Phi-0}
    \tilde{\Phi}_0 = 2 \sqrt{2} \frac{r}{(M-2r)^2} \(F_{0\theta} + i \frac{1}{\sin\theta} F_{0\phi}\) - 8 \sqrt{2} \frac{r^3}{(M-2r)(M+2r)^3} \(F_{r\theta} + i \frac{1}{\sin\theta} F_{r\phi}\) ~.
\end{equation}
In the quasi-static approximation we can ignore the temporal components. 
Let us 
focus on the magnetic part of the vector potential
satisfying $\partial_i A^i = 0$. It is convenient to rewrite is as
\begin{equation}
    A_i = \epsilon_{ijk} x^j \partial^k \Psi ~,
\end{equation}
which is automatically transverse. In the spherical coordinates, the vector potential can be written as 
\begin{equation}
    \begin{aligned}
        A_r &= 0 ~, \quad A_\theta = -r^2 \sin\theta\times r \frac{1}{r^2 \sin^2\theta} \partial_\phi \Psi = - \frac{r}{\sin\theta} \partial_\phi \Psi ~, \\
        A_\phi &= r^2 \sin\theta \times r \times \frac{1}{r^2} \partial_\theta \Psi = r \sin\theta \partial_\theta \Psi ~.
    \end{aligned}
\end{equation}
Plugging this into Eq.~\eqref{Full Expression of Phi-0} and taking the long-distance limit, we obtain 
\begin{equation}
    \tilde{\Phi}_0 \sim \frac{1}{r} \eth^0 \({A}_0 + i \Psi \) \,.
\end{equation}
This gives us the simplification in the matching procedure because the response of the electric and magnetic fields are separated. 
When applying the electric source, there is no ambiguity: the real part of the radial function of $\tilde{\Phi}_0$ corresponds to the electric conservative response, i.e., the Love number, while the imaginary part corresponds to the dissipation number.

\section{Matching Dissipation Numbers From Amplitudes}\label{app:abs}

In this appendix, we compute the EFT absorption cross section
for the spin-2 perturbations.

\subsection{Fluctuation-dissipation relation}

Let us first prove the fluctuation-dissipation theorem for 
Schwarzschild black holes~\cite{Goldberger:2005cd}.
To do this, we will use positive and negative frequency Wightman functions defined as 
\begin{equation}
    \begin{aligned}
        {W_+}^{L'}_{L}(\tau-\tau') = \langle Q_L(\tau) Q^{L'}(\tau') \rangle ~, \\
        {W_-}^{L'}_{L}(\tau-\tau') = \langle Q^{L'}(\tau') Q_{L}(\tau) \rangle ~.
    \end{aligned}
\end{equation}
With this definition, we can easily rewrite the Feynman Green function and the retarded Green functions in terms of Wightman functions,
\begin{align}
    {G_{\rm Fey}}^{L'}_L(\tau-\tau') &=  \theta(\tau-\tau') {W_+}^{L'}_{L}(\tau-\tau') + \theta(\tau'-\tau) {W_-}^{L'}_L(\tau-\tau') ~, \\
    {G_{\rm ret}}^{L'}_L(\tau-\tau') &= i \theta(\tau-\tau') \({W_+}^{L'}_L(\tau-\tau') - {W_-}^{L'}_L(\tau-\tau')\) ~.
\end{align}
We assume that $Q_{L}$ is a Hermitian operator, 
\begin{equation}
    {W_+}^{L'}_{L}(\tau-\tau')^* = {W_-}^{L'}_{L}(\tau-\tau') ~,
\end{equation}
and thus 
\begin{equation}
    \langle T Q_L(\tau) Q^{L'}(\tau') \rangle^* = \langle \bar{T} Q_L(\tau) Q^{L'}(\tau') \rangle ~,
\end{equation}
where $\bar{T}$ is the anti-time ordering operator. With the above definitions, we can also rewrite the positive frequency Wightman function in terms of the Feynman Green function,
\begin{equation}\label{Positive Wightman in Feynman}
    {W_+}_L^{L'}(\tau-\tau') = \theta(\tau-\tau'){G_{\rm Fey}}^{L'}_L(\tau-\tau') + \theta(\tau'-\tau){G_{\rm Fey}}_L^{L'}(\tau-\tau')^*
\end{equation}
In the frequency space, we will use the dispersive representation 
\begin{equation}
   {G_{\rm ret}}_L^{L'}(\omega) = i \int \frac{d\omega'}{2\pi} \frac{{W_+}^{L'}_L(\omega') - {W_-}^{L'}_{L}(\omega')}{\omega - \omega' + i\epsilon} ~.
\end{equation}
After expanding this relation into real and imaginary parts, we get 
\begin{equation}
    {\rm Re} {G_{\rm ret}}^{L'}_L (\omega) = -\frac{1}{2} {\rm Im}\[ {W_+}^{L'}_L(\omega) - {W_-}^{L'}_{L}(\omega) \] - {\rm Pr} \int_0^{\infty} \frac{\omega' d\omega'}{\pi} \frac{{\rm Re}\[ {W_+}^{L'}_L(\omega') - {W_-}^{L'}_{L}(\omega') \]}{\omega^2-{\omega'}^2} ~,
\end{equation}
and 
\begin{equation}\label{Retarded Imaginary Part}
    {\rm Im} {G_{\rm ret}}^{L'}_L (\omega) = \frac{1}{2} {\rm Re}\[ {W_+}^{L'}_L(\omega) - {W_-}^{L'}_{L}(\omega) \] - \omega {\rm Pr}\int_0^\infty \frac{d\omega'}{\pi} \frac{{\rm Im} \[ {W_+}^{L'}_L(\omega') - {W_-}^{L'}_{L}(\omega') \]}{\omega^2 - {\omega'}^2} ~.
\end{equation}
Here, we mention clearly that the above equation holds for a generic tensor structure in $L$ and $L'$. 
For the special situation of Schwarzschild BHs
the tensorial structure consistent with 
spherical symmetry is only $\delta^{\langle L \rangle}_{\langle L' \rangle}$,
i.e., 
\be 
{W_+}^{L'}_L(\omega)=w_+(\omega)\delta^{\langle L \rangle}_{\langle L' \rangle}\,.
\ee
In this case one can show that ${{W_{+}}_L^{L'}}(\omega)^* = {{W_{+}}_L^{L'}(\omega)}$, and thus ${W_{+}}_L^{L'}(\omega)$ is real. 

Now we make use of Eq.~\eqref{Positive Wightman in Feynman}, and express
the positive frequency Wightman function as,
\be
\label{eq:fd1}
\begin{split}
{W_+}_L^{L'}(\omega) &= \int_0^{+\infty} d\tau e^{i\omega \tau} \langle T Q_L(\tau) Q^{L'}(0) \rangle + \int_{-\infty}^0 d\tau e^{i\omega \tau} \langle T Q_{L}(\tau) Q^{L'}(0) \rangle^* \\
        & =  \int_0^{+\infty} d\tau e^{i\omega \tau} \langle T Q_L(\tau) Q^{L'}(0) \rangle + \int_0^{+\infty} d\tau e^{-i\omega \tau} \langle T Q_L(-\tau) Q^{L'}(0) \rangle^* \\
        & = \int_0^{+\infty} d\tau e^{i\omega \tau} \langle T Q_L(\tau) Q^{L'}(0) \rangle + \int_0^{+\infty} d\tau e^{-i\omega \tau} \langle T Q_L(0) Q^{L'}(\tau) \rangle^* \\
        & = 2 {\rm Im} \[ i \int_{0}^{+\infty} d\tau e^{i \omega \tau} \langle T Q_{L}(\tau) Q^{L'}(0) \rangle \] \,,
\end{split}
\ee
where we have used the translation invariance of the Green function
\be
 \langle T Q_L(-\tau) Q^{L'}(0) \rangle 
 = \langle T Q_L(0) Q^{L'}(\tau) \rangle\,,
\ee
and the definition of the Feynman Green function that implies 
invariance under the exchange 
of the time argument, 
\be
 \langle T Q_L(\tau) Q^{L'}(0) \rangle = \langle T Q_L(0) Q^{L'}(\tau) \rangle\,.
\ee
Since we are interested in classical black holes, 
we assume the Boulware state, so that the response is purely absorptive, 
\be
w_+(\omega<0) = 0\,. 
\ee
Thus, for $\omega>0$
we have from~\eqref{eq:fd1}:
\be
\begin{split}
0 & = {W_+}_L^{L'}(-\omega)=
2 {\rm Im} \[ i \int_{0}^{+\infty} d\tau e^{-i \omega \tau} \langle T Q_{L}(\tau) Q^{L'}(0) \rangle \] = 2 {\rm Im} \[ i \int_{-\infty}^{0} d\tau e^{i \omega \tau} \langle T Q_{L}(\tau) Q^{L'}(0) \rangle \] \,,
\end{split} 
\ee
which then allows us to complete the integral in~\eqref{eq:fd1}
and finally obtain
 \be
\label{eq:fd3}
\begin{split}
{W_+}_L^{L'}(\omega) = 2 {\rm Im} \[ i \int_{-\infty}^{+\infty} d\tau e^{i \omega \tau} \langle T Q_{L}(\tau) Q^{L'}(0) \rangle \]~, \quad \omega>0\,.
\end{split}
\ee
Now we can establish the relationship between the Feynman and retarded Green functions in the EFT. Using Eq.~\eqref{Retarded Imaginary Part}
we get
\be
\begin{split}
& \text{Im}G_{\rm ret}{}^{L'}_L (\omega)= \text{Re}G_{\rm Fey}{}^{L'}_L(\omega) \,,
% & \text{Re}G_{\rm ret}{}^{L'}_L = \text{Re}G_{\rm Feyn}{}^{L'}_L \,.
\end{split}
\ee
which fixes the odd frequency terms in Eq.~\eqref{eq:Fey}.

\subsection{Spin-2 Absorption Cross Section in the EFT}

% In order to show the equivalence of computing dissipation number from in-out and in-in formalism, in this appendix, we will use the spin-2 quadrupole case as an example to match the dissipation number from BH absorption cross section. In the EFT side, according to the finite size effective action in Eq.\eqref{Finite Size Action} and the definition of electric field 
We start with the 
definition of the electric-type tidal field,
\begin{equation}\label{Gravtitation Electric Field}
    E_{ij} = 2\sqrt{2} M_{\rm pl} C_{0i0j} \simeq - \sqrt{2} M_{\rm pl} \partial_0^2 h_{ij} \,.
\end{equation}
The electric part of the absorption cross section can be 
obtained from the optical theorem:
it is given by the imaginary part of the forward amplitude
\begin{equation}
    \sigma_{\rm abs}^{E}(\omega) = 
    \frac{2}{\omega} \times \frac{1}{2} \times 2{\rm Im} i \int d\tau e^{i\omega \tau} \[\omega^4 \epsilon^{*ij}(\boldsymbol{k},h)\epsilon_{kl}(\boldsymbol{k},h) \langle T Q_{ij}(\tau) Q^{kl}(0) \rangle \]  \,,
\end{equation}
where $1/\omega$ is the phase space factor and $1/2$ comes from the Taylor expansion of the S-matrix.
$\epsilon_{ij}(\boldsymbol{k},h)$ above is the polarization tensor. 
The first factor of $2$ comes from the definition of the tidal field~\eqref{Gravtitation Electric Field}, and the second factor of $2$ comes from the symmetry factor of the Feynman diagrams.
We choose $\boldsymbol{k}//\hat{\boldsymbol{z}}$, and then the polarization tensor takes the form
\begin{equation}
        \epsilon_{ij}(\boldsymbol{k},\pm 2) = 
    \begin{pmatrix}
        \frac{1}{2} & \pm \frac{i}{2} & 0 \\
        \pm \frac{i}{2} & - \frac{1}{2} & 0 \\
        0 & 0 & 0
    \end{pmatrix}\,.
\end{equation}
Using and the explicit expressions~\eqref{eq:Fey} and  
\begin{equation}
    \delta^{\langle kl \rangle}_{\langle ij \rangle} = \frac{1}{2} \[\delta_{i}^k \delta_j^l + \delta_i^l \delta_j^k - \frac{2}{3}\delta_{ij}\delta^{kl} \] ~,
\end{equation}
we find the electric part of the total cross section
\begin{equation}
    \sigma_{\rm abs}^{E}(\omega) =  4\omega^4 \lambda_{1}^{\text{non-loc.}}|_{\ell=s=2} ~, \quad \omega>0 \,.
\end{equation}
The magnetic part of the total cross section is the same as the electric part thanks to the ``electric-magnetic" duality \cite{Goldberger:2005cd,Porto:2007qi,Hui:2020xxx} of the Schwarzschild black holes.
Thus, the total cross section is given by
\begin{equation}
    \sigma_{\rm abs}(\omega) = 2 \sigma_{\rm abs}^{E}(\omega) = 8 \omega^4 \lambda_{1}^{\text{non-loc.}}|_{\ell=s=2}~.
\end{equation}
% Comparing this to the BH perturbation theory result~\cite{starobinskii1973amplification, page1976particle},
% \begin{equation}
%     \sigma_{\rm abs}(\omega) = \frac{1}{45} 4\pi (2M)^6 \omega^4 ~,
% \end{equation}
% we get the dissipation number~
% \begin{equation}
%     \lambda_{1}^{\text{non-loc.}}|_{\ell=s=2}= \frac{16\pi \times 4!}{5!!\times(3!!)^2}M^6 ~,
% \end{equation}
% which is consistent with the 1-point function matching result~\eqref{eq:s2diss}.

\bibliographystyle{JHEP}
\bibliography{references}
\end{document}